\documentclass{aa}  
 \usepackage[varg]{txfonts}

\usepackage{orcidlink}

\usepackage{natbib}
\bibpunct{(}{)}{;}{a}{}{,}
\usepackage{graphicx}
\usepackage{subcaption}
\usepackage{txfonts}
\usepackage{url}
\usepackage{hyperref}
\hypersetup{
     colorlinks=true, 
     linkcolor=red,  
     filecolor=magenta,   
     citecolor=blue,   
     urlcolor=blue
     }
\begin{document} 

   \title{Cluster Ages to Reconstruct the Milky Way Assembly (CARMA). IV.}
\subtitle{Chrono-dynamics of 7 old LMC star clusters and the peculiar origin of NGC~1841 
}

      \author{F. Niederhofer\inst{1}\orcidlink{0000-0002-4341-9819}
          \and
          D. Massari\inst{2}\orcidlink{0000-0001-8892-4301}
          \and 
          F. Aguado-Agelet\inst{3}\fnmsep\inst{4}\orcidlink{0000-0003-2713-1943}
          \and
          S. Cassisi\inst{5}\fnmsep\inst{6}\orcidlink{0000-0001-5870-3735}
          \and
          A. Bellini\inst{7}\orcidlink{0000-0003-3858-637X}
          \and
          V. Kozhurina-Platais\inst{8}\orcidlink{0000-0003-0218-386X}
          \and 
          \\
          M. Libralato\inst{9}\orcidlink{0000-0001-9673-7397}
          \and
          N. Kacharov\inst{1}\orcidlink{0000-0002-6072-6669}
          \and
          A. Mucciarelli\inst{10}\fnmsep\inst{2}\orcidlink{0000-0001-9158-8580}
          \and 
          M. Monelli\inst{11,5}
          \and
          N. Bastian\inst{12,13}\orcidlink{0000-0001-5679-4215}
          \and
          I. Cabrera-Ziri\inst{14}\orcidlink{0000-0001-9478-5731}
          \and
          \\
          E. Ceccarelli\inst{2,10}
          \and 
          M.-R. L. Cioni\inst{1}\orcidlink{0000-0002-6797-696X}
          \and
          F. Dresbach\inst{15}\orcidlink{0000-0003-0808-8038}
          \and
          M. Häberle\inst{16}\orcidlink{0000-0002-5844-4443}
          \and 
          S. Martocchia\inst{17}\orcidlink{0000-0001-7110-6775}
          \and
          S. Saracino\inst{18,19}\orcidlink{0000-0003-4746-6003}
      }

   \institute{Leibniz-Institut für Astrophysik Potsdam, An der Sternwarte 16, D-14482 Potsdam, Germany\\
        \email{fniederhofer@aip.de}
        \and INAF – Osservatorio di Astrofisica e Scienza dello Spazio di Bologna, Via Gobetti 93/3, 40129 Bologna, Italy\
        \and atlanTTic, Universidade de Vigo, Escola de Enxeñaría de Telecomunicación, 36310 Vigo, Spain\
        \and Universidad de La Laguna, Avda. Astrofísico Fco. Sánchez, 38205 La Laguna, Tenerife, Spain\
        \and INAF – Osservatorio Astronomico d' Abruzzo, Via M. Maggini, 64100 Teramo, Italy\
        \and INFN – Sezione di Pisa, Universitá di Pisa, Largo Pontecorvo 3, 56127 Pisa, Italy\
        \and Space Telescope Science Institute, 3700 San Martin Drive, Baltimore, MD 21218, USA\
        \and Eureka Scientific, Inc., 2452 Delmer Street Suite 100, Oakland, CA 94602-3017, USA\
        \and INAF - Osservatorio Astronomico di Padova, Vicolo dell'Osservatorio 5, Padova I-35122, Italy\
        \and Dipartimento di Fisica e Astronomia, Università degli Studi di Bologna, via Gobetti 93/2, I-40129 Bologna, Italy
        \and INAF – Osservatorio Astronomico di Roma, Via Frascati 33, 00078 Monte Porzio Catone, Roma, Italy
        \and Donostia International Physics Center (DIPC), Paseo Manuel de Lardizabal, 4, 20018 Donostia-San Sebastián, Guipuzkoa, Spain
        \and IKERBASQUE, Basque Foundation for Science, 48013 Bilbao, Spain
        \and Vyoma GmbH, Karl-Theodor-Straße 55, 80803 Munich, Germany
        \and Lennard-Jones Laboratories, School of Chemical and Physical Sciences, Keele University, Keele ST5 5BG, UK
        \and Max-Planck-Institut für Astronomie, Königstuhl 17, D-69117 Heidelberg, Germany
        \and Aix Marseille Universit{\'e}, CNRS, CNES, LAM, Marseille, France
        \and INAF – Osservatorio Astrofisico di Arcetri, Largo E. Fermi 5, 50125 Firenze, Italy
        \and Astrophysics Research Institute, Liverpool John Moores University, IC2 Liverpool Science Park, 146 Brownlow Hill, Liverpool
        L3 5RF, UK
}

   \date{Received date /
Accepted date}

 
  \abstract{In this study, we report conclusive evidence for an ancient star cluster that has been accreted by the Large Magellanic Cloud (LMC). By leveraging observations from the \textit{Hubble} Space Telescope (\textit{HST}), we investigate the chrono-dynamical structure of a sample of seven old star clusters within the LMC in a self-consistent way. The multi-epoch nature of the dataset allowed the determination of high-precision proper motions for the clusters. Employing an isochrone-fitting methodology, we additionally infer from the deep high-resolution \textit{HST} data homogeneous and robust estimates for their distances, ages and metallicities. Supplementing these data with literature line-of-sight velocities, we investigate the full 3-dimensional dynamics of the clusters within the frame of the LMC. With respect to the other clusters in our sample, NGC~1841 depicts a peculiar case. Its position in the age-metallicity plane, that makes it about 1~Gyr younger
  than the other metal-poor LMC clusters, but also its dynamical properties with a radial orbit almost perpendicular to the LMC disc plane, clearly advocates for a different origin. We thus conclude that NGC~1841 has likely been accreted by the LMC from a smaller galaxy. The other clusters in our sample show disc-like kinematics, with the case of NGC~2210 being peculiar, based on its inclined orbit. Their coherent age-metallicity relation closely resembles that of Gaia-Sausage-Enceladus globular clusters, thus suggesting a similar early evolution for the two dwarf galaxies.
  We do not find clear-cut chrono-kinematic evidence that NGC~2005 has been accreted by the LMC as suggested by a previous study based on its chemical abundance pattern. Regardless of its nature, its very old age illustrates that peculiar chemical evolutions already emerge at very early times. 
  }

   \keywords{proper motions -- stars: kinematics and dynamics -- Magellanic Clouds -- galaxies: star clusters: general -- galaxies: interactions -- techniques: photometric
               }

   \maketitle
%

\section{Introduction}\label{sec:intro}

According to the $\Lambda$ cold dark matter cosmological model, galaxy assembly and evolution is mainly driven by hierarchical mergers, meaning that the halos of larger galaxies build up and grow by the accretion of smaller systems. Thus, galaxies should be accompanied by a population of smaller satellites, as widely observed. The largest satellite of our Milky Way is the Large Magellanic Cloud (LMC), with a total mass of $\sim1-2\times10^{11}$~M$_{\sun}$ \citep[e.g.][]{Erkal19, Erkal21, Vasiliev21b, Kacharov24, Watkins24}. Given its large mass, which is only a factor of ten less than that of our Galaxy, the LMC should have brought its own system of dwarf galaxy satellites as it entered the halo of the Milky Way \citep[e.g.][]{Lynden-Bell76, D'Onghia08}. 
Its most noticeable companion is the Small Magellanic Cloud (SMC). Both galaxies, the LMC and SMC, form a pair of interacting dwarf galaxies and are currently undergoing the early stages of a minor merger event. Beyond the SMC, a group of ultra-faint dwarf (UFD) galaxies is further expected to be associated with the LMC \citep{Sales11}.

Using multi-epoch observations with the \textit{Hubble} Space Telescope (\textit{HST}), \citet{Kallivayalil06a, Kallivayalil06b, Kallivayalil13} measured precise proper motions (PMs) of the LMC and SMC, and constrained for the first time their orbits around the Milky Way.
The high tangential velocity of the galaxies indicates they are most likely on their first \citep[e.g.][]{Besla07, Patel17} or second \citep{Vasiliev24} passage around our Galaxy. As a consequence, the system of LMC satellites should still be intact without being stripped by tidal forces of the Milky Way. Indeed, within the last years, several UFD galaxies have been identified to be satellites of the LMC \citep[see, e.g.][]{Jethwa16, Sales17, Kallivayalil18, Erkal20, Vasiliev24}.

The ongoing interaction between the Clouds offers us the unique opportunity to witness the cosmological process of hierarchical structure formation close up. It is also natural to assume that the LMC has already experienced accretion events with smaller satellites in its past. Thus, reconstructing the interaction history of the LMC will provide us with valuable insights in the assembly of systems on the dwarf-galaxy scale. In retracing the evolution of our Galaxy, globular clusters (GCs) have proven to be a powerful tool \citep[e.g.][]{Searle78, Forbes10, Massari19, Kruijssen19}. Compact stellar systems such as GCs are able to survive the merging event during which the host dwarf galaxy is dissolved by tidal forces. These clusters keep the kinematic signature of the merger event, and their characteristics, such as their chemical composition, reflect the properties of the environment they formed in, making them distinguishable from GCs formed in-situ \citep[e.g.][]{Horta20, Ceccarelli24}. 

Our understanding of the assembly history of the Milky Way has benefited from the precise astrometric measurements provided by the \textit{Gaia} mission \citep{Gaia16, Gaia18b, Gaia21b}. With PMs being for a long time the missing piece of kinematic information, it has now become possible to trace the orbits of GCs within the potential of the Milky Way \citep[e.g.][]{Gaia18, Baumgardt19, Vasiliev21}. In combination with determinations of their chemical abundances and ages, most of the Galactic GC population can now be associated with different progenitors that form the present Milky Way \citep[e.g.][]{Massari19, Myeong19, Callingham22, Malhan22}. Based on the most-likely properties of the progenitors, resulting from the characteristics of the relic GCs, \citet{Kruijssen20} constructed the first merger tree of our Galaxy. Chemo-dynamical information alone, though, is not enough to unambiguously disentangle the origin of each Milky Way GC \citep[e.g.][]{Minelli21, Carretta22}. For this reason, \citet{Massari23} started the Cluster Ages to Reconstruct the Milky Way Assembly (CARMA) project to determine in a homogeneous way the ages of the Galactic GCs with the aim to unambiguously assign each GC to a merging event or an in-situ formation, based on the age-metallicity relations defined by the different progenitors.

With the availability of precise PM measurements even at the distance of the Clouds \citep[e.g.][]{Massari21, Niederhofer22, Niederhofer24}, we are now at the advent of studying in detail the chemo-dynamics of their star cluster systems, providing us with additional clues about the formation, assembly and evolution of the galaxies, as well as the origins of their cluster population. 
Based on line-of-sight velocity measurements, the ancient ($>$10~Gyr) star clusters in the LMC seem to follow a flattened rotating disc-like structure \citep[e.g.][]{Freeman83, Grocholski06, Sharma10}, suggesting that the LMC does not harbour a population of halo GCs. Such a lack would be in strong contrast to the system of Milky Way GCs, which is composed of disc and halo clusters with different characteristics, raising the question whether the LMC star-cluster system assembled in a different way than the one of our Galaxy. \citet{Bekki07} discussed the implications of these observations and argued that given the halo of the LMC formed from accreted low-mass systems, akin the build-up of the Milky Way halo, the ancient star clusters and halo field stars should share the same kinematics. As a possible formation scenario, they proposed that there might exist a lower mass limit for a low-mass galaxy to form GCs. Thus, the LMC halo might have formed solely from systems below this threshold and the ancient star clusters arose in the early stages of the formation of the disc. 
\citet{Wagner-Kaiser17} estimated the ages of a sample of six ancient LMC clusters, by determining the brightness of their main-sequence turn-off points. Their results suggested that their sample of clusters being coeval to Galactic GCs, indicating they both formed at similar times. 

\citet{Piatti19} were the first to study the full 3D velocity structure of the old LMC star-cluster system by combining line-of-sight velocities with PM measurements from the second data release (DR2) from \textit{Gaia} \citep{Gaia18}. Unlike the findings based on 1D kinematics, they claimed to have found evidence for two distinctive populations of clusters, one of which showing halo-like kinematics and spatial distributions. This result was later questioned by \cite{Bennet22}, who combined \textit{Gaia} and \textit{HST} data to measure the PMs of a sample of 32 young ($\lesssim3$~Gyr) and old ($>$10~Gyr) LMC star clusters. Including literature line-of-sight velocities and distances, the resulting dynamic structure they found does not show any indication of a bi-modality and is consistent with a single population with disc-like kinematics. 

In order to investigate the origin of the LMC star-cluster system from a chemical point-of-view, \citet{Mucciarelli21} analysed in detail high-resolution spectra of a sample of 11 old LMC clusters. They discovered that NGC~2005 displays a peculiar chemical abundance pattern, and argued that this cluster likely formed in a low-mass galaxy that has been accreted by the LMC in a past merger event. 

The inconclusive results regarding the nature of the old LMC star-cluster system illustrate the need for a homogeneous derivation and analysis of its chemo-dynamical properties. In this paper of the CARMA project we employ our recently published \textit{HST} astro-photometric catalogues \citep{Niederhofer24} to derive in a robust and self-consistent way the PMs, ages, metallicities and distances of a sample of seven old LMC star clusters with available multi-epoch HST data. 
This work represents the first successful attempt to combine dynamical information with accurate determinations of the cluster properties (age, metallicity) to investigate the formation of the LMC cluster system.

The paper is organised as follows. In Section~\ref{sec:data}, we describe the used data sets and provide a brief outline of the photometric and astrometric reduction procedures. In Section~\ref{sec:isochrone_fitting}, we describe the isochrone fitting routine and present the results for our sample of clusters. We derive the kinematics of the clusters within the frame of the LMC in Section~\ref{sec:kinematic_structure} and discuss the results and its implications regarding the different clusters. In Section \ref{sec:conclusions}, we provide a summary of the paper and draw final conclusions.


\section{Data sets and reduction\label{sec:data}}

For the seven clusters analysed in this study we use multi-epoch \textit{HST} observations taken with the Ultraviolet-Visible (UVIS) channel of the Wide Field Camera 3 (WFC3) and the Wide-Field Channel (WFC) of the Advanced Camera for Surveys (ACS). The astro-photometric catalogues of six of those clusters were presented in \citet{Niederhofer24}\footnote{The astro-photometric catalogues are publicly available as a High Level Science Product
at MAST under: \href{https://archive.stsci.edu/hlsp/hamsters}{https://archive.stsci.edu/hlsp/hamsters}}. A detailed list of all observations used to create these catalogues can be found in the appendix of \citet{Niederhofer24} and in the Mikulski Archive for Space Telescopes (MAST) under the following DOI: \href{https://doi.org/10.17909/7d5e-s940}{10.17909/7d5e-s940}. In addition to these six clusters, we also include the old LMC cluster Reticulum, which we found to have archival long time-baseline data, suitable for PM determinations. The list of observations for Reticulum is given in Table~\ref{tab:reticulum_obs}.

\begin{table}\footnotesize
\centering
\caption{List of observations for Reticulum\label{tab:reticulum_obs}}
\begin{tabular}{@{}l@{ }c@{ }c@{ }c@{ }c@{ }c@{ }c@{ }}
\hline\hline
\noalign{\smallskip}
Programme &~~~~~~~PI~~~~~~~& ~~~~~Epoch~~~~& ~~Camera~~~& ~~~Filter~~~&~~~~Exposures~\\
ID& & (yyyy/ & & & (N\,$\times$\,t$_{\mathrm{exp}}$) \\
& & mm) & & &  \\

\noalign{\smallskip}
\hline
\noalign{\smallskip}

GO-9891 & G. Gilmore & 2003/09 & ACS/ & F555W & 1\,$\times$\,330\,s\\
        &            &         & WFC  &       &                    \\
 \noalign{\smallskip}
        &            & 2003/09 & ACS/ & F814W & 1\,$\times$\,200\,s\\
        &            &         & WFC  &       &                    \\
 \noalign{\smallskip}
\hline
\noalign{\smallskip}
GO-13435 & M. Monelli & 2014/08 & WFC3/ & F336W & 2\,$\times$\,989\,s\\
         &            &         & UVIS  &       &                    \\
 \noalign{\smallskip}
         &            & 2014/08 & WFC3/ & F438W & 2\,$\times$\,200\,s\\
         &            &         & UVIS  &       &                    \\
 \noalign{\smallskip}
         &            & 2014/08 & WFC3/ & F814W & 1\,$\times$\,100\,s\\
         &            &         & UVIS  &       &                    \\
 \noalign{\smallskip}
\hline
\noalign{\smallskip}
GO-14164 &  A. Sarajedini & 2017/01 & WFC3/ & F336W & 12\,$\times$\,750\,s \\
         &                &         & UVIS  &       &                      \\
 \noalign{\smallskip}
        &                 & 2016/07 & ACS/ & F606W & 2\,$\times$\,50\,s \\
        &                 &         & WFC  &       & 9\,$\times$\,353\,s\\
        &                 &         &      &       & 2\,$\times$\,525\,s\\
 \noalign{\smallskip}
        &                 & 2016/07 & ACS/ & F814W & 2\,$\times$\,70\,s \\
        &                 &         & WFC  &       & 3\,$\times$\,353\,s\\
        &                 &         &      &       & 6\,$\times$\,360\,s\\   
        &                 &         &      &       & 2\,$\times$\,525\,s\\
\noalign{\smallskip}
\hline

\end{tabular}

\end{table}

The photometric reduction is described in \citet{Niederhofer24} and follows the prescriptions and methods presented in \citet{Bellini17b, Bellini18}. 
In brief, we performed our analysis on the un-resampled \texttt{\_flc} images that have been corrected for the effects of imperfect charge transfer efficiency (CTE; see \citealt{Anderson18, Anderson21}). The reduction routine is a combination of a first- and second-pass photometric run, using \texttt{hst1pass} \citep{Anderson22} and \texttt{KS2} \citep[see][for details]{Sabbi16, Bellini17b}. To fit the PSFs of the stars, we used the focus-diverse PSF models\footnote{\href{https://www.stsci.edu/~jayander/HST1PASS/LIB/PSFs/STDPBFs/}{https://www.stsci.edu/$\sim$jayander/HST1PASS/LIB/PSFs/STDPBFs/}}, which we perturbed to adjust them to each exposure. We corrected the measured positions of the stars for geometric distortions, applying the solutions by \citet{Anderson10} for ACS/WFC\footnote{We also used a look-up table of residuals to account for changes in the original distortion solution of the ACS/WFC that occurred during the \textit{Hubble} Service Mission 4 in 2009.} and \citet{Bellini09} and \citet{Bellini11} for WFC3/UVIS. Finally, we calibrated the instrumental magnitudes to the VEGA-mag system, as described in \citet{Bellini17b}.

We measured relative PMs in an iterative way following the techniques developed by \citet{Bellini14, Bellini18, Libralato18b, Libralato22}. We refer the readers to these papers for a detailed description of the methods and outline the main steps below. We transformed the positions of the stars in the individual exposures to a common frame of reference by means of a general six-parameter transformation, using well-measured cluster members as reference objects. Thus, the PMs are relative to the bulk motions of the clusters.
Within each iteration, this list of reference stars is refined, excluding sources that are not in agreement with the motion of the cluster (i.e. that do not cluster around 0). The transformed positions of the stars as a function of observing epoch are then fitted with a straight line. The slopes of the fitted lines directly correspond to the PMs of the stars. We inspected the PMs and corrected them for spatially variable systematic effects, as described in \citet{Bellini14}. Finally, we transformed the relative PMs to an absolute scale using stars from the \textit{Gaia} DR3 catalogue \citep{Gaia23}. Table~\ref{tab:clusters_params} presents the absolute PMs of the star clusters, together with their RA and Dec positions and literature line-of-sight velocity measurements.


\begin{table*}\small
\centering
\caption{Positions and velocities of our sample of old LMC star clusters. \label{tab:clusters_params}}
\begin{tabular}{lcccccccr}
\hline\hline
\noalign{\smallskip}
Cluster ID & RA$_0$ & $\Delta$RA$_0$ & Dec$_0$ & $\Delta$Dec$_0$ & $\mu_{\alpha}$cos($\delta$) & $\mu_{\delta}$ & LOS Velocity & Ref\\
& [h:m:s] & [arcsec] & [\degr:\arcmin:\arcsec] & [arcsec] & [mas\,yr$^{-1}$] & [mas\,yr$^{-1}$] & [km\,s$^{-1}$] &\\
\noalign{\smallskip}
\hline
\noalign{\smallskip}
Hodge 11 & 06:14:22.83 & 0.38 & $-$69 50 50.0 & 0.38 & 1.545 $\pm$ 0.041 & 0.986 $\pm$ 0.046 & 245.1 $\pm$ 1.0 & (1)\\
NGC 1841 & 04:45:22.60 & 0.30 & $-$83 59 54.6 & 0.29 & 1.967 $\pm$ 0.032  &  0.010 $\pm$ 0.035 & 210.8 $\pm$ 0.3 & (2)\\
NGC 1898 & 05:16:42.04 & 0.18 & $-$69 39 24.2 & 0.17 &  1.962 $\pm$ 0.036  &  0.274 $\pm$ 0.041 & 209.0 $\pm$ 1.5 & (3)\\
NGC 2005 & 05:30:10.24 & 0.10 & $-$69 45 10.0 & 0.09 &  1.847 $\pm$ 0.038  &  0.469 $\pm$ 0.043 & 277.7 $\pm$ 7.3 & (4)\\
NGC 2210 & 06:11:31.65 & 0.11 & $-$69 07 18.4 & 0.10 &  1.533 $\pm$ 0.038  &  1.303 $\pm$ 0.033 & 335.6 $\pm$ 0.3 & (3)\\
NGC 2257 & 06:30:12.50 & 0.42 & $-$64 19 37.5 & 0.38 &  1.467 $\pm$ 0.036  &   0.970 $\pm$ 0.041 & 301.8 $\pm$ 0.3 & (2)\\
Reticulum & 04:36:10.75 & 1.06 & $-$58:51:37.8 & 1.16 & 1.917 $\pm$ 0.059  & $-$0.300 $\pm$ 0.059 & 247.5 $\pm$ 1.5 & (1)\\
\noalign{\smallskip}
\hline

\end{tabular}
\tablefoot{References: (1): \citet{Grocholski06}; (2): \citet{Song21}; (3): \citet{Usher19}; (4): \citet{Mucciarelli21}.
}
\end{table*}


\section{Isochrone fitting}\label{sec:isochrone_fitting}

In order to obtain homogeneous, self-consistent ages, metallicities and distances of our sample of clusters, we employed the isochrone-fitting routine used in the CARMA project. This code was originally developed by \citet{Saracino19} and updated and refined by \citet{Massari23}. In the following we provide a brief overview of the employed method and refer the interested reader to \citet{Massari23} for a detailed description.

\subsection{The isochrone-fitting methodology}\label{subsec:isochrone_code}

The isochrone fitting routine developed within CARMA allows for inferring simultaneously the metallicity [M/H], reddening $E(B-V)$, distance modulus $(m-M)$ and age of the clusters. It is based on a Bayesian statistical framework and implements a Markov Chain Monte Carlo (MCMC) sampling technique to explore the posterior probability distribution. The code performs a star-by-star comparison between an observed colour-magnitude diagram (CMD) in a given filter combination and a set of theoretical isochrone models, determining the associated log-likelihood. The log-likelihood function is the sum of two terms and is given by:
\begin{equation}  \label{eq:log_l}
\mathrm{ln}\mathcal{L}_{\mathrm{tot}} = \mathrm{ln}\mathcal{L}_{\mathrm{prior}} + 
0.3 \times\mathrm{ln}\mathcal{L}_{\mathrm{fit}}.
\end{equation}
The first term is the likelihood associated to the initial priors that we provide for the parameters [M/H], $E(B-V)$ and $(m-M)$. For each of the three parameters, we opted for a Gaussian form, thus $\mathrm{ln}\mathcal{L}_{\mathrm{prior}}$ can be written as:
\begin{equation}\label{eq:log_l_prior}
   \mathrm{ln}\mathcal{L}_{\mathrm{prior}} = -0.5\times\sum^3_{i=1} \frac{(x_i-x_{i,\mathrm{prior}})^2}{x^2_{i,\mathrm{std}}},
\end{equation}
where $x$ corresponds to [M/H], $E(B-V)$ and $(m-M)$, and $x_{\mathrm{prior}}$ and $x_{\mathrm{std}}$ are the provided priors and their associated uncertainties, respectively. 
The second term in Equation~\ref{eq:log_l} ($\mathrm{ln}\mathcal{L}_{\mathrm{fit}}$) gives the probability of the observed CMD for a given model. It can be written as: 
\begin{equation}\label{eq:log_l_fit}
    \mathrm{ln}\mathcal{L}_{\mathrm{fit}} = -0.5 \times \sum^N_{i=1} \frac{[\mathrm{min}(dist)]^2}{\sigma_i^2},
\end{equation}
where $N$ is the total number of stars, min($dist_i$) gives the minimum distance between star $i$ and the model and $\sigma$ is the photometric uncertainty. 
In Equation~\ref{eq:log_l}, a factor of 0.3 is introduced in front of $\mathrm{ln}\mathcal{L}_{\mathrm{fit}}$ to prevent the code to get stuck in local minima \citep[see][]{Massari23}. 

To infer the best-fitting values for [M/H], $E(B-V)$, $(m-M)$ and age, and its associated uncertainties, the code explores the four-dimensional posterior probability space using the MCMC sampler \texttt{emcee} \citep{Foreman-Mackey13}, which is a \texttt{python} implementation of the aﬃne-invariant
MCMC ensemble sampler \citep{Goodman10}. 

As theoretical stellar models to which the observed CMDs are compared, the routine adopts the Bag of Stellar Tracks and Isochrones (BaSTI) database \citep{Hidalgo18, Pietrinferni21}. We used the newest release that also includes the effects of diffusive processes. As discussed in \citet{Massari23} we made use of solar scaled models and fitted for the global metallicity [M/H], rather than the iron content [Fe/H]\footnote{This approach is valid since the effects of the specific $\alpha$-element abundance on the magnitudes in optical/near-infrared bands at fixed global metallicity [M/H] is negligible, as shown by \citet{Cassisi04}.}. Since measurements of the $\alpha$-element abundance do not exist for all clusters in our sample, this choice allows for homogeneous results that are not prone to any assumptions on the [$\alpha$/Fe] content of the clusters. 

To create the models that are fitted to the observed CMDs, we built a grid of theoretical BaSTI isochrones with ages ranging from 6~Gyr to 15~Gyr in steps of 100~Myr, and with metallicities going from [M/H]=0.0~dex to [M/H]=$-$2.0~dex with a step of 0.01~dex. We further interpolated the models in age for a finer sampling. To apply interstellar reddening to the models, we determined the extinction coefficients, $A_{\lambda}$, in the different ACS/WFC and WFC3/UVIS filter bands, assuming a \citet{Cardelli89} extinction law. To take into account the dependence of $A_{\lambda}$ on the stellar spectral type, we created for each filter a grid of $A_{\lambda}$ values over a large range of stellar effective temperatures and for $E(B-V)$ values ranging between 0.0~mag and 1.0~mag, in steps of 0.1~mag, and linearly interpolated this grid. 

\subsection{Selection criteria}

\begin{figure*}
\centering
\includegraphics[width=1.8\columnwidth]{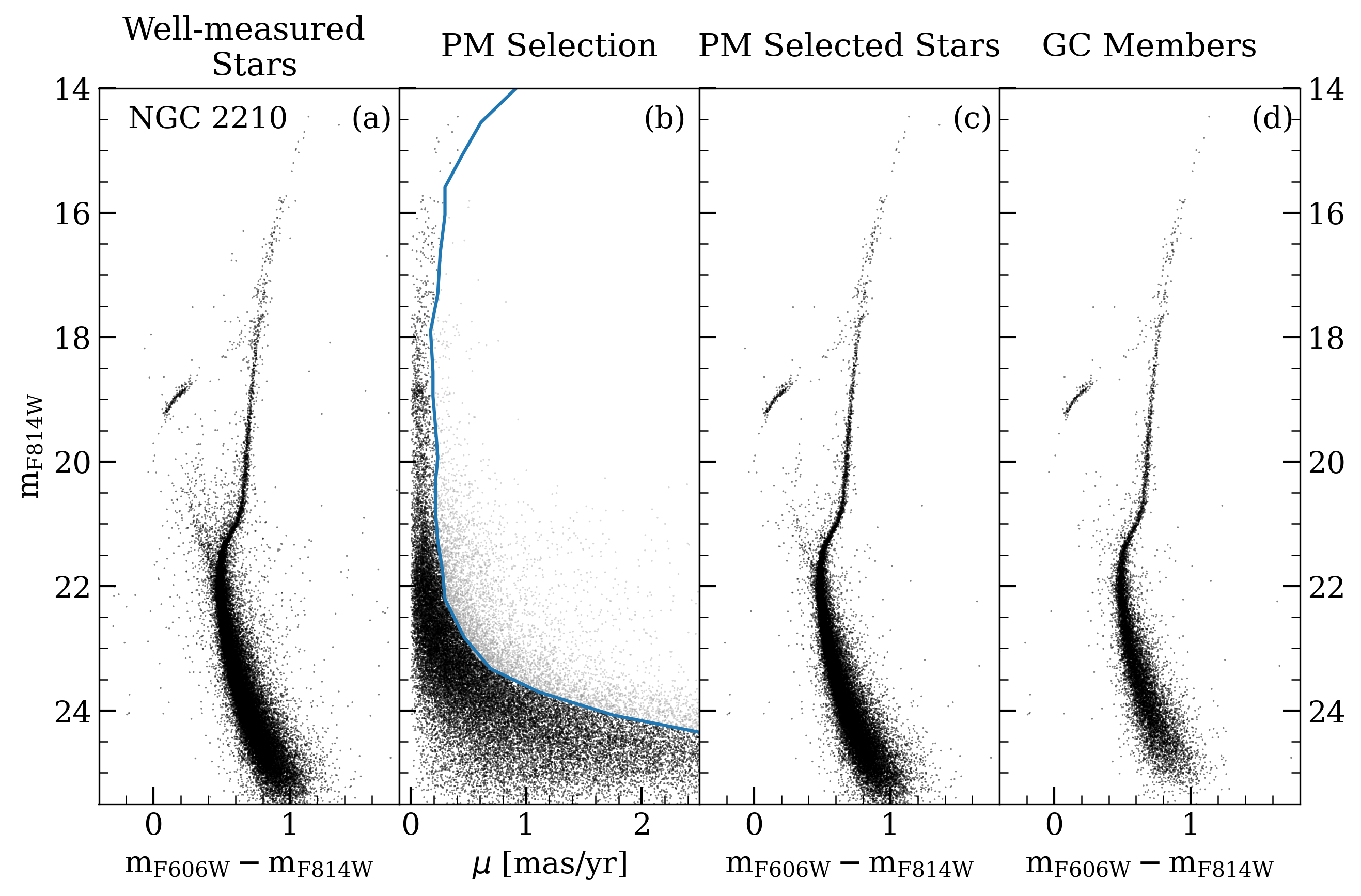}
\caption{Illustration of the selection of cluster member stars for NGC~2210. (a) The $m_{\rm F814W}$ vs $m_{\rm F606W}-m_{\rm F814W}$ CMD of well-measured stars in the field of the cluster for which PMs have been determined (b) 1-D 
relative
PMs as a function of the $m_{\rm F814W}$ magnitude. The blue line (drawn by hand) follows our selection of likely cluster members (black dots) based on their motions. (c) CMD of PM-selected cluster members. (d) CMD of PM-selected cluster stars, including only stars within one effective radius from the cluster centre. \label{fig:cluster_star_selection}}
\end{figure*}

To ensure to run the isochrone-fitting routine only on sources with the best measurements, we applied several photometric and astrometric selection criteria to our catalogues. To achieve this, we closely follow the prescriptions in \citet{Niederhofer24} and applied the same quality cuts. Specifically, we used the following photometric diagnostic parameters, which are provided by the \texttt{KS2} output: the quality-of-fit or $\texttt{QFIT}$ parameter (which indicates how well a source was fitted by the PSF model), the photometric rms error, the shape parameter \texttt{RADXS} (which indicates how extended a source is with respect to the PSF model, see \citealt{Bedin08}), the isolation parameter of a source (which gives the fraction of flux within the PSF fitting aperture that comes from neighbouring sources, before neighbour subtraction) and the fraction of good measurements of a source with respect to the total number of detections. The astrometric quality parameters are based on the PM measurements and include the reduced $\chi^2$ of the PM fit, the fraction of data points of a source actually used to determine its PM and the uncertainty in the PM measurement.

We further took advantage of the available kinematic information to minimize the contribution from field stars within our sample of well-measured stars.
Although in most cases, there is a considerable overlap between the PM distribution of cluster and field stars, we are able to remove a large fraction of the field stars based on their motions, given the larger velocity dispersion of the field stars. This selection is illustrated in Fig.~\ref{fig:cluster_star_selection} for the example of NGC~2210. Panel (a) shows the CMD of all well-measured stars in the field of NGC~2210. We selected (by hand) stars that follow the bulk of cluster stars in the $m_{\rm F814W}$ vs. PM diagram (blue line in panel b). 
Note that we are exploiting here the relative PMs, thus cluster stars have motions close to zero, whereas field stars show larger motions. 
Panel (c) shows the CMD of the PM-selected cluster stars. We further refined our selection by restricting it to stars that are within one effective radius of the clusters (as determined by \citealt{Niederhofer24}). This final selection of cluster members is displayed in panel (d) of Fig.~\ref{fig:cluster_star_selection}.

\subsection{Differential reddening correction}
Before running the isochrone fitting routine, we corrected the photometry of the seven clusters for the effects of differential reddening across the cluster field, applying the widely-used techniques as described by \citet{Milone12} and \citet{Bellini17c}. Briefly, we first selected for each cluster a sample of reference stars along the main sequence, excluding stars on the binary sequence. We then determined the fiducial line for these stars. For this we divided the magnitude range covered by the reference stars into bins of 0.2~mag and calculated the sigma-clipped median colour within each bin. These median colours are then fit with a cubic spline model. For each reference star we subsequently determined its offset from the fiducial line in the direction of the reddening vector. As extinction coefficients in the different filter bands, we assumed the $A_{\lambda}$ values as explained in Section~\ref{subsec:isochrone_code}. Finally, we determined the reddening of each star as the 2.5$\sigma$-clipped median value of the closest 75 reference stars, excluding the target star itself. 
Note that this procedure only corrects for any variable reddening across the cluster, thus the photometry is still affected by the mean global $E(B-V)$.
We found moderate reddening variations across the fields of Hodge~11, NGC~1841 ($-$0.02~mag$\lesssim\delta E(B-V)\lesssim$0.02~mag for both clusters) and NGC~1898 ($-$0.04~mag$\lesssim\delta E(B-V)\lesssim$0.04~mag) and will use the corrected photometry of these clusters for the analysis. The remaining four clusters show negligible effects of differential reddening.

\subsection{Results}\label{sec:results}

For each cluster, we ran the code on two different CMDs, resulting from the combination of the available filters for each cluster. For Hodge~11, NGC~1841, NGC~2210, NGC~2257 and Reticulum, these are the ($m_{\rm F814W}$ vs $m_{\rm F606W}-m_{\rm F814W}$) and ($m_{\rm F606W}$ vs $m_{\rm F606W}-m_{\rm F814W}$) CMDs. For NGC~1898 the ($m_{\rm F438W}$ vs $m_{\rm F438W}-m_{\rm F814W}$) and ($m_{\rm F814W}$ vs $m_{\rm F438W}-m_{\rm F814W}$) CMDs and for NGC~2005 the ($m_{\rm F435W}$ vs $m_{\rm F435W}-m_{\rm F814W}$) and ($m_{\rm F814W}$ vs $m_{\rm F435W}-m_{\rm F814W}$) CMDs. 
We did not use observations in the F336W filter for the fitting, since the bolometric corrections show an increasing sensitivity on the exact chemical abundance pattern of the stars in the ultra-violet wavelengths \citep[e.g.][]{Cassisi04, Pietrinferni24}, thus might have an impact on the final results when using only solar-scaled models. 
Although for Reticulum observations in more filter bands are available, we opted to only use data in the ACS/WFC F606W and F814W filters to be more consistent with the other clusters in our sample. 
Before fitting the isochrones, we had to tailor the input catalogues for the code to give reliable results. Specifically, we selected stars within each CMD that follow the cluster sequence, i.e. excluding stars on the binary sequence, blue straggler stars and left-over field interlopers. For this, we first defined a magnitude bin of 0.1~mag, starting at the brightest magnitude in the catalogue. If this bin contained more than five stars, we determined the sigma-clipped median colour of the stars within this bin and kept all stars that are within 1.5--2.0 times (depending on the cluster) the median colour uncertainty in the bin. We then moved this bin by 0.001~mag to fainter magnitudes and repeated the process until we reached the faintest magnitude. Finally, we also defined a bright and faint limit to the list of stars. The main reason for this is to prevent giving stars on the main sequence too much weight in the fitting. Tests have shown that including stars that are within +2.5~mag and $-$3.5~mag from the main-sequence turn-off yields the most robust results in terms of age and metallicity. 

For each cluster, we set up Gaussian priors for the metallicity, distance modulus and reddening. For [M/H] we chose the spectroscopically determined mean values and uncertainties as given by \citet{Grocholski06}  and \citet{Mucciarelli21}, whereas for the distance modulus and $E(B-V)$ values, we used the measurements provided by \citet{Wagner-Kaiser17} and \citet{Milone23a}. 
The spectroscopic metallicity measurements will provide the most robust prior information for the fitting and also break the age-metallicity degeneracy. 
The isochrone fitting routine does not require any priors on the ages to freely explore the parameter space. 
We created an ensemble of 100 walkers and ran the MCMC for a total of 1\,000 steps, using the first 200 steps as burn-in phase. The inferred best-fit solutions and their associated uncertainties correspond to the 50th, 16th and the 84th percentiles of the posterior distributions, respectively. In Appendix~\ref{sec:appendix} we provide for each of the seven clusters the CMDs with the best-fitting isochrone model along with the corresponding corner plots showing the posterior probability distribution.

\begin{table}\small
\centering
\caption{Results of the isochrone fits. \label{tab:isochrone_fits}}
\begin{tabular}{lcccc}
\hline\hline
\noalign{\smallskip}
Cluster ID & [M/H] & $E(B-V)$ & $(m-M)$ & Age\\
& [dex] & [mag] & [mag] & [Gyr]\\
\noalign{\smallskip}
\hline
\noalign{\smallskip}
Hodge 11  & $-1.63^{+0.02}_{-0.01}$& $0.05^{+0.01}_{-0.01}$& $18.49^{+0.01}_{-0.01}$& $13.8^{+0.2}_{-0.2}$\\
 \noalign{\smallskip}
NGC 1841 & $-1.76^{+0.02}_{-0.01}$& $0.18^{+0.01}_{-0.01}$& $18.32^{+0.01}_{-0.01}$& $12.8^{+0.2}_{-0.1}$\\
 \noalign{\smallskip}
NGC 1898  & $-1.26^{+0.03}_{-0.04}$& $0.07^{+0.01}_{-0.01}$& $18.53^{+0.01}_{-0.01}$& $12.4^{+0.2}_{-0.2}$\\
 \noalign{\smallskip}
NGC 2005  & $-1.72^{+0.05}_{-0.06}$& $0.08^{+0.01}_{-0.01}$& $18.44^{+0.01}_{-0.01}$& $14.0^{+0.2}_{-0.2}$\\
 \noalign{\smallskip}
NGC 2210  & $-1.43^{+0.04}_{-0.04}$& $0.05^{+0.01}_{-0.01}$& $18.35^{+0.01}_{-0.01}$& $12.2^{+0.2}_{-0.2}$\\
 \noalign{\smallskip}
NGC 2257  & $-1.44^{+0.06}_{-0.04}$& $0.04^{+0.01}_{-0.01}$& $18.31^{+0.02}_{-0.02}$& $12.5^{+0.2}_{-0.2}$\\
 \noalign{\smallskip}
Reticulum & $-1.33^{+0.04}_{-0.04}$& $0.02^{+0.01}_{-0.01}$& $18.37^{+0.03}_{-0.05}$& $12.3^{+0.9}_{-0.7}$\\
\noalign{\smallskip}
\hline

\end{tabular}

\end{table}

\begin{figure*}
\begin{tabular}{cc}
\includegraphics[width=0.45\textwidth]{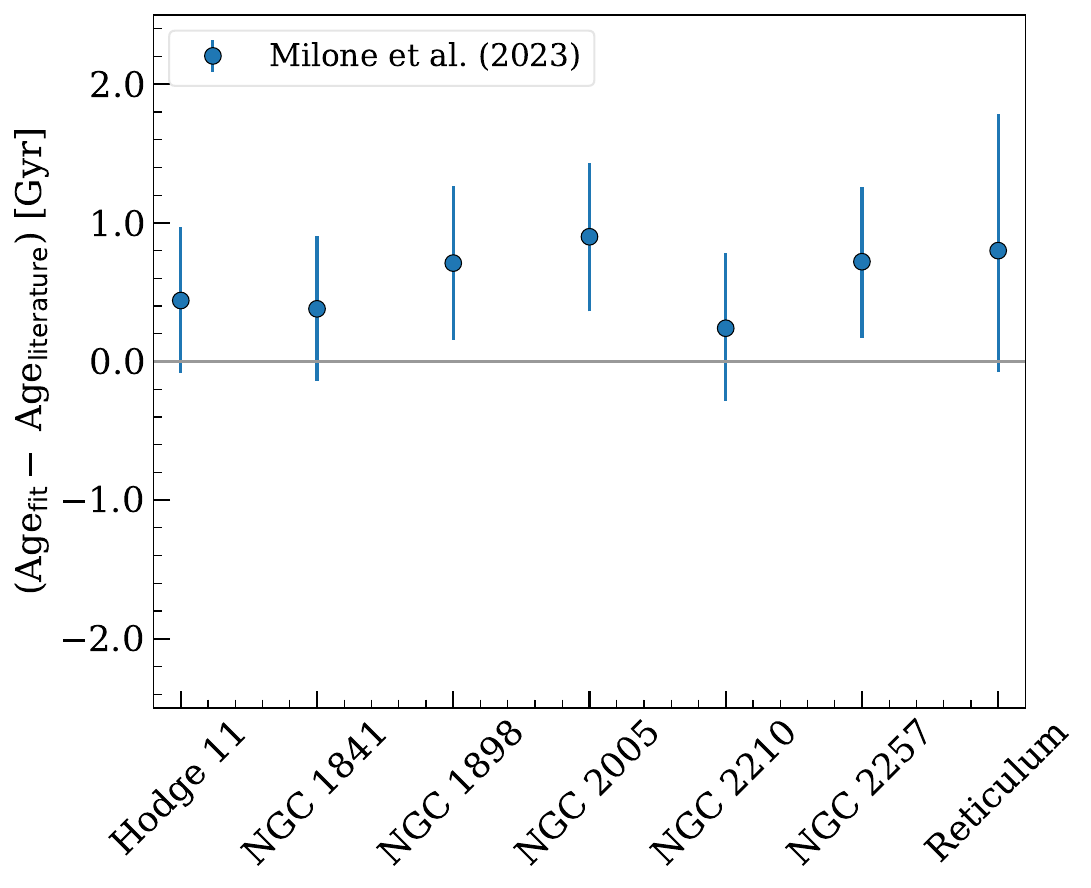} &
\includegraphics[width=0.45\textwidth]{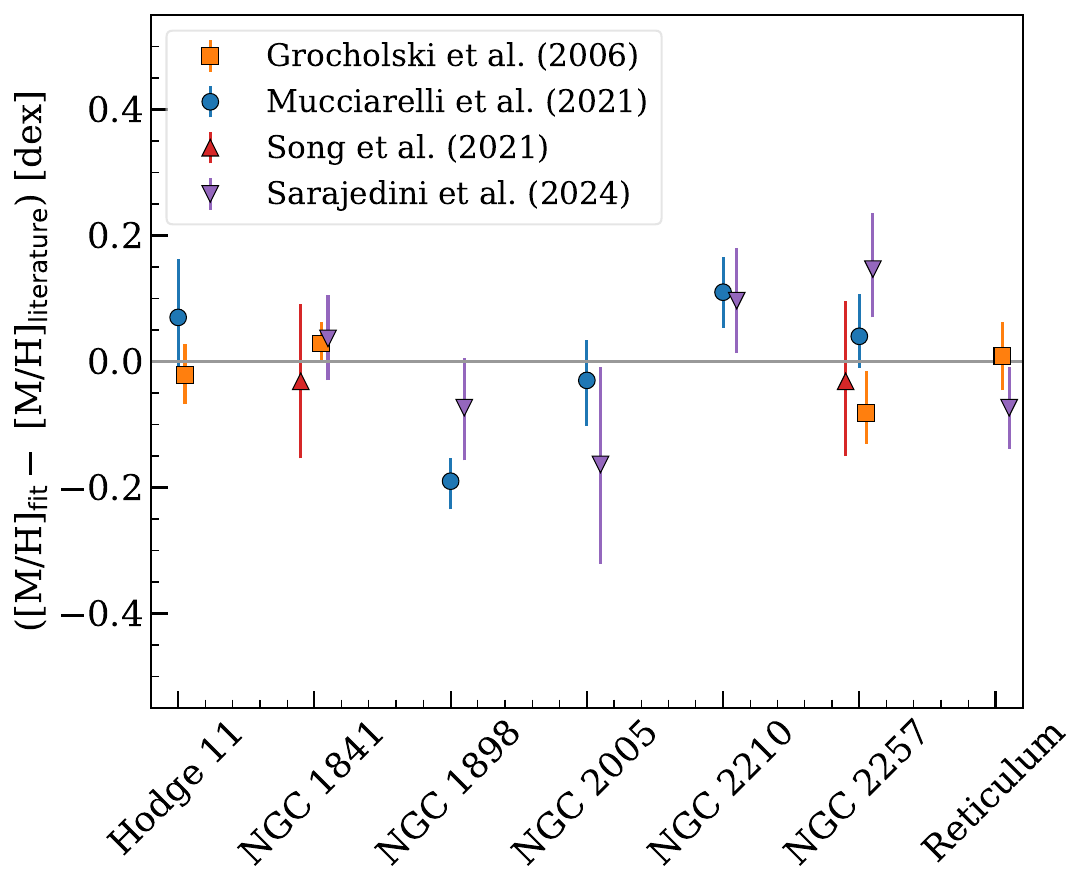} \\
\includegraphics[width=0.45\textwidth]{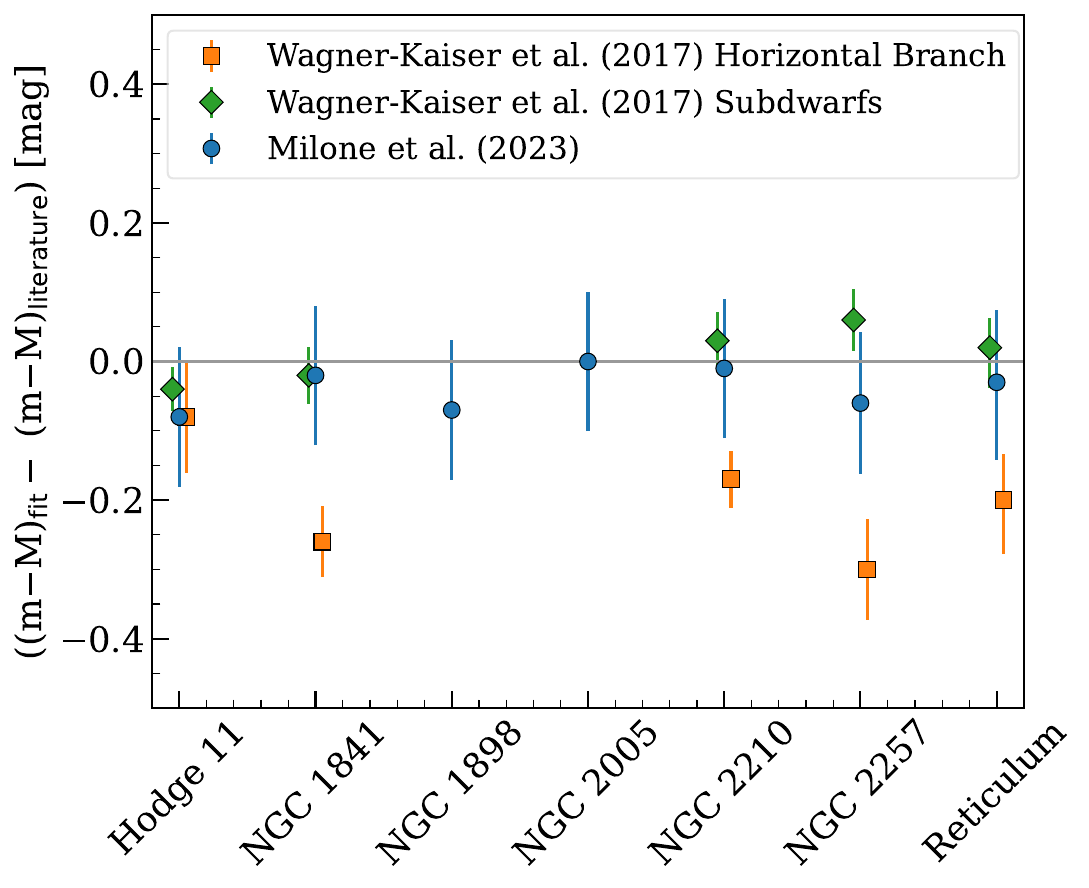} &
\includegraphics[width=0.45\textwidth]{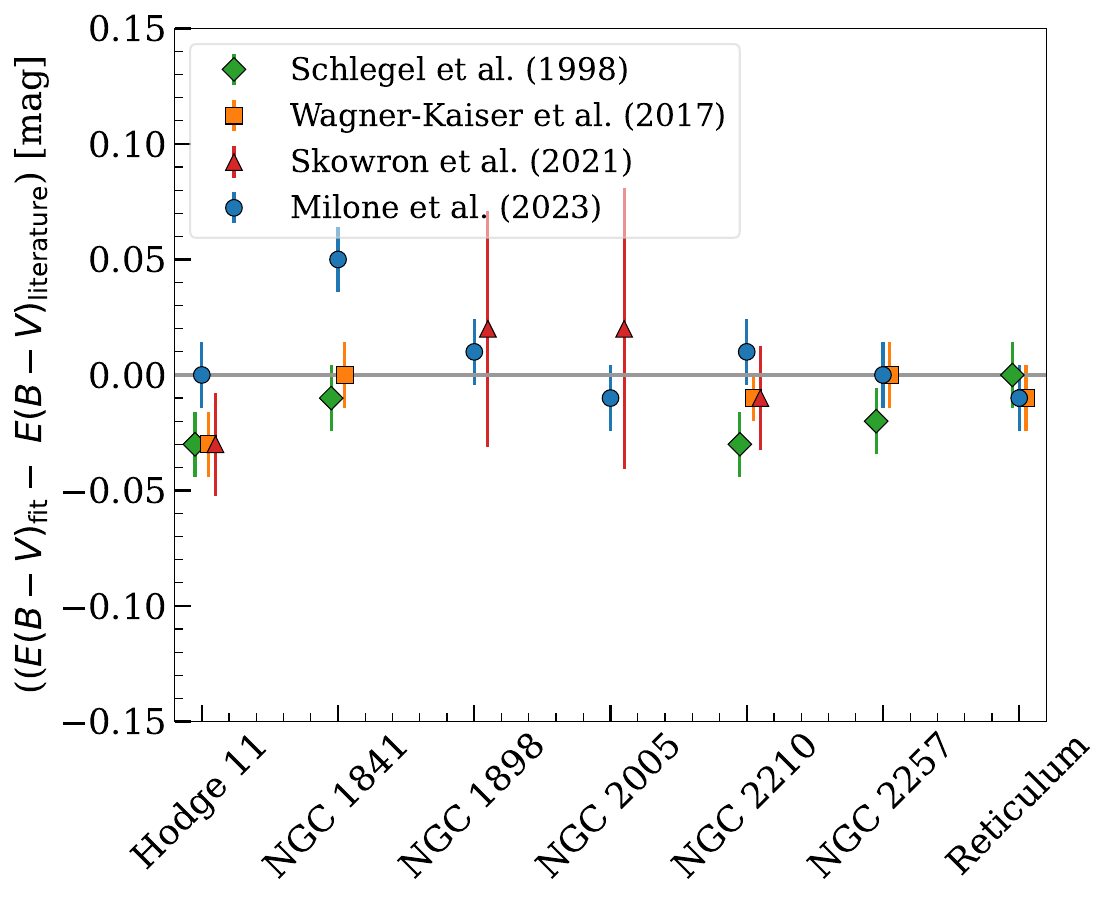} \\
\end{tabular}
\caption{Comparison between the cluster parameters derived in this work and literature values. Top-left panel: Differences between the measured cluster ages and ages derived by \citet{Milone23a}. Top-right panel: Differences between the inferred metallicity and the results from \citet{Grocholski06}, \citet{Mucciarelli21}, \cite{Song21} and \citet{Sarajedini24}. 
Bottom-left panel: Differences between the distance modulus derived in this study and the measurements from \citet{Wagner-Kaiser17} and \citet{Milone23a}. Bottom-right panel: Differences between the derived $E(B-V)$ values and the reddenings from \citet{Wagner-Kaiser17} and \citet{Milone23a}, as well as the resulting values from the reddening maps from \citet{Schlegel98} and \citet{Skowron21}.
\label{fig:clusters_compar}
}

\end{figure*}

The results of fitting process are reported in Table~\ref{tab:isochrone_fits}. The values provided here are the average values resulting from the individual fits to the two CMDs. The uncertainties associated with each value are derived accounting for the upper and lower limits of the inferred quantities from both fits. We will be using these values for our further analysis. We want to highlight here that the values we present in this study (and the associated uncertainties) are to be intended in a relative sense, rather than in an absolute one. 

As a test of consistency of our results, we compared our inferred values for age, metallicity, distance and reddening to values from the literature. The top-left panel in Fig.~\ref{fig:clusters_compar} shows the difference in cluster ages between our values and the ones from \citet{Milone23a}, which are derived from isochrone fitting, as well. We note that there is a mean offset of $\sim$0.6 ~Gyr, that most likely originates from the use of different theoretical isochrone models. While the BaSTI set of isochrones is used in this work, \citet{Milone23a} employed the database of Padova models \citep{Marigo17}. However, since we are only interested in relative ages of the clusters, and not absolute ages, that offset is not an issue. More importantly, we found a good concordance in the relative ages, with a spread around the mean difference of only 0.2~Gyr.

In the top-right panel of Fig.~\ref{fig:clusters_compar}, we present the comparison of our inferred metallicities with literature spectroscopic measurements from \citet{Grocholski06}, \citet{Song21} and \citet{Mucciarelli21}, as well as with metallicity estimates based on periods and amplitudes of fundamental mode RR Lyrae stars \citep{Sarajedini24}. Since we determined the global metallicity [M/H], in order to perform a proper comparison, we need to convert the measured literature [Fe/H] values to [M/H], taking into account the respective $\alpha$-element abundances of the clusters. \citet{Salaris93} showed that the global metallicity can be derived with the following relation:
\begin{equation}\label{eq:mh}
    \rm [M/H] = [Fe/H] + log(0.694 \times 10^{[\alpha/Fe]} + 0.301).
\end{equation}
Note that here the coefficients have been slightly modified with respect to the original solution by \citet{Salaris93}, taking into account the reference solar mixture used in BaSTI models \citep[see also][]{Massari23}. We converted the [Fe/H] measurements from \citet{Mucciarelli21} to [M/H], using their determined $\alpha$-element abundances. Since the other literature studies only measured the iron content for their samples of clusters, we determined an average $\alpha$-abundance for the clusters in common by least-squares fitting Equation~\ref{eq:mh} for [$\alpha$/Fe] using the [Fe/H] values from the different literature studies and the [M/H] inferred by us. We found mean $\alpha$-abundances for the literature values of [$\alpha$/Fe]=0.32$\pm$0.05~dex. The shown error bars in the top-right panel of Fig.~\ref{fig:clusters_compar} are the combined uncertainties from our measurements, the literature values, as well as the uncertainties from the $\alpha$-abundance estimates. 
The comparison with the literature [M/H] measurements shows a good agreement for most of the clusters, with an average difference between our estimates and the literature of $-0.01$~dex, with $\sigma$=0.09. For Hodge~11, NGC~1841, NGC~2005, NGC~2257 and Reticulum, the different measurements from the literature agree well within the uncertainties with our inferred values. For NGC~2210, our value is slightly higher than the measurements by \citet{Mucciarelli21} and \citet{Sarajedini24}. Only for NGC~1898, the metallicity inferred by us shows a discrepancy of more than 2$\sigma$ with respect to the value determined by \citet{Mucciarelli21}, but it agrees within the 1$\sigma$ uncertainties with the metallicity derived by \citet{Sarajedini24}.

The bottom-left panel of Fig.~\ref{fig:clusters_compar}, presents the comparison of our results for the distance modulus with the findings from \citet{Wagner-Kaiser17} and \citet{Milone23a}. \citet{Wagner-Kaiser17} determined the distances to a sample of six old LMC clusters using two different methods. The first method is based on comparing the horizontal-branch magnitudes to the ones of a sample of Galactic GCs with known distances. The second method involves fitting the colours and magnitudes of several reference subdwarfs to the main-sequence fiducial lines of the LMC clusters. The comparison reveals an excellent agreement with the results from \citet{Milone23a} and the subdwarf-based distances from \citet{Wagner-Kaiser17}, whereas there are significant discrepancies from the values derived using the horizontal-branch brightness. These discrepancies are likely due to the number of assumptions involved when estimating the distances based on the horizontal branch brightness. 
As an additional consistency check on the distances and the sensitivity of the fitting on the assumed priors, we used literature RR Lyrae-based distances \citep[from][]{Walker89, Walker90, Walker92a, Walker92b, Kuehn13, Cusano21} to several of the clusters in our sample as new priors and performed the fitting again. The differences in distance modulus, compared to the original results, are all less than 0.01~mag ($\lesssim$0.2~kpc, at the distance of the LMC), suggesting the fitting is not very sensitive to the exact choice of the distance prior.

The bottom-right panel of Fig.~\ref{fig:clusters_compar}, finally, shows the comparison of our our derived values for the interstellar reddening with measurements from \citet{Wagner-Kaiser17} and \citet{Milone23a}, as well as with values resulting from the extinction maps from \citet{Schlegel98} and \citet{Skowron21}. Overall, we find a good agreement with the derived values for the clusters from the literature studies. Only for NGC~1841, \citet{Milone23a} derived a considerably smaller $E(B-V)$ (about 0.05~mag lower) compared to our measurement. Further, our reddenings match very well the values from the two extinction maps. The large error bars for NGC~1898 and NGC~2005 from the comparison with the \citet{Skowron21} reddening map result from the high uncertainties within the central regions of the LMC that are present in their map.

\begin{figure}[h]
\centering
\includegraphics[width=\columnwidth]{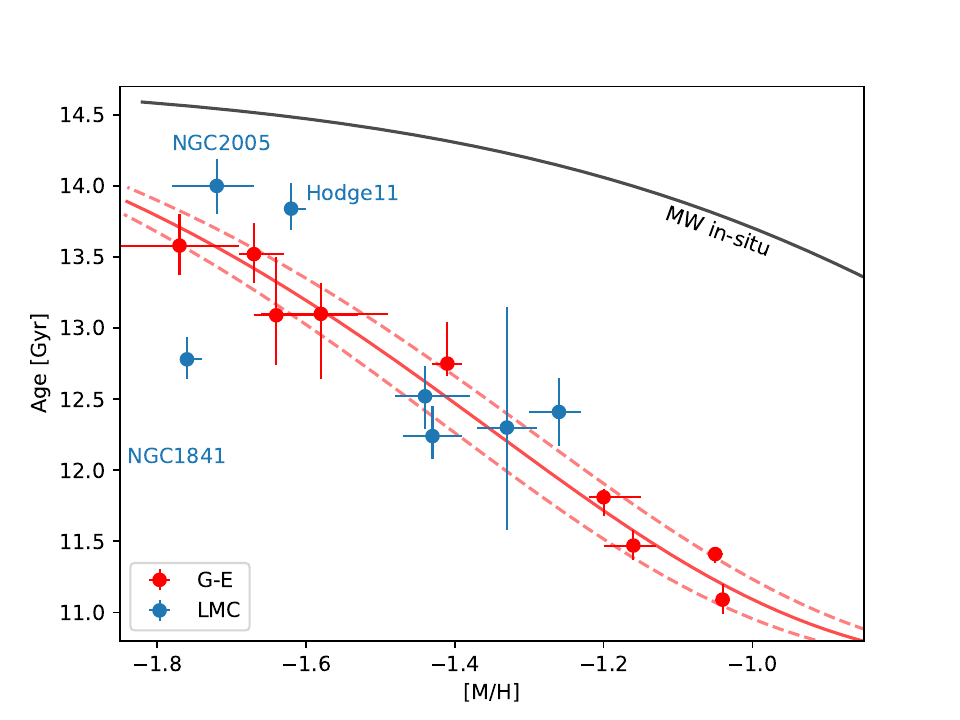}
\caption{Resulting age-metallicity relation of the seven old LMC clusters studied in this work (blue symbols). For a comparison, nine Milky Way GCs selected as genuine members of GSE (red symbols) are also shown. The red and black solid lines follow the parametric age-metallicity model derived by \cite{Massari19} for GSE and the MW, respectively. The red dashed lines illustrate the associated uncertainties of the GSE model. \label{fig:amr}}
\end{figure}

The above consistency checks have demonstrated that our isochrone-fitting method is able to produce reliable results for old star clusters in the LMC. With these results in hand we first focus on the ages and metallicities of the clusters in our sample. Figure~\ref{fig:amr} shows the positions of the seven LMC clusters (blue symbols) in the age-metallicity space, compared to that of the nine Milky Way GCs selected as genuine members of the Gaia-Sausage-Enceladus \citep[GSE,][]{helmi18, belokurov18} merger event by \citet[][red symbols]{Aguado-Agelet25}. 
We chose GSE for the comparison, since its estimated mass ($\sim6\times10^{10}$~M$_{\odot}$, \citealt{helmi18}; $\sim9\times10^{10}$~M$_{\odot}$, \citealt{Kruijssen20}
) is comparable to the one of the LMC \citep[$1-2\times10^{11}$~M$_{\sun}$][]{Vasiliev23}.
The first feature that is immediately evident is that our sample of LMC clusters can be divided into two age groups. NGC~2005 and Hodge~11 with ages of 14.0~Gyr and 13.8~Gyr, respectively, are the two oldest clusters within our sample. The remaining five clusters (NGC~1841, NGC~1898, NGC~2210, NGC~2257 and Reticulum) are significantly younger with similar ages. Their mean age is 12.4~Gyr with a small dispersion of $\sigma_t$=0.2~Gyr. GSE clusters describe two epochs of formation as well, but at clearly different times.
The second remarkable feature is that the general trend of the age-metallicity relation (AMR) of the LMC clusters closely follows that shown by the GSE GCs. This is also evident when comparing the location of the LMC clusters with the parametric AMR model derived by \cite{Massari19} for the GSE GCs (red solid line). The shape of this model ultimately depends on the mass of the progenitor and on its star-formation rate, assumed to be constant \citep[][]{prantzos08}. The similarity of the two observed AMRs points towards a similar early evolution of the two dwarf galaxies, as possibly indicated by the chemistry of their metal-poor populations, too \citep[e.g.,][]{matsuno21}. One small difference is given by the location of the two oldest LMC clusters, which seem to be located systematically above the theoretical model, even when considering the associated uncertainty (red dashed lines).  
Given the shape of the AMR \citep[as described by][]{prantzos08}, this might indicate that the mass, or the star-formation efficiency, of the LMC at the time of the formation of these clusters was slightly higher than that of GSE.
We note here that, despite the similar [M/H] values, Hodge~11 and NGC~2005 have substantially different chemical compositions \citep{Mucciarelli21}. NGC~2005 shows chemical abundance ratios in several elements that are in strong contrast to the ones of the other ancient LMC clusters. 
Thus, Hodge~11 and NGC~2005 formed from distinct gas clouds with different enrichment histories. We will discuss the origin of NGC~2005 in more detail in Sec.~\ref{sec:orbits}.
For the sake of comparison, in Fig.~\ref{fig:amr} we also show the AMR model used by \cite{Massari19} to describe in-situ MW GCs: the much larger early mass and star-formation rate of the MW manifest in a significantly different global trend (see the black solid line) compared to the other two dwarf galaxies \citep[see also e.g.,][]{marinfranch09, leaman13, souza24}.
The final striking feature is the unique position of NGC~1841 in the age-metallicity space. While the other clusters broadly follow the theoretical AMR model (suggesting they formed within the LMC), NGC~1841 has a significantly lower metallicity than the other clusters in the younger age group. Instead, its [M/H] value is consistent with the one of NGC~2005, the oldest cluster in our sample. This anomalous position of NGC~1841 with respect to the other studied clusters prompts us to the speculation that NGC~1841 might not have formed within the LMC, but rather originates from an environment with a lower star-formation efficiency, and thus slower chemical enrichment rate.

As an independent test to verify the relative ages among the LMC star clusters in our sample, we employed the horizontal method \citep[e.g.][]{VandenBerg90, Salaris98}, that is based on measuring the distance in a given colour between the main-sequence turn-off point and the base of the red giant branch. This difference in colour, for a given metallicity, is dependent on the age of a cluster, with older clusters having smaller distances. 
Since there is a strong metallicity dependence in the horizontal method, we only performed direct comparisons between Hodge~11 and NGC~1841, as well as between NGC~2210, NGC~2257 and Reticulum. For NGC~2005 and NGC~1898, a direct comparison with the other clusters is not possible, since they have observations in different filters (F435W and F814W for NGC~2005, and F438W and F814W for NGC~1898). The tests verified that Hodge~11 is indeed older than NGC~1841 and, further, that NGC~2210, NGC~2257 and Reticulum are approximately coeval.

Additionally, in Appendix~\ref{sec:isoc_comparsion} we illustrate the CMDs of Hodge~11, NGC~2005 and NGC~1841 together with isochrones of different age and metallicities.

\section{Kinematic structure of the old LMC clusters}\label{sec:kinematic_structure}

To build on our hypothesis regarding the origin of NGC~1841 and to reach an unequivocal conclusion, we now take into account the kinematics of the clusters within the LMC, as well. Combining the observed positions on the sky and 3D velocities from Table~\ref{tab:clusters_params} with the determined distances from the isochrone fitting (Table~\ref{tab:isochrone_fits}), we have the full 6D phase-space information for the clusters, allowing to examine in detail the velocities and positions of the clusters within the LMC and estimate their orbits within the potential of the galaxy.

\begin{table*}\small
\centering
\caption{Positions and velocity components of our sample of old star clusters projected
into the reference frame of the LMC. \label{tab:clusters_lmc_params}}
\begin{tabular}{lcccccc}
\hline\hline
\noalign{\smallskip}
Cluster ID & R & $\phi$ & Z & V$_{\phi}$ & V$_{R}$ & V$_{Z}$\\
& [kpc] & [deg] & [kpc] & [km\,s$^{-1}$] & [km\,s$^{-1}$] & [km\,s$^{-1}$]\\
\noalign{\smallskip}
\hline
\noalign{\smallskip}
 Hodge 11 & 3.98 $\pm$ 0.05 & 338.55 & $-$1.22 $\pm$ 0.20 &   79.3 $\pm$  9.3 & $-$25.5 $\pm$  9.9 &    15.4 $\pm$ 5.7 \\
 NGC 1841 & 8.71 $\pm$ 0.09 &  36.77 &    9.32 $\pm$ 0.15 & $-$7.5 $\pm$  6.1 &    41.9 $\pm$  6.7 &    62.7 $\pm$ 5.4 \\
 NGC 1898 & 1.05 $\pm$ 0.13 &  89.02 & $-$0.88 $\pm$ 0.19 &   27.8 $\pm$  9.3 & $-$43.0 $\pm$  8.0 &    31.9 $\pm$ 5.5 \\
 NGC 2005 & 1.02 $\pm$ 0.07 & 325.82 &    0.52 $\pm$ 0.19 &    8.4 $\pm$  9.1 &  $-$7.9 $\pm$  9.2 & $-$12.9 $\pm$ 8.1 \\
 NGC 2210 & 4.57 $\pm$ 0.08 & 311.44 &    1.18 $\pm$ 0.18 &  111.8 $\pm$  7.6 & $-$27.6 $\pm$  7.4 & $-$85.7 $\pm$ 4.0 \\
 NGC 2257 & 7.91 $\pm$ 0.17 & 289.39 & $-$0.04 $\pm$ 0.38 &   71.3 $\pm$  8.2 &     5.1 $\pm$  8.3 &     4.3 $\pm$ 4.0 \\
Reticulum & 9.87 $\pm$ 0.18 & 207.35 &    0.88 $\pm$ 0.56 &   41.1 $\pm$ 12.4 &  $-$9.7 $\pm$ 13.4 &     1.2 $\pm$ 7.4 \\

\noalign{\smallskip}
\hline

\end{tabular}

\tablefoot{$\phi$ denotes the position angle within the LMC disc, measured anti-clockwise from the positive X-axis. A positive tangential velocity V$_{\phi}$ follows the clockwise rotation pattern of the LMC.
}
\end{table*}

\subsection{Velocity and coordinate transformation}

Before we can analyse the kinematics of the seven old LMC clusters in our sample and determine their orbits within the galaxy, we first need to transform their observed on-sky positions, distances, as well as their line-of-sight velocities and PMs into the reference frame of the LMC. This frame is defined as a right-handed orthogonal coordinate system, centred on the dynamical centre of the galaxy. The X--Y plane is aligned with the disc of the LMC, where the X axis is along the line of nodes (the line where the LMC-disc plane intersects the sky plane). To project the observed motions and positions into
the coordinate system of the LMC we applied the formalism derived by \citet{vanderMarel01b} and \citet{vanderMarel02}. To solve the transformation equations, we adopt the following values for the position, orientation and motion of the LMC disc: the coordinates of the LMC dynamical centre ($\alpha_0$, $\delta_0$) = (79.95$\degr$, $-69.31\degr$), the PM of the LMC centre-of-mass ($\mu_{\alpha}$cos($\delta$)$_0$, $\mu_{\delta, 0}$) = (1.867, 0.314)~mas\,yr$^{-1}$, the inclination angle $i$ = 33.5$\degr$ and the angle of the line-of-nodes $\Theta$ = 129.8$\degr$ \citep[all taken from][]{Niederhofer22}. Additionally, we adopt a distance to the LMC of 49.59~kpc \citep{Pietrzynski19} and a line-of-sight velocity of 262.2~km\,s$^{-1}$ \citep{vanderMarel02}. The transformed positions and velocities in the LMC frame, expressed in cylindrical coordinates, are presented in Table~\ref{tab:clusters_lmc_params}. The associated errors are calculated by propagating the measurement uncertainties throughout the transformation equations. We like to stress here that the exact values of the reported positions and velocities are specific to the adopted parameters of the LMC. In the literature there exist a broad variety of measurements of these parameters, employing different tracers and methods \citep[see, e.g.][for an overview]{Niederhofer22}. However, any other choice of the adopted orientation, position and motion of the LMC would not have significantly affected our final results. We verified this by performing the dynamic analysis described below again assuming the parameters of the LMC as determined by \citet{vanderMarel14}.

\subsection{Kinematics within the LMC}

\begin{figure*}
\begin{tabular}{cc}
\includegraphics[width=0.95\columnwidth]{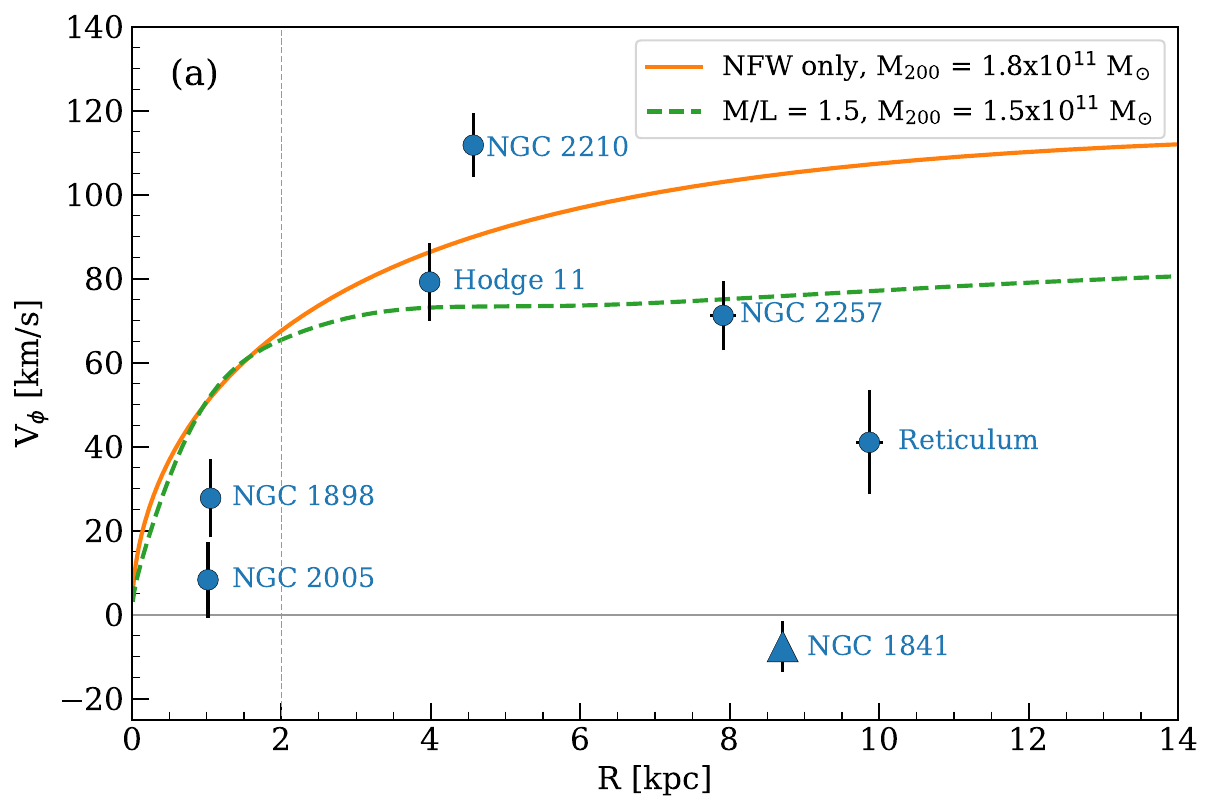} &
\includegraphics[width=0.95\columnwidth]{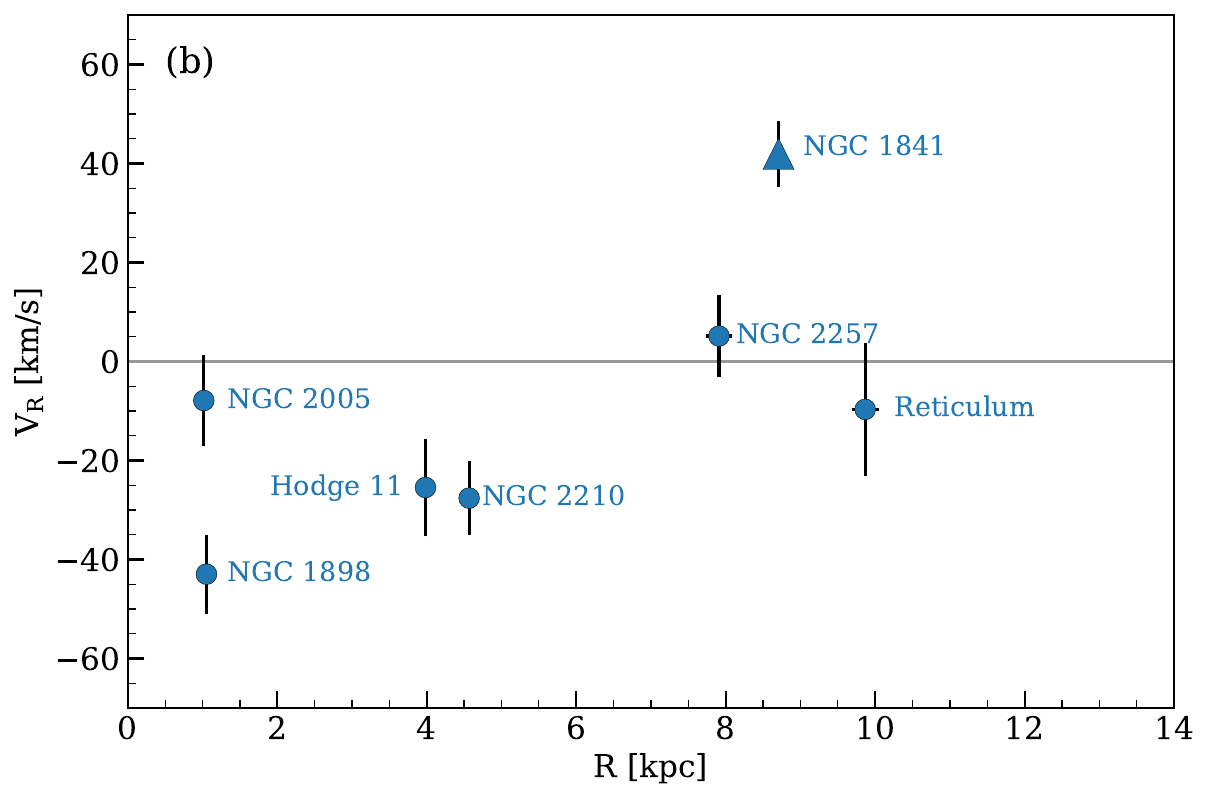} \\
\includegraphics[width=0.95\columnwidth]{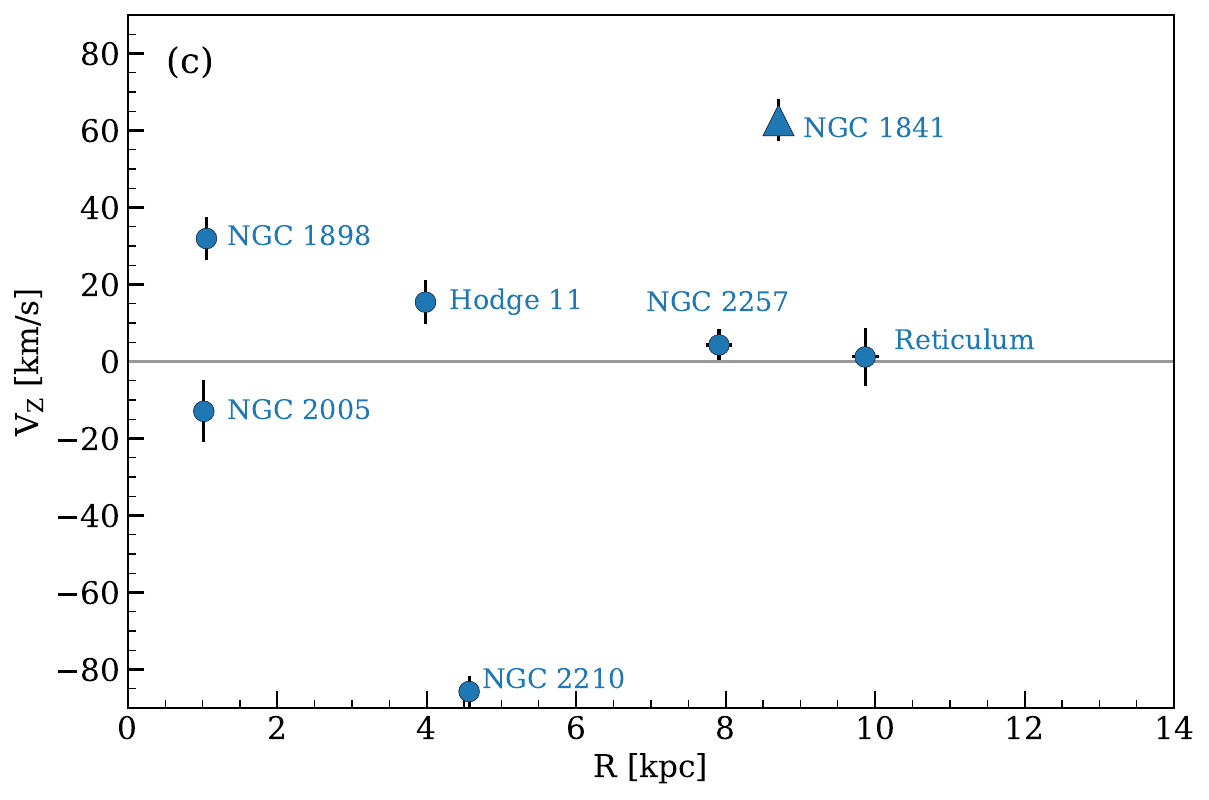} &
\includegraphics[width=0.95\columnwidth]{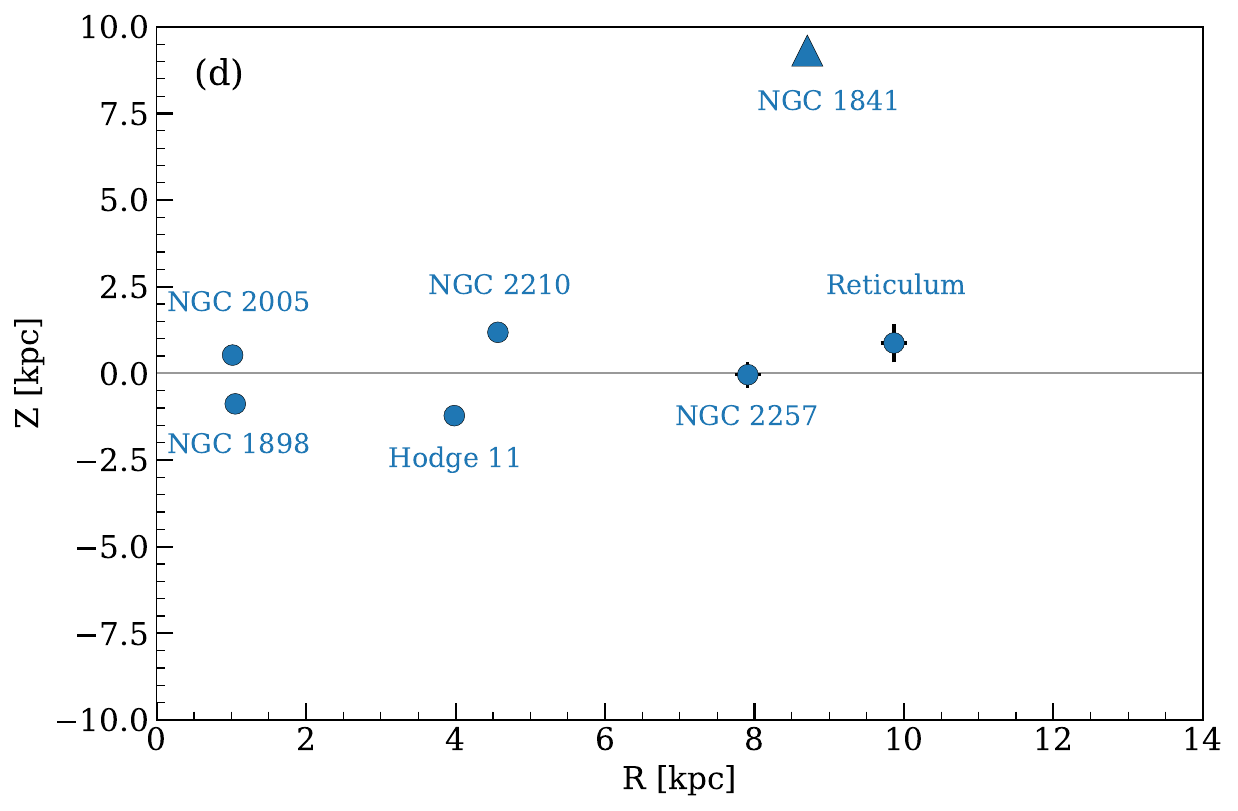} \\
\end{tabular}
\caption{Outline of the clusters' velocities and positions within the LMC. In each panel, NGC~1841 is highlighted with triangle symbol. Panel (a) shows the tangential velocity V$_{\phi}$ as a function of the radial distance from the LMC centre R. Also shown are model circular velocities resulting from a pure NFW profile (orange solid line) and a model composed of a dark matter halo and a stellar disc and bar (green dashed line). The size of the stellar bar is indicated by the vertical dashed line. The radial velocity V$_{R}$ as a function of R is presented in panel (b). Panel (c) illustrates the out-of-plane velocity V$_{Z}$ as a function of R and panel (d) shows the vertical distance from the plane Z as a function of R. 
\label{fig:clusters_kinematics}
}
\end{figure*}

In Fig.~\ref{fig:clusters_kinematics}, panels (a) and (b) present, for each cluster, the tangential (V$_{\phi}$, panel a) and radial (V$_{R}$, panel b) velocity components, respectively, as a function of the cylindrical galactocentric radius R. Also shown in panel (a) are two rotation curves of the LMC as determined by \citet{Kacharov24}, based on dynamical models. For the first model (orange line), the authors fitted axisymmetric Jeans dynamical models to stars in the \textit{Gaia} DR3 catalogue with measured PM and line-of-sight velocities. In this model, it is assumed that the mass distribution of the LMC follows a pure dark-matter halo with a spherical \citet*[][NFW]{Navarro97} mass profile. Their best-fit model gives a virial mass of the LMC of M$_{200}$ = $1.81\times10^{11}$~M$_{\sun}$ at a virial radius r$_{200}$ of 75~kpc, with an overdensity $\Delta_{\rm c}$=200 relative to the critical density of the Universe. The second model (green dashed line) is based on a Schwarzschild orbit superposition method and is fitted to the line-of-sight velocity field of the LMC. In this model \citet{Kacharov24} also included the contribution from the stellar component of the galaxy, which is described as a triaxial bar and an axisymmetric disc. The shown model assumes a mass-to-light ratio M/L of 1.5~M$_{\sun}$/L$_{\sun}$ and a total luminosity of the LMC of $1.3\times10^{9}$~L$_{\sun}$.

We can see from panels (a) and (b) of Fig.~\ref{fig:clusters_kinematics} that most clusters have tangential velocities that are smaller than the predicted model rotation curves and also have non-zero radial velocities, suggesting the clusters are on non-circular, elongated orbits. A particular case depicts NGC~1841, highlighted with a triangle symbol in all panels of Fig.~\ref{fig:clusters_kinematics}. It has a small negative tangential velocity, that is almost consistent with being zero (V$_{\phi} = -7.5\pm6.1$~km\,s$^{-1}$) and a relatively large radial velocity (V$_{R} = 41.9\pm6.7$~km\,s$^{-1}$), which leads us to the conclusion that NGC~1841 is on a highly eccentric, retrograde orbit.

Panel (c) of Fig.~\ref{fig:clusters_kinematics} shows the vertical, out-of-plane velocity component (V$_{Z}$) of the clusters as a function of R. Studying the PMs of 15 old LMC star clusters using data from the \textit{Gaia} DR2, \citet{Piatti19} claimed the existence of two kinematically distinct populations of clusters, a disc and a halo population. Their separation criterion was solely based on a sharp limit in $|$V$_{Z}|$. According to their classification, NGC~1841, NGC~2005 and NGC~2210 would belong to the disc family of clusters, whereas Hodge~11, NGC~1898, NGC~2257 and Reticulum would be halo clusters \citep[see table~1 by][]{Piatti19}. However, as it will become apparent below when we discuss the orbits of the clusters (see Section~\ref{sec:orbits}), our data does not support this categorisation. Our measurements indicate that most clusters in our sample have absolute vertical velocities smaller than 40~km\,s$^{-1}$. The two remarkable exceptions with extreme out-of-plane velocities are NGC~1841 with V$_{Z} = 62.7\pm5.4$~km\,s$^{-1}$ and NGC~2210 with V$_{Z} = -85.7\pm4.0$~km\,s$^{-1}$. From their study of the kinematics of LMC star clusters, \citet{Bennet22} determined an unusually high vertical velocity of NGC~2210 ($-111\pm8$~km\,s$^{-1}$), as well. Their measured total velocity of NGC~2210 ($\sim$160~km\,s$^{-1}$) is close to the local escape velocity of the LMC \citep[$\sim$166~km\,s$^{-1}$, as determined by][based on LMC runaway stars]{Boubert17}, thus they speculated that NGC~2210 might not physically be associated with the LMC, but rather originated from the Milky Way halo. Our measurements result in a smaller total velocity of NGC~2210 of $143.6\pm6.8$~km\,s$^{-1}$ (mainly due to a closer distance to NGC~2210 and a smaller assumed line-of-sight velocity, compared to what has been used by \citealt{Bennet22}).
Its position in the age-metallicity space further coincides with the other LMC clusters of similar ages. Thus, we conclude that NGC~2210 is likely a genuine member of the LMC.

Finally, panel (d) of Fig.~\ref{fig:clusters_kinematics} illustrates the vertical distance of the seven clusters from the disc plane of the LMC, as a function of the cylindrical radius R. Also in this plot, NGC~1841 stands out as an outlier with respect to the other clusters in the sample. While the current positions of most clusters are close to the LMC disc plane ($|$Z$|\lesssim$1.2~kpc), NGC~1841 is currently located more than 9~kpc above the plane of the galaxy, suggesting NGC~1841 is on a highly inclined orbit around the LMC.

\begin{figure}
\begin{tabular}{c}
\includegraphics[width=0.43\textwidth]{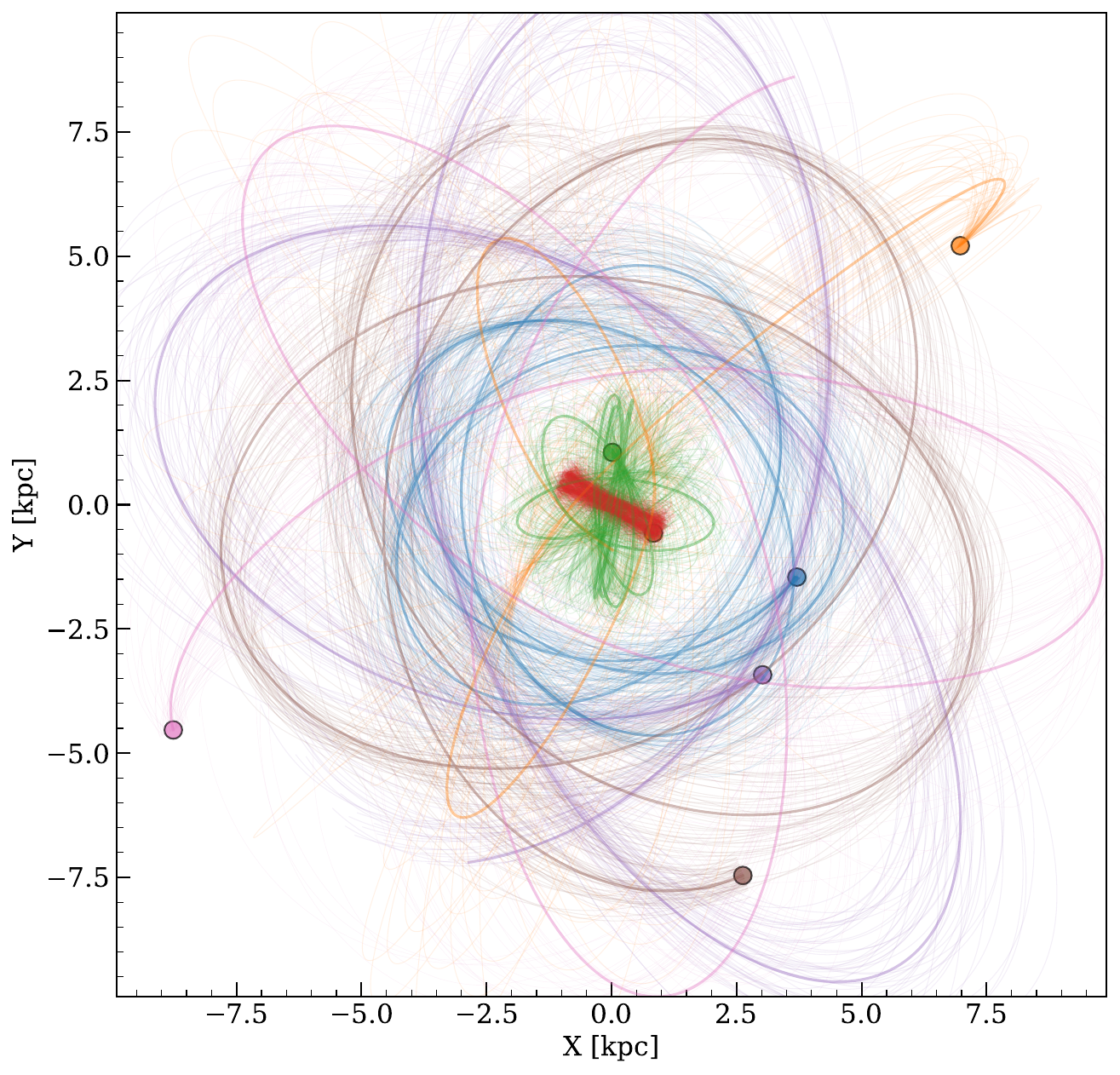} \\
\includegraphics[width=0.43\textwidth]{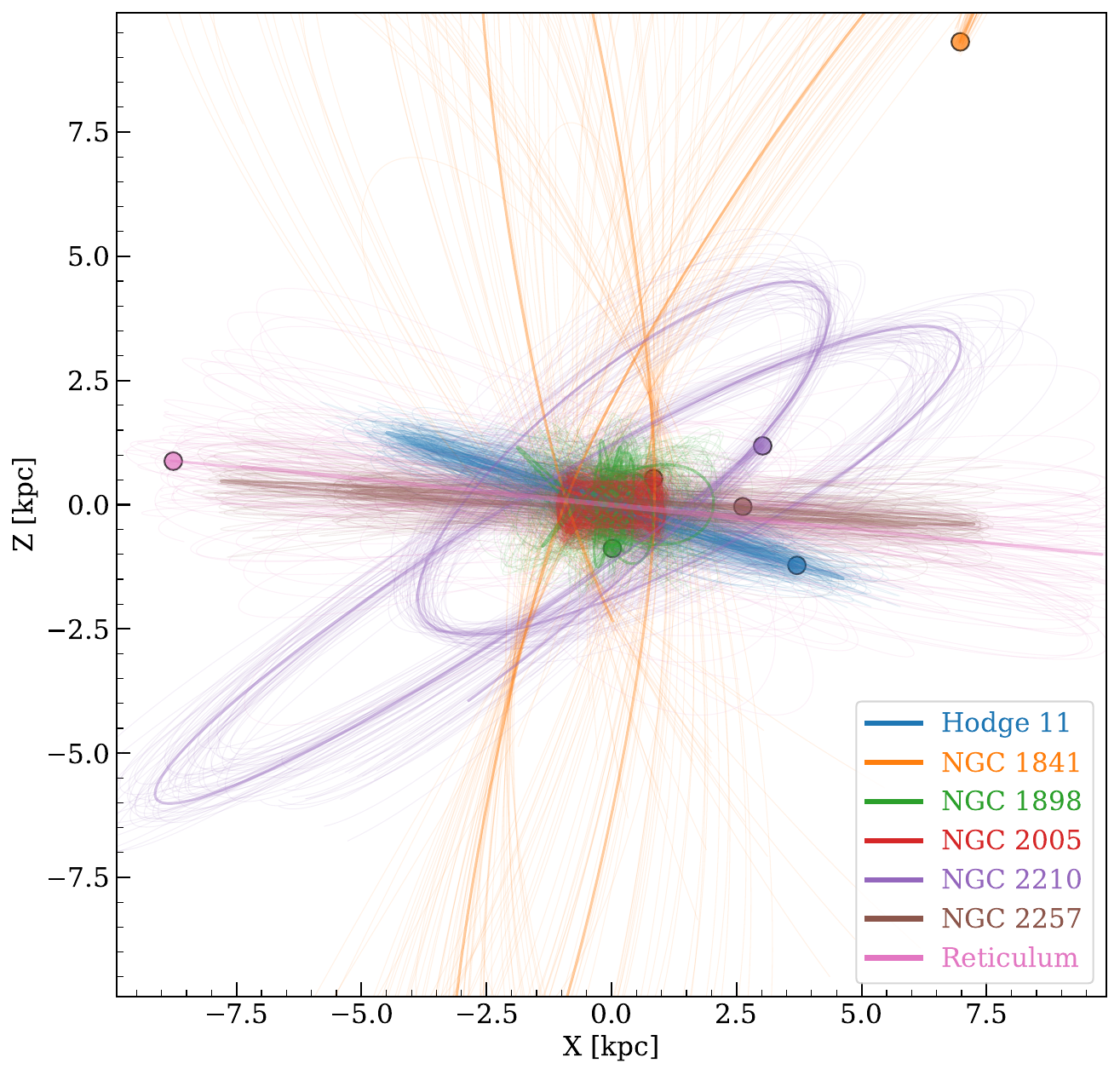} \\
\includegraphics[width=0.43\textwidth]{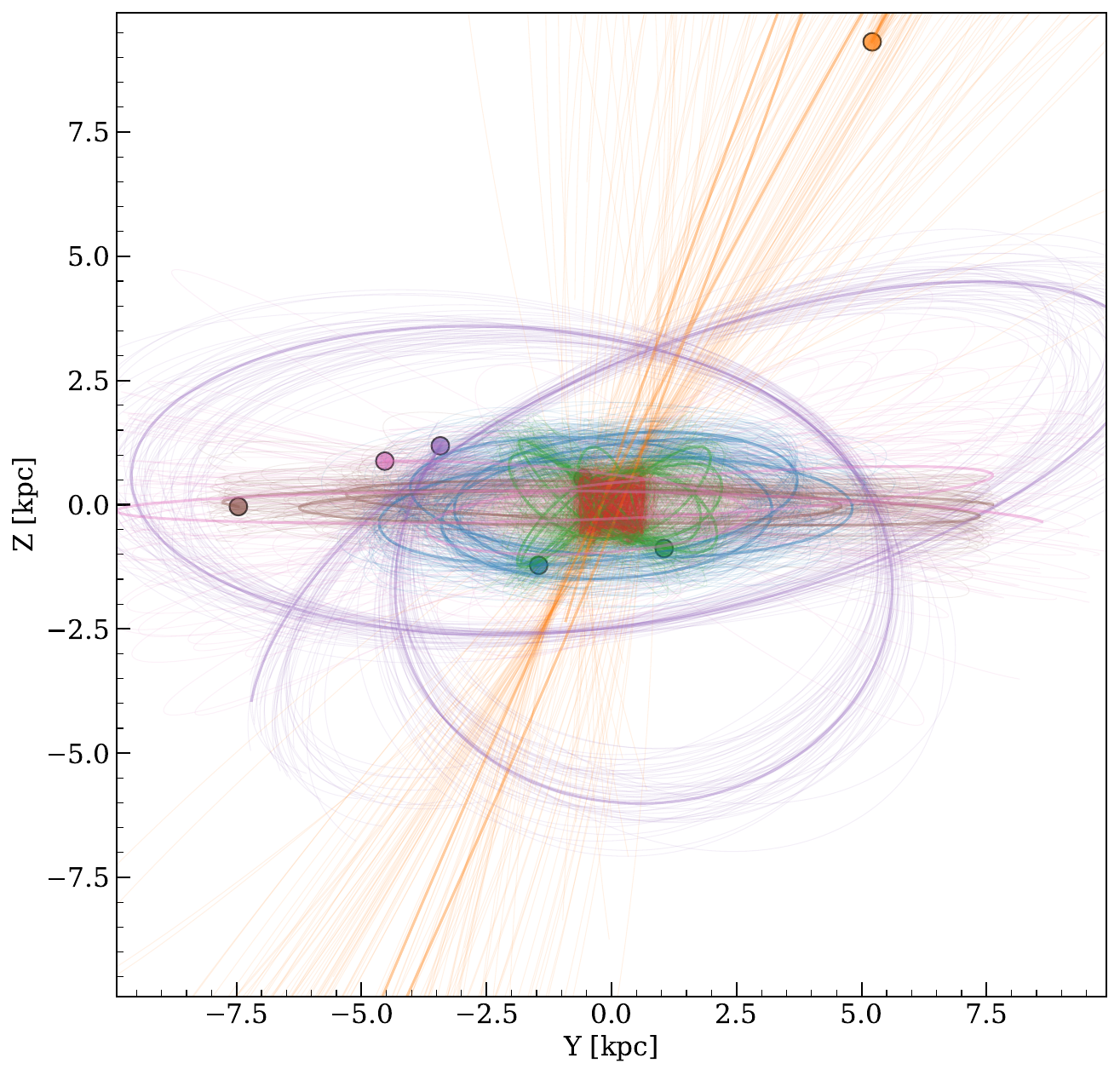} \\

\end{tabular}
\caption{Orbits of the seven clusters within the LMC, integrated forward for 1~Gyr. The three panels show projections in the XY-- (top), XZ-- (middle) and YZ--plane (bottom). The current positions of the clusters are indicated by a circle. 
\label{fig:clusters_orbits}
}

\end{figure}

\subsection{Cluster orbits}\label{sec:orbits}

The sum of the evidence collected above strongly suggests that NGC~1841 differs from the other studied LMC clusters also from a dynamical point of view. The measured velocities and positions of the clusters, however, only form a snapshot of the momentary state of the cluster system. To investigate the kinematic structure of the cluster system in a more descriptive way, we now model and examine the orbits of the clusters within the LMC. To trace the orbits, we used two of the LMC potentials that are derived by \cite{Kacharov24} based on Jeans and Schwarzschild models, with the following considerations in mind: For clusters within the inner 2~kpc of the LMC (the region where the dynamics are dominated by the bar), we employed the multi-component model that is composed of a spherical NFW dark matter halo, as well as an axisymmetric stellar disc and a triaxial stellar bar component. As pointed out by \citet{Kacharov24}, the pure NFW model fails to properly reproduce the stellar kinematics within the central parts of the LMC. We note here that their best-fit Schwarzschild model results in an un-physically low mass-to-light ratio of M/L = 0.3~M$_{\sun}$/L$_{\sun}$, and a virial mass of $4\times10^{12}$~M$_{\sun}$, more than a factor of 20 of what is now the widely accepted mass estimate of the LMC \citep[$1-2\times10^{11}$~M$_{\sun}$, see][for a recent review]{Vasiliev23}. Thus, we opt to use their model with the mass-to-light ratio fixed at M/L = 1.5~M$_{\sun}$/L$_{\sun}$, which is a more realistic value for LMC-like galaxies \citep{vanderMarel02}. For radii larger than 2~kpc, we used the pure NFW mass profile, resulting from the Jeans modelling. While within about 2~kpc, both the pure NFW model and the composite models with different assumptions about the mass-to-light ratios produce similar results \citep[see figure~17 in][]{Kacharov24}, they start to deviate beyond 2~kpc. Thus, the pure NFW model depicts the best representation of the outer parts of the LMC. This is also backed up by independent measurements of the LMC potential. \citet{Watkins24} used the sample of LMC clusters from \citet{Bennet22} as tracers of the LMC potential, assuming a NFW mass density profile as well. Although the radial limits of the dynamical model fits from \citet{Kacharov24} are only 6.2~kpc, and the dataset used by \citet{Watkins24} extends out to about 13~kpc, the profile obtained by \citet{Watkins24} is identical to the one determined by \citet{Kacharov24}. 
We want to emphasize here that the potential used here is a simplified (no spiral arms, central bar instead of off-centre bar) and static potential of an isolated (no tidal forces from the Milky Way and SMC) LMC galaxy. Thus, it reflects an idealized case and the interpretations below are made with these caveats in mind.

For the integration of the cluster orbits, we employed the
\texttt{python} package \texttt{galpy}\footnote{\href{http://github.com/jobovy/galpy}{http://github.com/jobovy/galpy}} \citet{Bovy15}. For each cluster, we integrated the orbit forward for 1~Gyr. To estimate the uncertainties in the orbital parameters of the clusters, we also created for each cluster 500 realisations of the initial conditions, by randomly drawing from Gaussian distributions centred around the measurements for their positions and velocities, and with standard deviations corresponding to the measurement uncertainties. 

Fig.~\ref{fig:clusters_orbits} presents the resulting orbits of the clusters, together with 50 of these random realisations. Each panel shows a different projection of the 3D orbits: in the XY-- (top panel), the XZ-- (middle panel) and the YZ-- plane (bottom panel). The current positions of the clusters are also indicated by coloured circles. It is immediately evident from the plots that NGC~1841 (orange-coloured lines) is on a peculiar orbit that is very different from the ones of the other clusters. With an inclination of $69\pm6\degr$ its orbital plane is highly inclined with respect to the LMC disc plane. Thereby, the cluster can reach a maximum vertical height Z$_{\rm max}=14.85\pm1.29$~kpc above the disc. Additionally, as already predicted from its measured velocity components, NGC~1841 is on a highly eccentric orbit, with an eccentricity of $0.90\pm0.04$. We note that also the Milky Way hosts several GCs with orbital properties similar to the ones of NGC~1841, where all of those clusters have an accreted origin. 
The combined properties of NGC~1841 (its unique position in the age-metallicity space and its peculiar kinematics) that we gathered in this work form the consistent picture that NGC~1841 is most likely not a genuine LMC cluster. Since the LMC is massive enough to be accompanied by its own system of dwarf satellite galaxies, including the SMC, of which low-mass ones have been identified within the last years \citep[see, e.g.][]{Erkal20, Patel20, Vasiliev24}, NGC~1841 might have been accreted by the LMC from a smaller companion during a minor merger event. Using E-MOSAICS simulations, \citet{Pfeffer20} analysed the relation between the observed orbital parameters of accreted clusters and the properties of the progenitor host galaxies. They found that either recent merger events or low-mass progenitors preferentially place clusters at orbits with large apocentres. 
Thus, NGC~1841 might have been brought in during an early accretion event of a dwarf spheroidal galaxy. The fact that four of the in-situ LMC clusters formed at similar times with comparable metallicities suggests that they emerged from a starburst event, which, potentially, could have been linked to that merger. 
NGC~1841 is about 0.5~Gyr older than these four clusters. Thus, if the merger that brought in NGC~1841 was indeed linked to this starburst event, then NGC~1841 has formed shortly before the merger.
Alternatively, NGC~1841 might have been stripped recently from the SMC during a close encounters with the LMC.

The orbits of the other clusters display several interesting features as well that we will discuss in the following. Hodge~11, NGC~2257 and Reticulum (blue, brown and pink lines, respectively, in Fig.~\ref{fig:clusters_orbits}) orbit the LMC within its disc plane. The orbits of NGC~2257 and Reticulum, with inclination angles of $3\pm3\degr$ and $6\pm12\degr$, respectively, are consistent, within the errors, of being coplanar with the plane of the disc. Only the orbit of Hodge~11 shows a moderate inclination ($18\pm3\degr$). So, they follow clear disc-like kinematics. This is in contrast to the results by \citet{Piatti19} who classified these three clusters as belonging to a halo population.

NGC~1898 and NGC~2005 (green and red lines, respectively, in Fig.~\ref{fig:clusters_orbits}) are situated close to the centre of the LMC, so their orbits are largely affected by the central bar structure of the galaxy. Both clusters seem to follow box-like orbits \citep{Binney82}, whereas
the orbit of NGC~1898 is more extended (apocentre distance of $2.25\pm0.38$~kpc, with a maximum vertical distance of $1.30\pm0.34$~kpc). NGC~2005 follows a very elongated orbit (eccentricity of $0.89\pm0.07$) along the direction of the bar. Based on the peculiar chemical composition of NGC~2005, \citet{Mucciarelli21} suggested that it once belonged to a small dwarf galaxy that has been accreted by the LMC. From the inferred orbit, NGC~2005 appears to be contained within the very central parts of the LMC, with an apocentre distance of $1.18\pm0.13$~kpc and a maximum height above the plane of $0.57\pm0.20$~kpc. Thus, the position and dynamic properties of NGC~2005 measured in this study do not display any obvious hint of the cluster being accreted to the LMC, unless the cluster was accreted so long ago that it had the time to sink in so deep in the potential well of the LMC. \citet{Mucciarelli21} noted as well that the line-of-sight velocity of NGC~2005 is similar to other clusters in its vicinity. 
One possible explanation might be that the orbit of NGC~2005 was affected by dynamical friction which caused the cluster to spiral to the central parts of the galaxy as it moved through the galactic disc. One the other hand, \citet{Piatti23} argued that the chemical abundance pattern of NGC~2005 is not a univocal sign of an ex-situ formation but might also be consistent with the cluster having formed within the LMC. In this case, NGC~2005 would constitute one of the oldest building blocks of the LMC, given its old age, that formed near the centre of the potential well of the galaxy and its chemical composition might resemble that of the still inhomogeneous gas at the earliest time of the formation of the LMC. As already noted in Sec.~\ref{sec:results}, the specific abundances in NGC~2005 result in a [M/H] value that is compatible with the AMR of the in-situ LMC clusters. Thus the position of NGC~2005 in the age-metallicity plane is not able to confirm or refute an ex-situ origin. 
Regardless of the nature of NGC~2005, the old age of the cluster demonstrates that peculiar chemical evolutions already emerge at very early times.

Apart from NGC~1841, also NGC~2210 (purple-coloured lines in Fig.~\ref{fig:clusters_orbits}) follows an extended orbit that is significantly inclined with respect to the LMC disc, although not as extreme as NGC~1841. We found an inclination angle of $37\pm2\degr$ with a maximum height above the plane of Z$_{\rm max}=6.01\pm0.48$~kpc and an apocentre distance of $11.02\pm0.98$ kpc. The current position of the cluster near its pericentre ($4.29\pm0.14$~kpc) results in the rather high measured velocity. Given the kinematic characteristics of NGC~2210, it is tempting to interpret it as a halo cluster. However, as discussed above, it shares the same region in the age-metallicity space as the other clusters with disc-like kinematics and similar ages. Moreover, NGC~2210 is the only cluster in our sample that shows this type of orbital properties. Thus, at the moment we cannot make any strong claim about its nature. An alternative interpretation might be that the orbit of NGC~2210 has been disturbed in the past by a close encounter with the SMC, tilting the orbital plane.


\section{Summary and Conclusions}\label{sec:conclusions}

In this work, we applied the isochrone-fitting algorithm developed within the CARMA project to a sample of seven old LMC star clusters to derive homogeneous and robust estimates for their distances, ages and metallicities. In combination with their full 3D kinematic data, resulting from multi-epoch \textit{HST} PM measurements and literature line-of-sight velocities, we determined the positions and velocity components of the clusters in the frame of the LMC galaxy. With this collection of information we explored any connection between the orbital properties of the clusters and their location in the age-metallicity space for the first time in a self-consistent way. 

We found that the AMR of the LMC clusters is remarkably similar to that described by Milky Way GCs associated to the GSE merger event. This suggests that during their early evolution, the two galaxies share similar mass and star-formation efficiency. Two of the clusters in our sample, Hodge~11 and NGC~2005, are significantly older than the other five clusters, which are approximately coeval. Of this younger group of clusters, NGC~1841 stands out as a clear outlier in the age-metallicity space, owing to its low [M/H] value. The kinematic data further underlined the peculiarity of NGC~1841. In contrast to the other clusters, NGC~1841 follows a highly eccentric orbit that is almost perpendicular to the plane of the LMC disc and reaches up to about 15~kpc above the galaxy. 
Based on the combined properties of NGC~1841, we concluded that this cluster has not formed within the LMC, but has been accreted by the LMC from a smaller dwarf galaxy with a lower star formation efficiency, likely a dwarf spheroidal galaxy. There is also the possibility that NGC~1841 might have been stripped from the SMC during one of the last close encounters between the LMC and SMC. 

We analysed the orbital properties of the other clusters as well and found that Hodge~11, NGC~2257 and Reticulum follow disc-like orbits. Our data suggest that NGC~2210 is likely bound to the LMC, but follows a more inclined orbit. Based on its orbit, this cluster might be a halo cluster, but since its position in the age-metallicity plane is consistent with the ones of the clusters with disc-like kinematics, we speculated that its orbit might have been disturbed.
Finally, we did not find any clear kinematic evidence that NGC~2005 has been accreted by the LMC. Instead, our data suggest the cluster, which is the oldest among the LMC ones, is constrained to the inner regions of the galaxy. We interpreted this result with two different scenarios: either NGC~2005 has an ex-situ origin and it has been dragged to the centre by, e.g. dynamical friction, or NGC~2005 is a genuine member and formed within the early LMC where its peculiar chemistry reflects the conditions at the time of formation of the galaxy. 

Our results based on about half of the old LMC cluster population, have already shown that precise PM measurements in combination with homogeneous determinations of the clusters' distances, ages and metallicities are a powerful tool to assess the nature of the LMC cluster population. A natural follow-up of this work would be to expand our sample to the remaining ancient LMC clusters to study the full population. Since most of them have already one existing epoch of \textit{HST} observations, such expanded study could be done with little observational effort in the future.


\begin{acknowledgements} 
We thank the anonymous referee for constructive comments and suggestions that improved the quality of our paper. 
This research was funded by DLR grant 50 OR 2216.
DM, and SC acknowledge financial support from PRIN-MIUR-22: CHRONOS: adjusting the clock(s) to unveil the CHRONO-chemo-dynamical Structure of the Galaxy” (PI: S. Cassisi). 
A.M .acknowledges support from the project "LEGO – Reconstructing the building blocks of the Galaxy 
by chemical tagging" (P.I. A. Mucciarelli). granted by the Italian MUR through contract PRIN 2022LLP8TK\_001. 
SS acknowledges funding from the European Union under the grant ERC-2022-AdG, {\em "StarDance: the non-canonical evolution of stars in clusters"}, Grant Agreement 101093572, PI: E. Pancino.
Support for this work was provided by NASA through grants for program GO-16478 from the Space Telescope Science Institute (STScI), which is operated by the Association of Universities for Research in Astronomy (AURA), Inc., under NASA contract NAS5-26555. This work is based on observations made with the NASA/ESA Hubble Space Telescope, obtained from the Data Archive at the Space Telescope Science Institute. This work has made use of data from the European Space Agency (ESA) mission \textit{Gaia} (\url{https://www.cosmos.esa.int/gaia}), processed by the Gaia Data Processing and Analysis Consortium (DPAC, \url{https://www.cosmos.esa.int/web/gaia/dpac/consortium}). Funding for the DPAC has been provided by national institutions, in particular the institutions participating in the Gaia Multilateral Agreement.
This research made use of \texttt{astropy},\footnote{\href{http://www.astropy.org}{http://www.astropy.org}} a community-developed core \texttt{python} package for Astronomy \citep{Astropy13, Astropy18}, \texttt{iphython} \citep{Perez07}, \texttt{Jupyter Notebook} \citep{Kluyver16}, \texttt{matplotlib} \citep{Hunter07}, \texttt{numpy} \citep{Harris2020} and \texttt{scipy} \citep{Virtanen20}.
\end{acknowledgements}



\bibliographystyle{aa}
\bibliography{references}


\begin{appendix} 

\section{Isochrone fitting results}\label{sec:appendix}
\begin{figure*}
     \centering
     \begin{subfigure}[b]{\columnwidth}
         \centering
         \includegraphics[width=\textwidth]{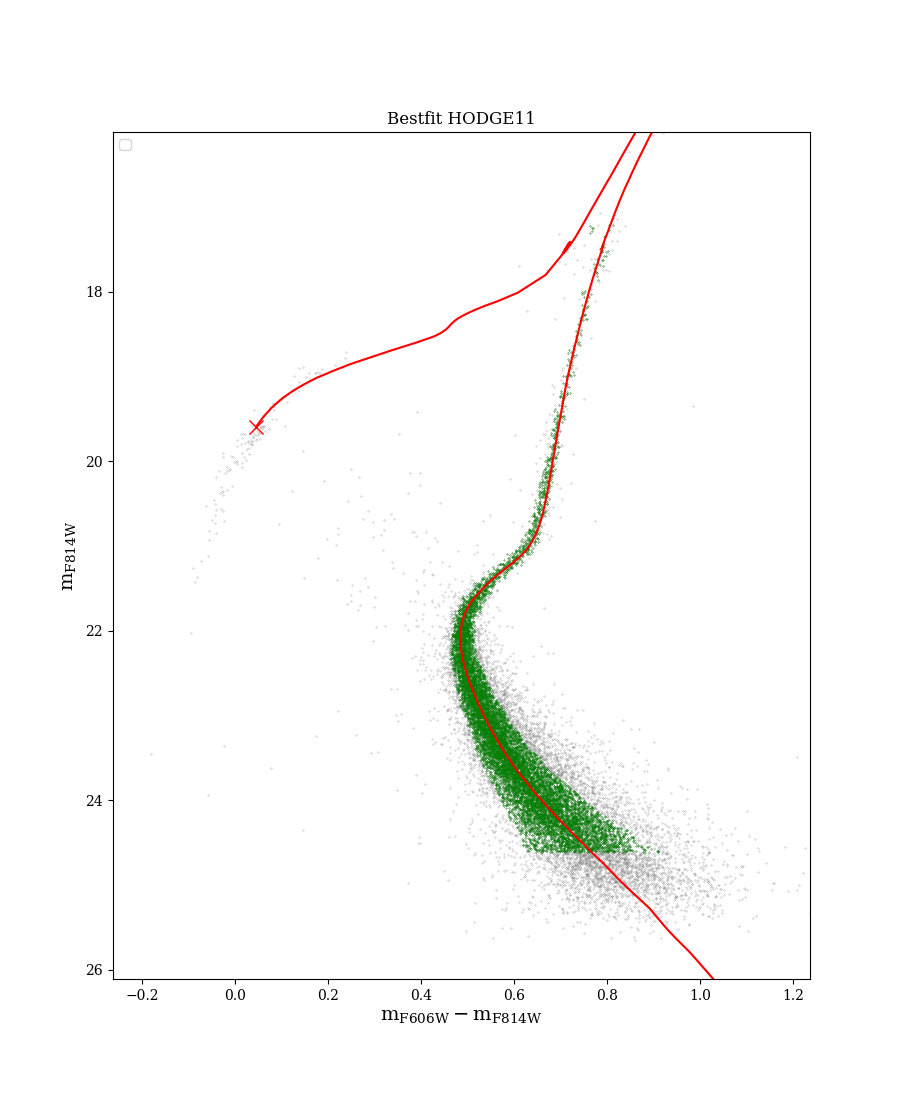}
         \caption{}
     \end{subfigure}
     \hfill
     \begin{subfigure}[b]{\columnwidth}
         \centering
         \includegraphics[width=\textwidth]{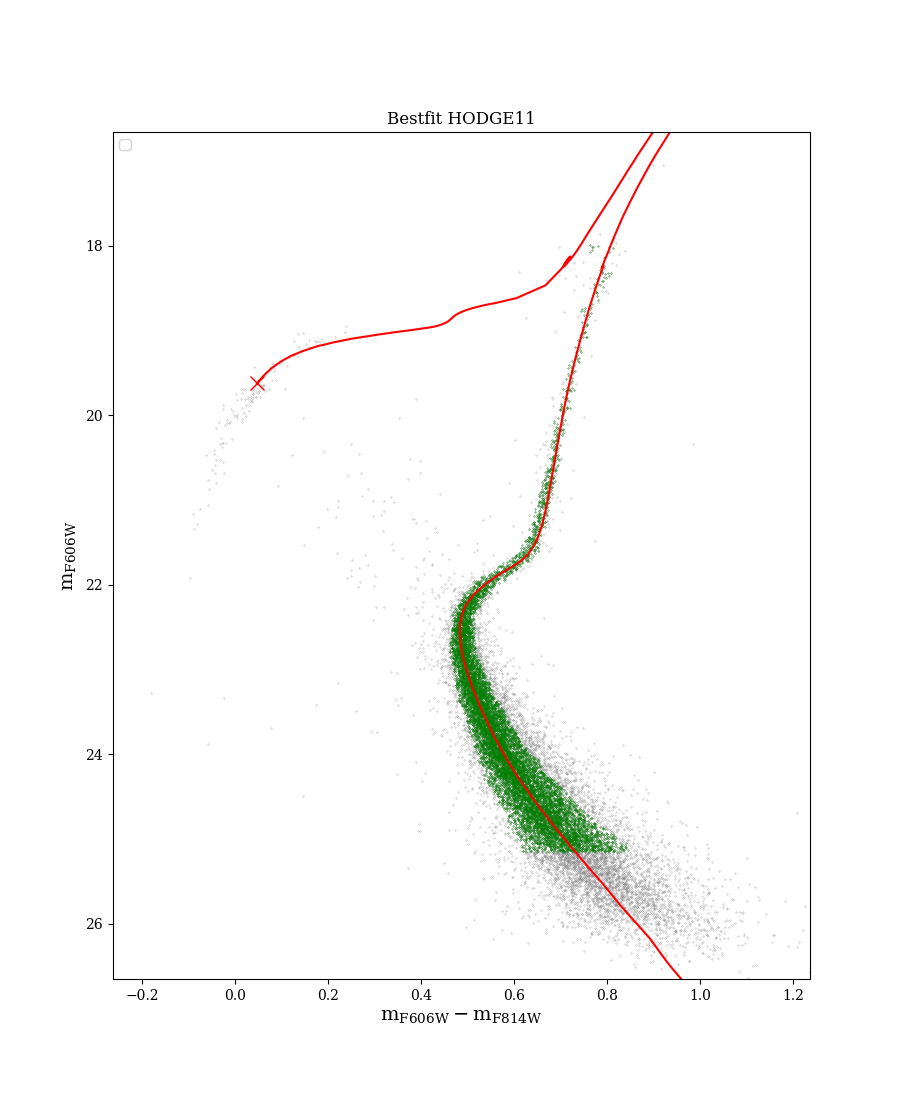}
         \caption{ }
     \end{subfigure}

     \begin{subfigure}[b]{\columnwidth}
         \centering
         \includegraphics[width=\textwidth]{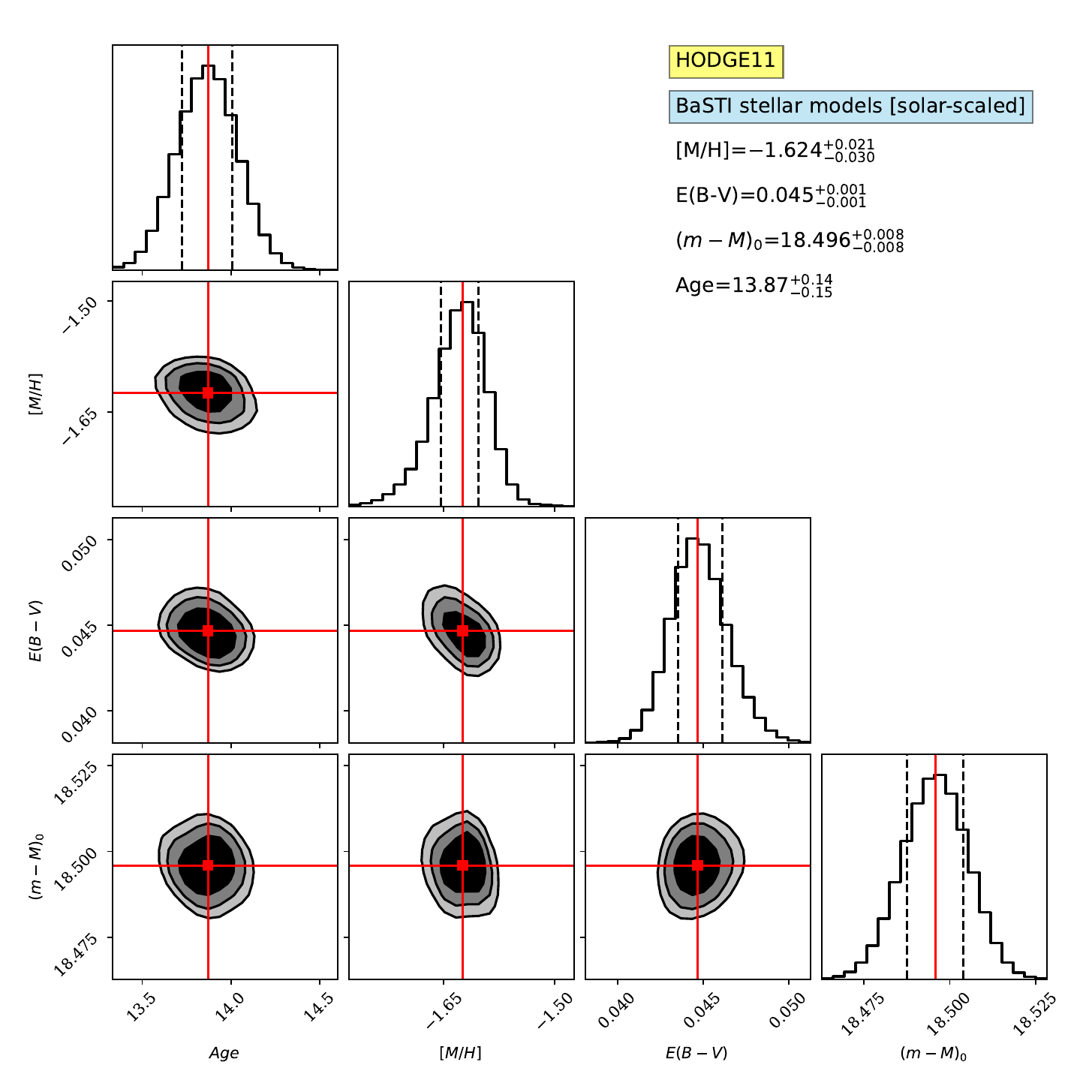}
         \caption{}
     \end{subfigure}
     \hfill
     \begin{subfigure}[b]{\columnwidth}
         \centering
         \includegraphics[width=\textwidth]{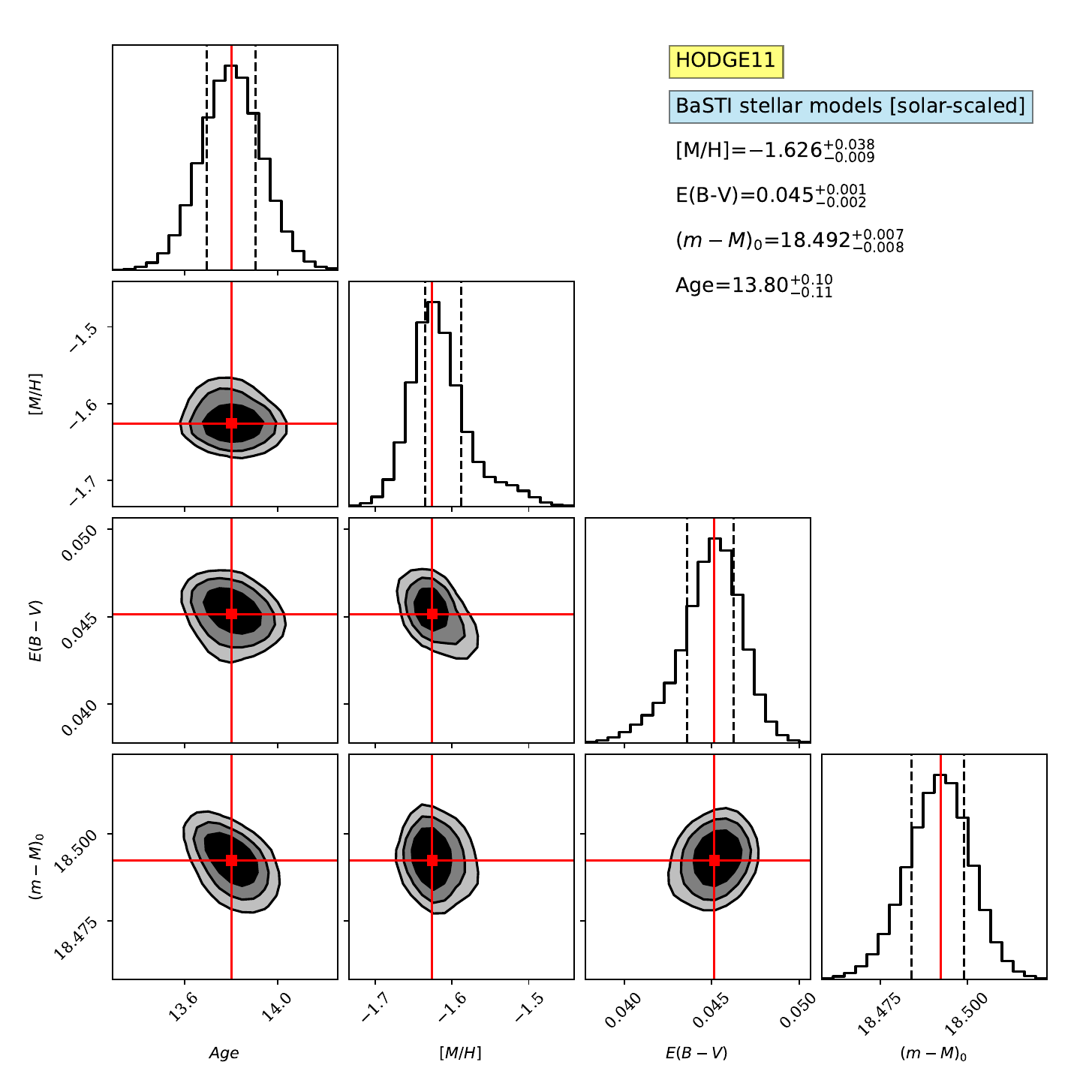}
         \caption{}
     \end{subfigure}
        \caption{Isochrone fitting results for Hodge~11. (a) Best-fit isochrone model in the $m_{\rm F814W}$ vs. $m_{\rm F606W}-m_{\rm F814W}$ CMD. (b) Best-fit isochrone model in the $m_{\rm F606W}$ vs. $m_{\rm F606W}-m_{\rm F814W}$ CMD. 
        The red X-symbol shown in both CMDs corresponds to the location of the zero-age horizontal branch.
        (c): Corner plot of the posterior probability distributions of pairwise model parameters for the $m_{\rm F814W}$ vs. $m_{\rm F606W}-m_{\rm F814W}$ CMD. The best-fit parameters are quoted in the labels. 
        (d): Corner plot of the posterior probability distributions of pairwise model parameters for the $m_{\rm F606W}$ vs. $m_{\rm F606W}-m_{\rm F814W}$ CMD. The best-fit parameters are quoted in the labels.
        }
        \label{fig:hodge11_isofit}
\end{figure*}

\begin{figure*}
     \centering
     \begin{subfigure}[b]{\columnwidth}
         \centering
         \includegraphics[width=\textwidth]{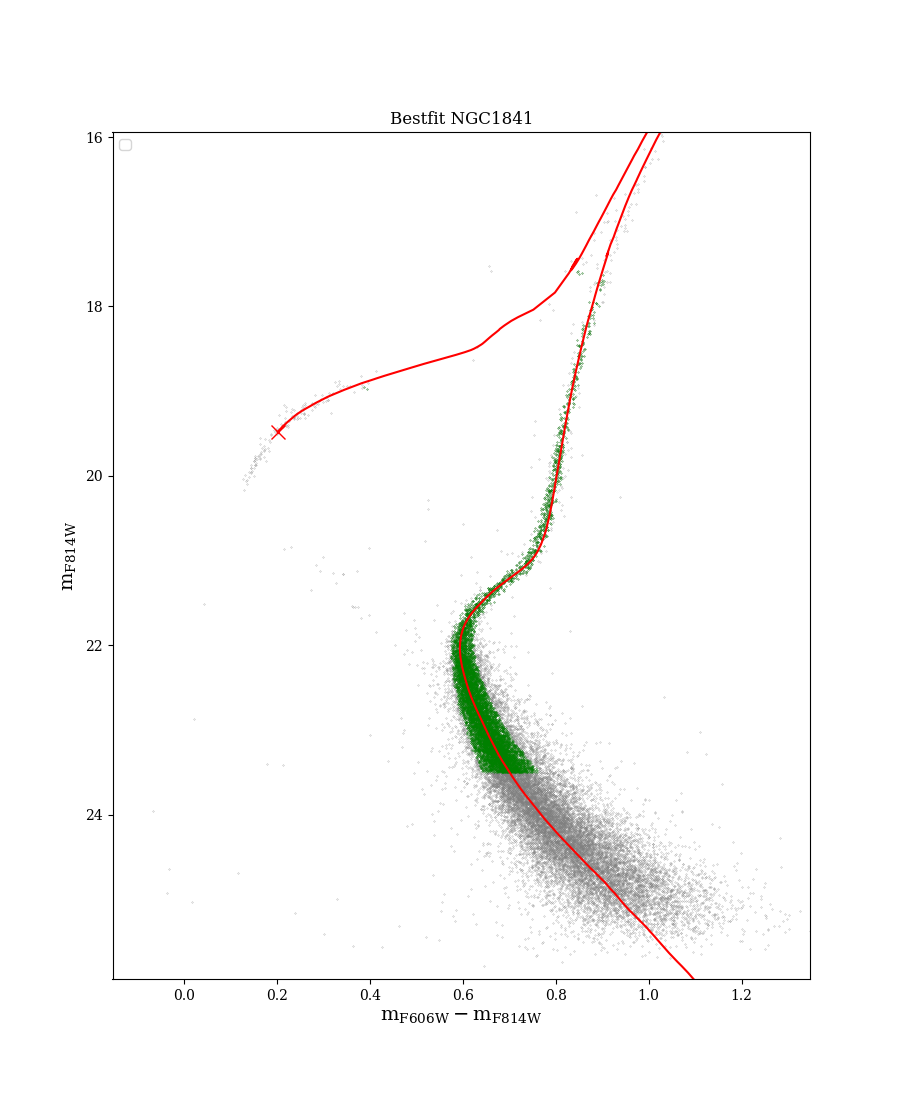}
         \caption{}
     \end{subfigure}
     \hfill
     \begin{subfigure}[b]{\columnwidth}
         \centering
         \includegraphics[width=\textwidth]{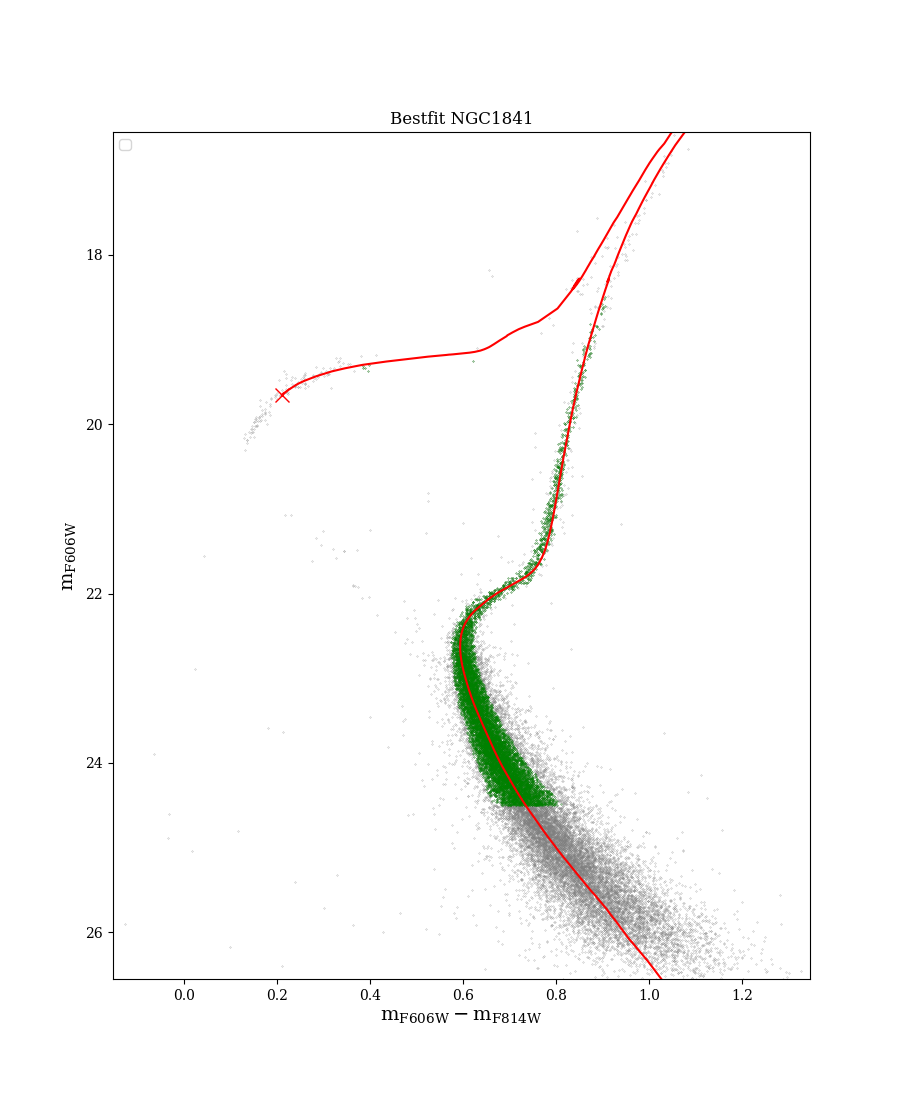}
         \caption{ }
     \end{subfigure}

     \begin{subfigure}[b]{\columnwidth}
         \centering
         \includegraphics[width=\textwidth]{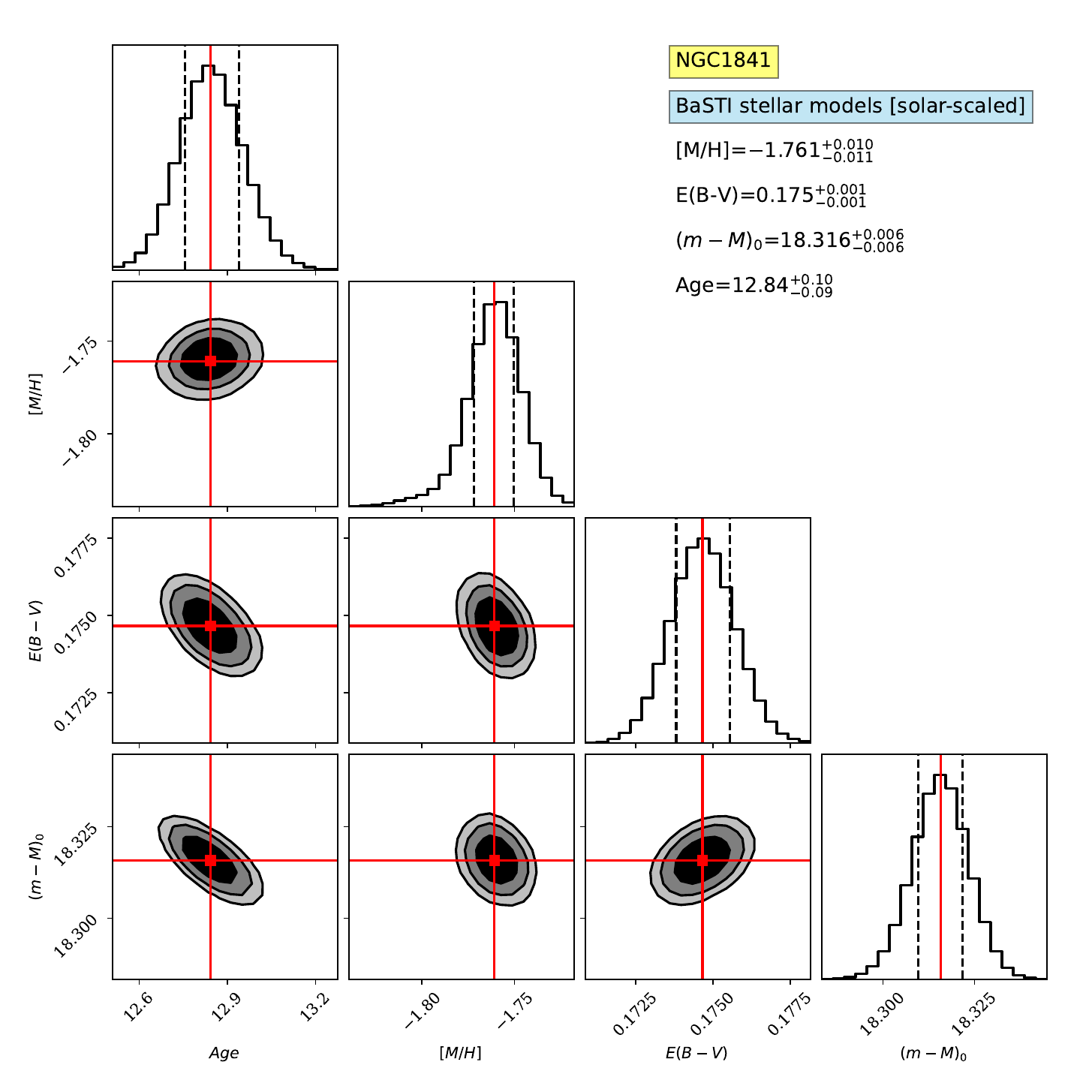}
         \caption{}
     \end{subfigure}
     \hfill
     \begin{subfigure}[b]{\columnwidth}
         \centering
         \includegraphics[width=\textwidth]{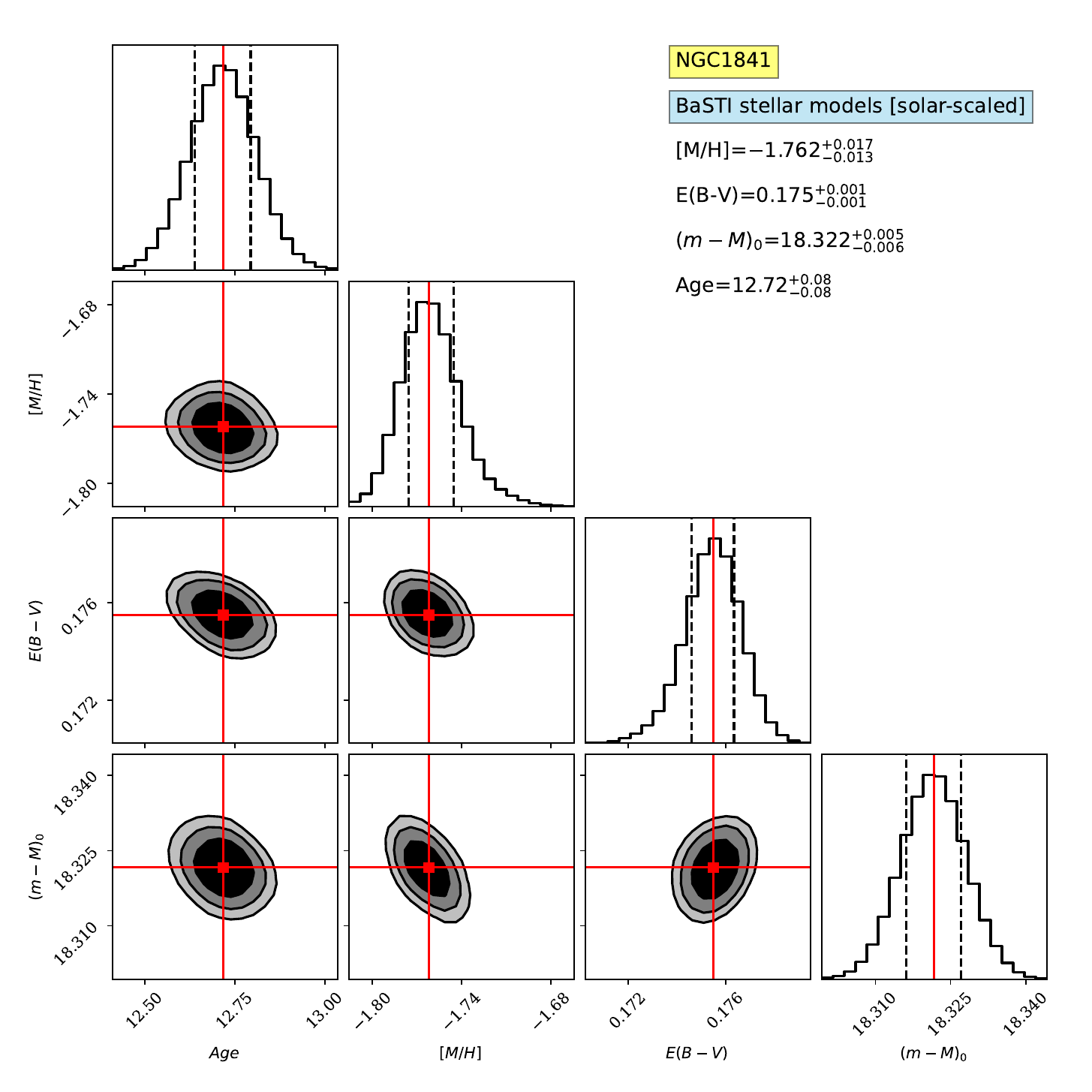}
         \caption{}
     \end{subfigure}
        \caption{Same as Fig.~\ref{fig:hodge11_isofit}, but now for NGC~1841.
        }
        \label{fig:ngc1841_isofit}
\end{figure*}

\begin{figure*}
     \centering
     \begin{subfigure}[b]{\columnwidth}
         \centering
         \includegraphics[width=\textwidth]{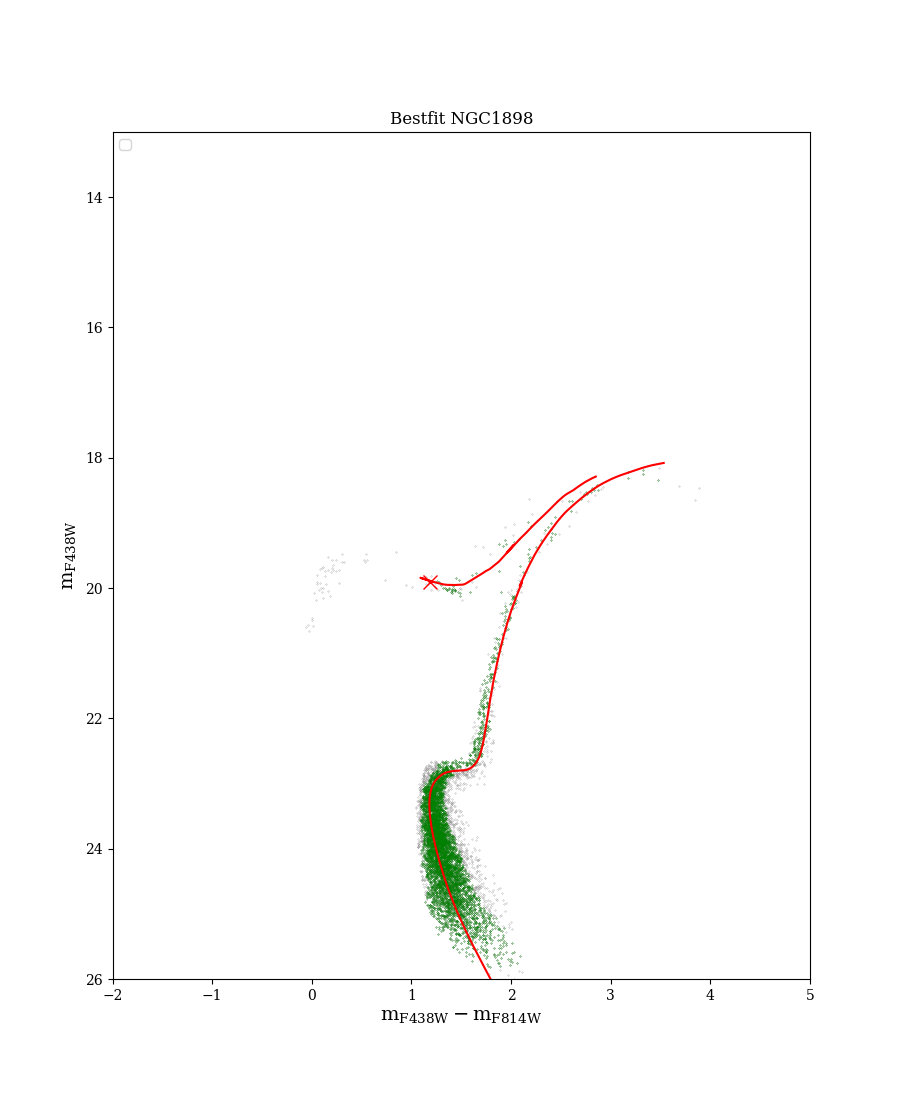}
         \caption{}
     \end{subfigure}
     \hfill
     \begin{subfigure}[b]{\columnwidth}
         \centering
         \includegraphics[width=\textwidth]{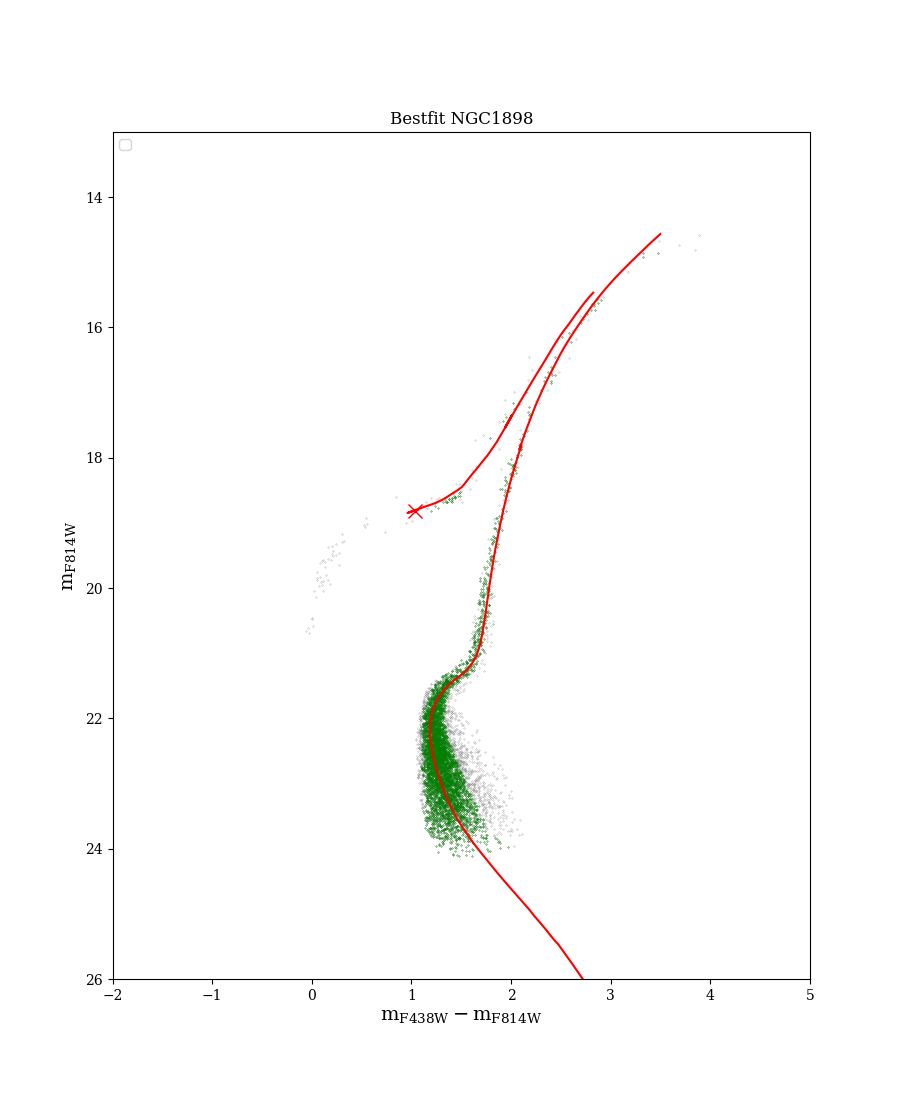}
         \caption{ }
     \end{subfigure}

     \begin{subfigure}[b]{\columnwidth}
         \centering
         \includegraphics[width=\textwidth]{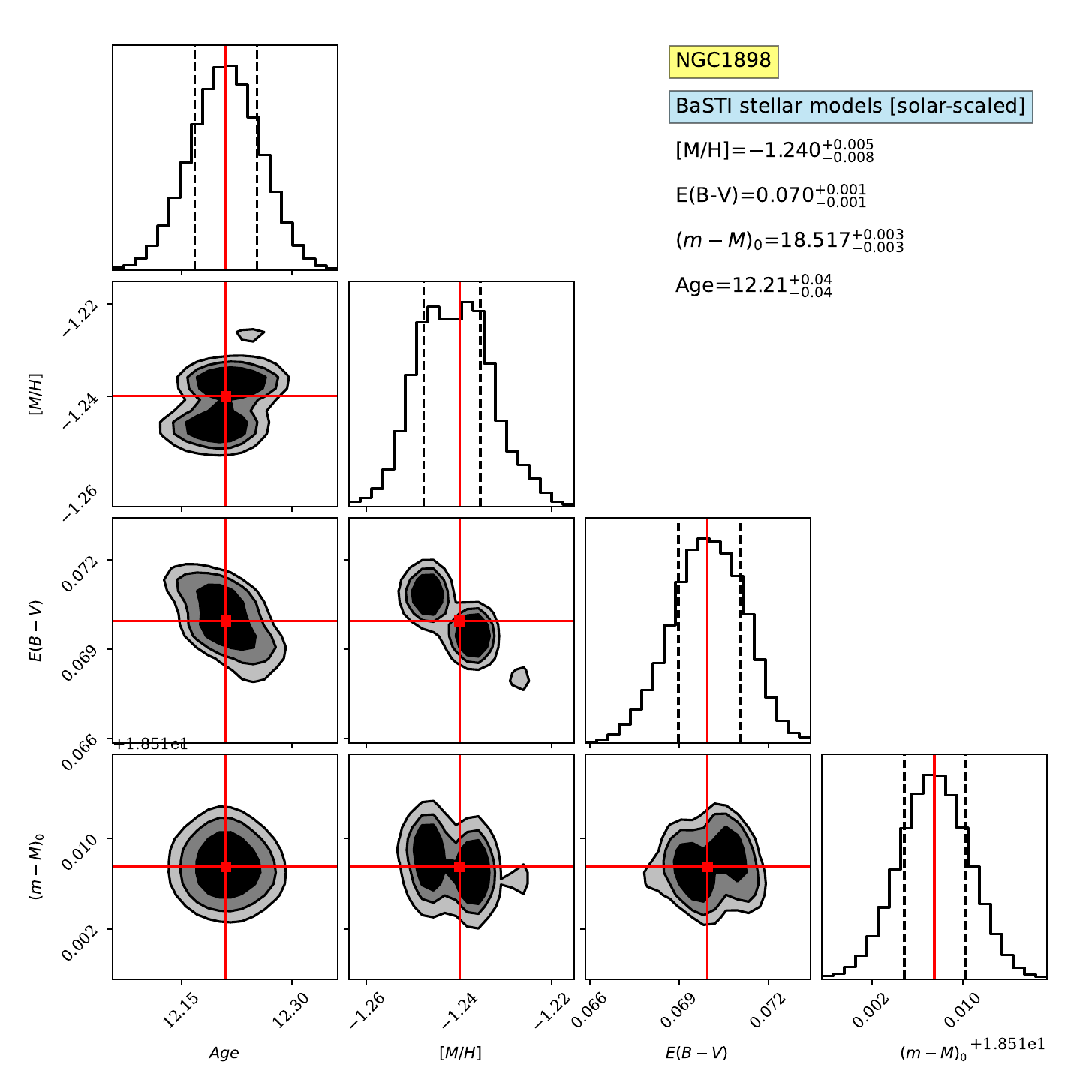}
         \caption{}
     \end{subfigure}
     \hfill
     \begin{subfigure}[b]{\columnwidth}
         \centering
         \includegraphics[width=\textwidth]{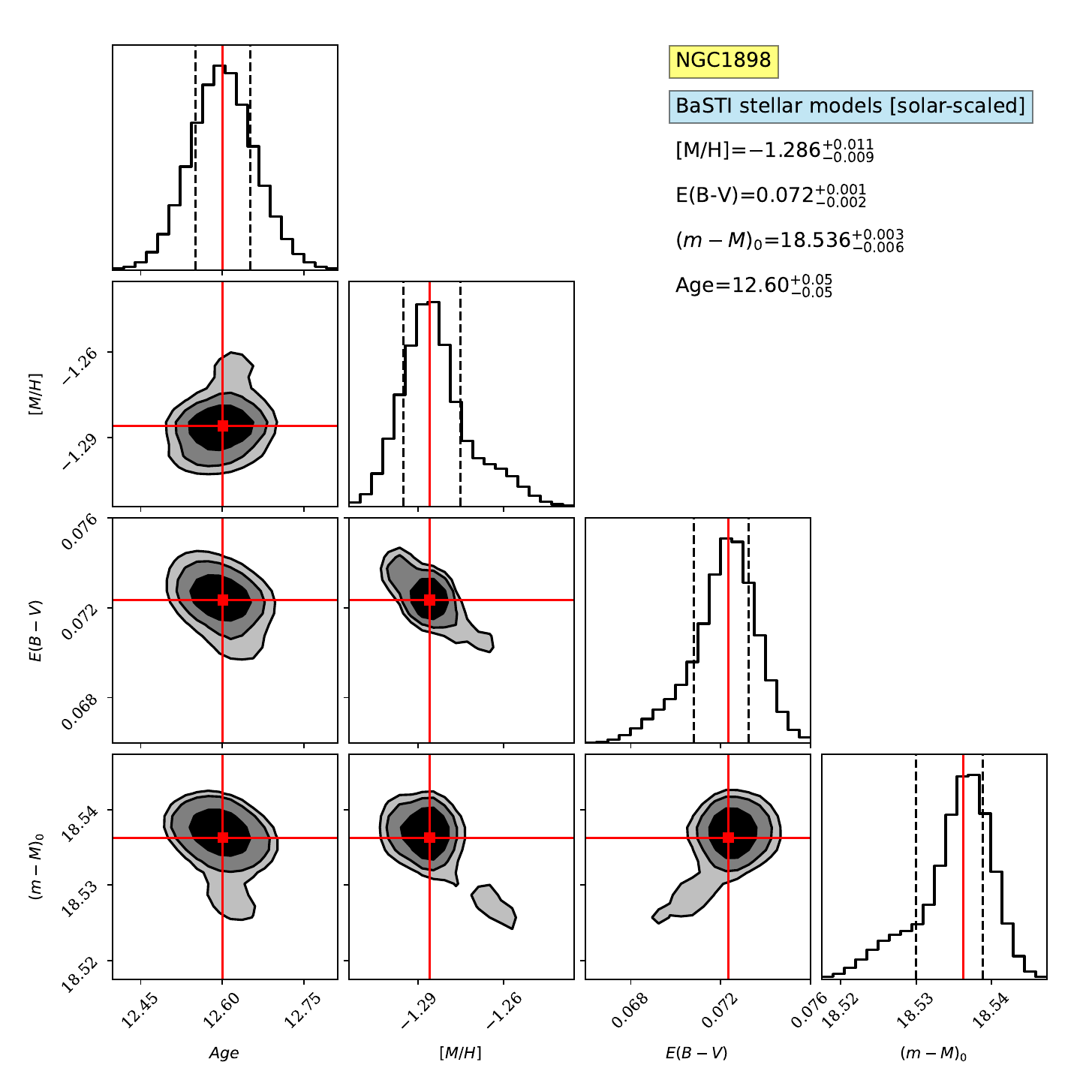}
         \caption{}
     \end{subfigure}
        \caption{
        Same as Fig.~\ref{fig:hodge11_isofit}, but now for NGC~1898 and for the $m_{\rm F438W}$ vs. $m_{\rm F438W}-m_{\rm F814W}$ (left panels) and $m_{\rm F814W}$ vs. $m_{\rm F438W}-m_{\rm F814W}$ (right panels) colour-magnitude combinations. 
        }
        \label{fig:ngc1898_isofit}
\end{figure*}

\begin{figure*}
     \centering
     \begin{subfigure}[b]{\columnwidth}
         \centering
         \includegraphics[width=\textwidth]{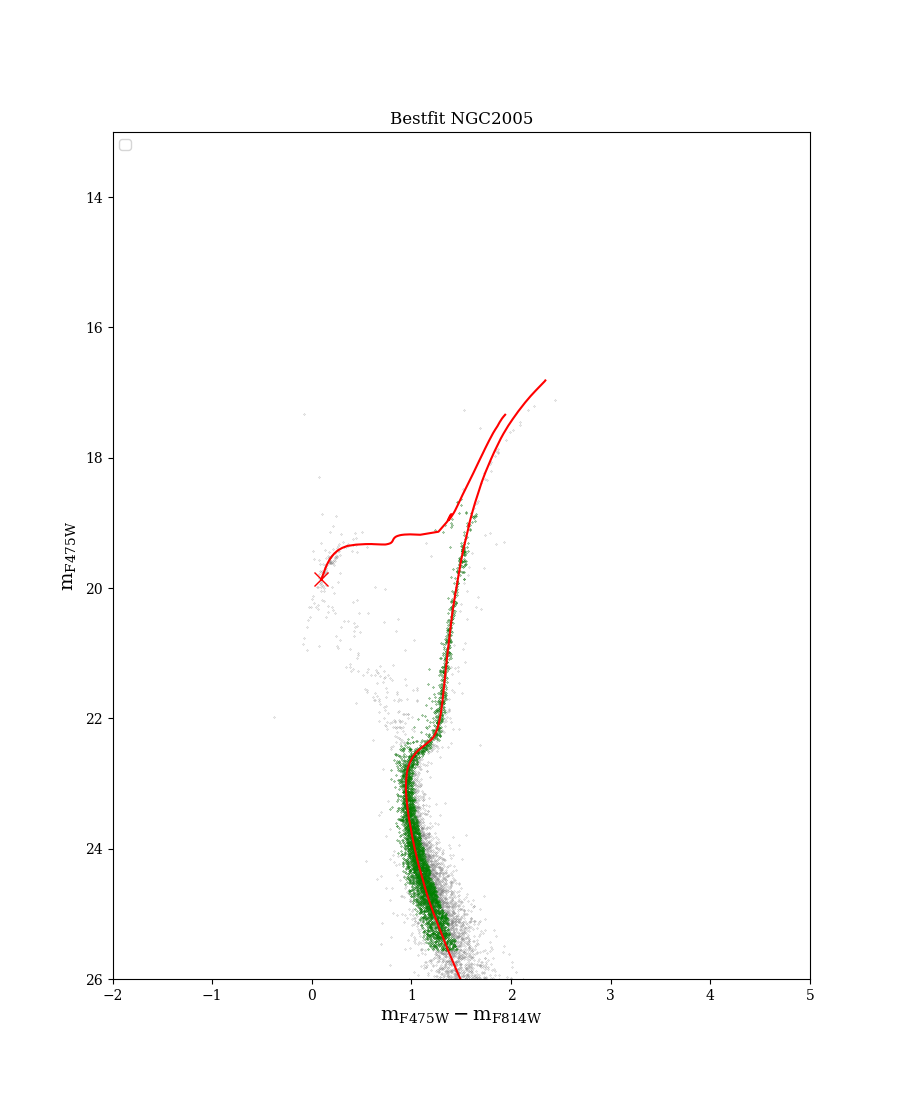}
         \caption{}
     \end{subfigure}
     \hfill
     \begin{subfigure}[b]{\columnwidth}
         \centering
         \includegraphics[width=\textwidth]{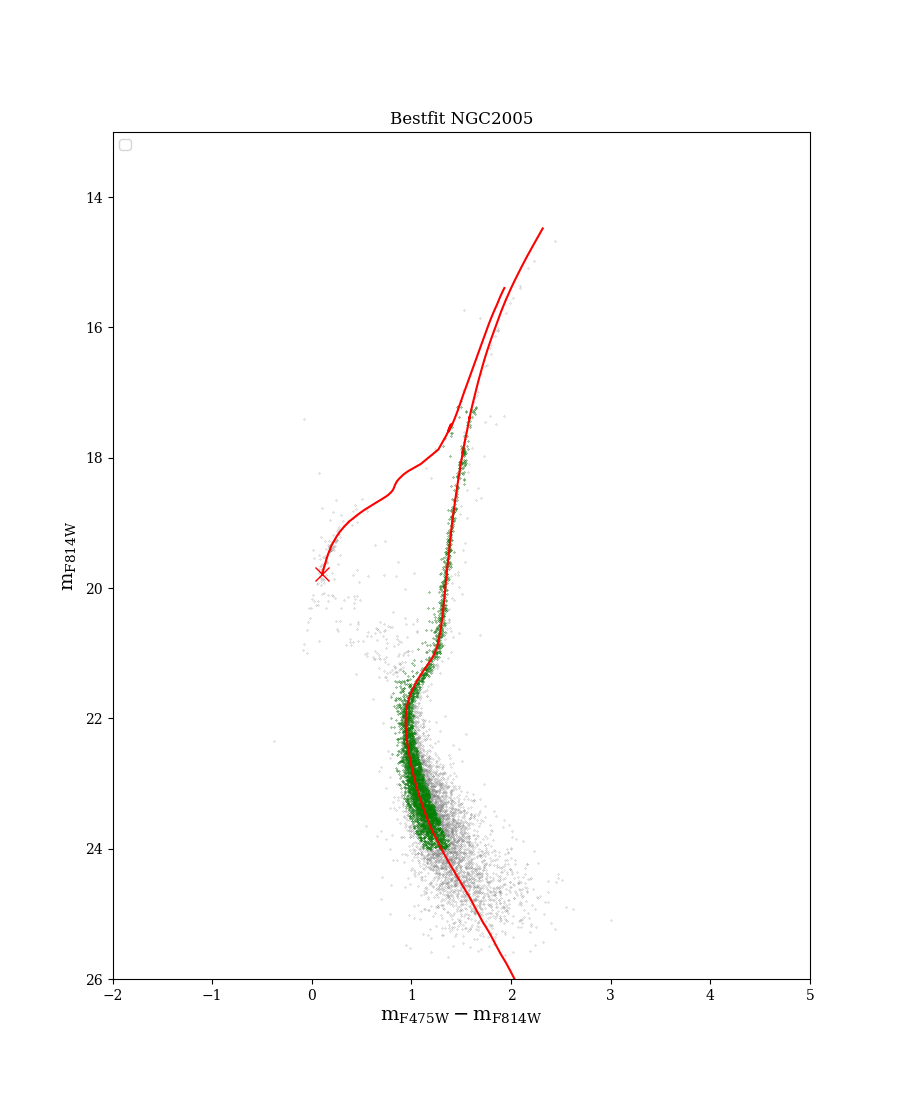}
         \caption{ }
     \end{subfigure}

     \begin{subfigure}[b]{\columnwidth}
         \centering
         \includegraphics[width=\textwidth]{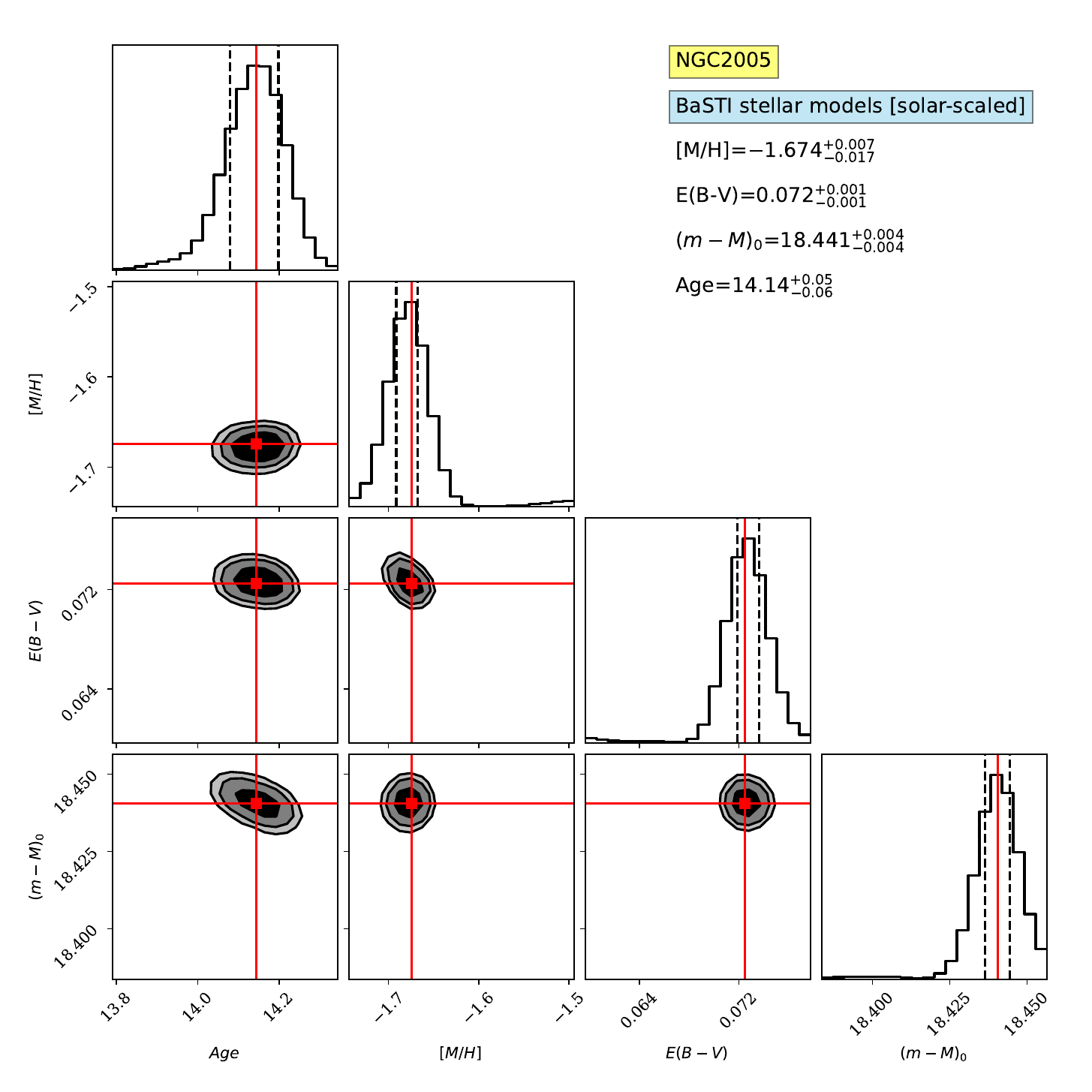}
         \caption{}
     \end{subfigure}
     \hfill
     \begin{subfigure}[b]{\columnwidth}
         \centering
         \includegraphics[width=\textwidth]{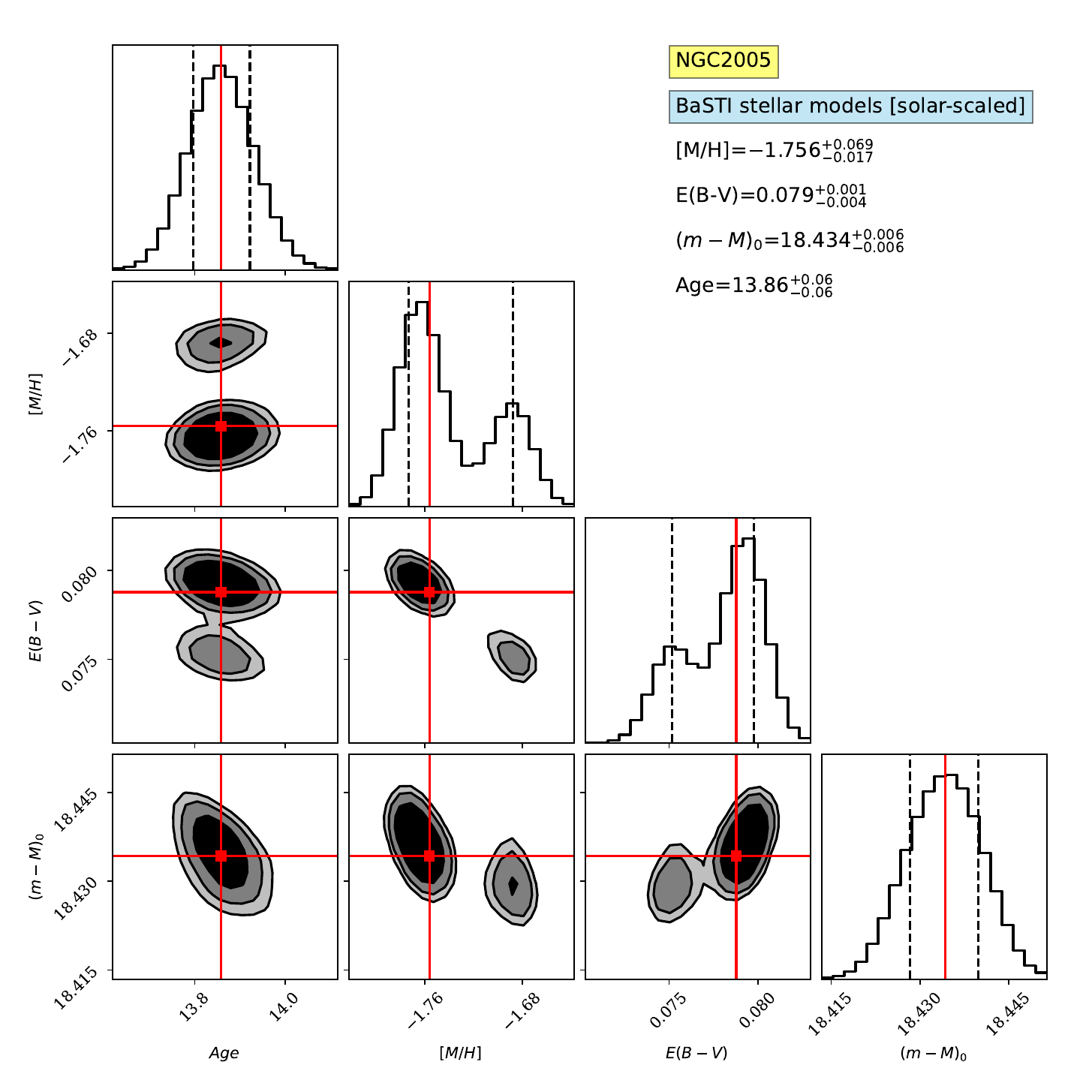}
         \caption{}
     \end{subfigure}
        \caption{
        Same as Fig.~\ref{fig:hodge11_isofit}, but now for NGC~2005 and for the $m_{\rm F475W}$ vs. $m_{\rm F475W}-m_{\rm F814W}$ (left panels) and $m_{\rm F814W}$ vs. $m_{\rm F475W}-m_{\rm F814W}$ (right panels) colour-magnitude combinations. 
        }
        \label{fig:ngc2005_isofit}
\end{figure*}

\begin{figure*}
     \centering
     \begin{subfigure}[b]{\columnwidth}
         \centering
         \includegraphics[width=\textwidth]{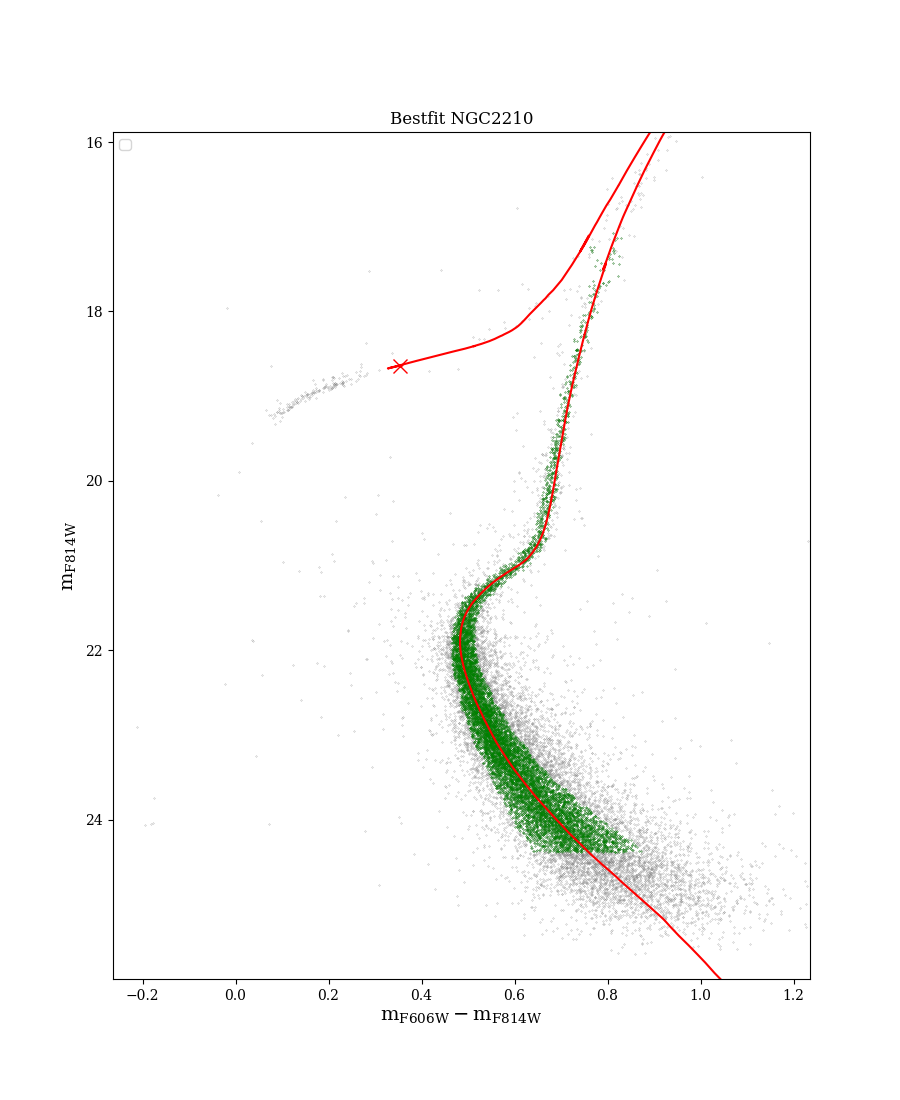}
         \caption{}
     \end{subfigure}
     \hfill
     \begin{subfigure}[b]{\columnwidth}
         \centering
         \includegraphics[width=\textwidth]{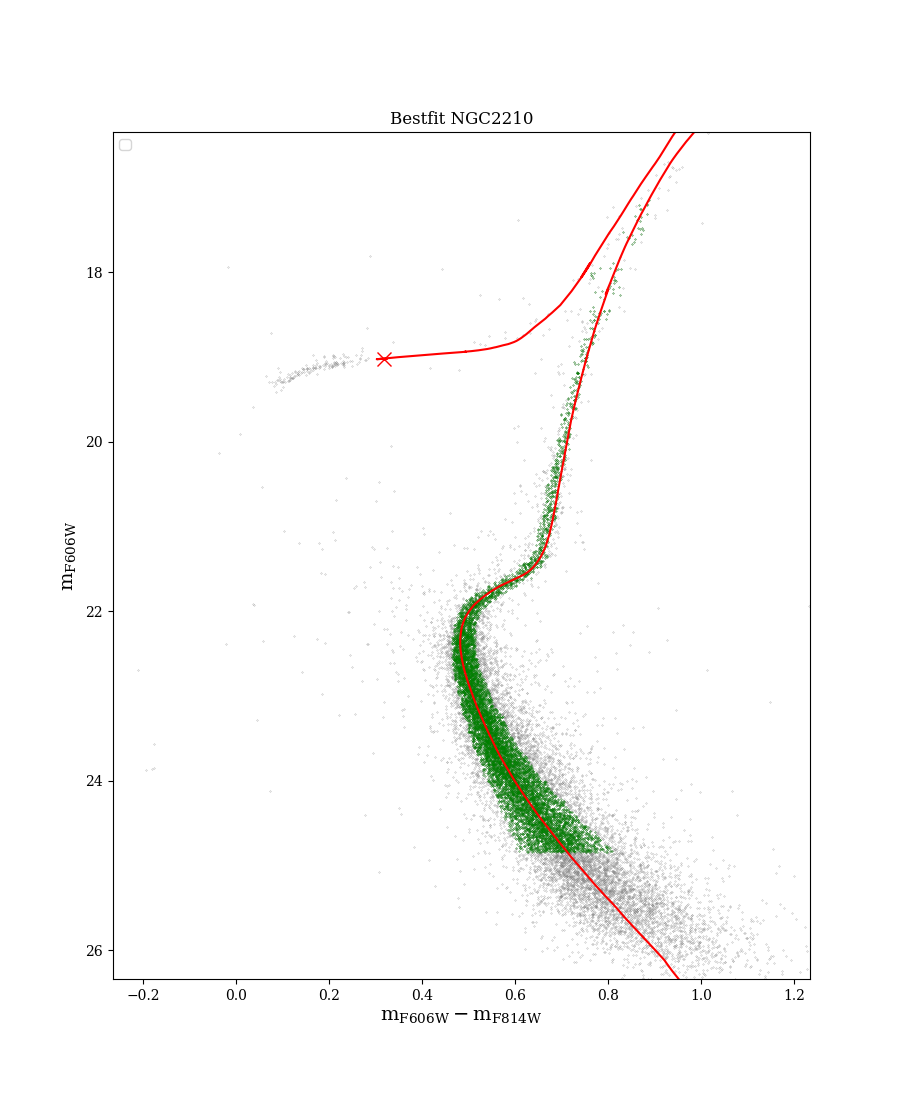}
         \caption{ }
     \end{subfigure}

     \begin{subfigure}[b]{\columnwidth}
         \centering
         \includegraphics[width=\textwidth]{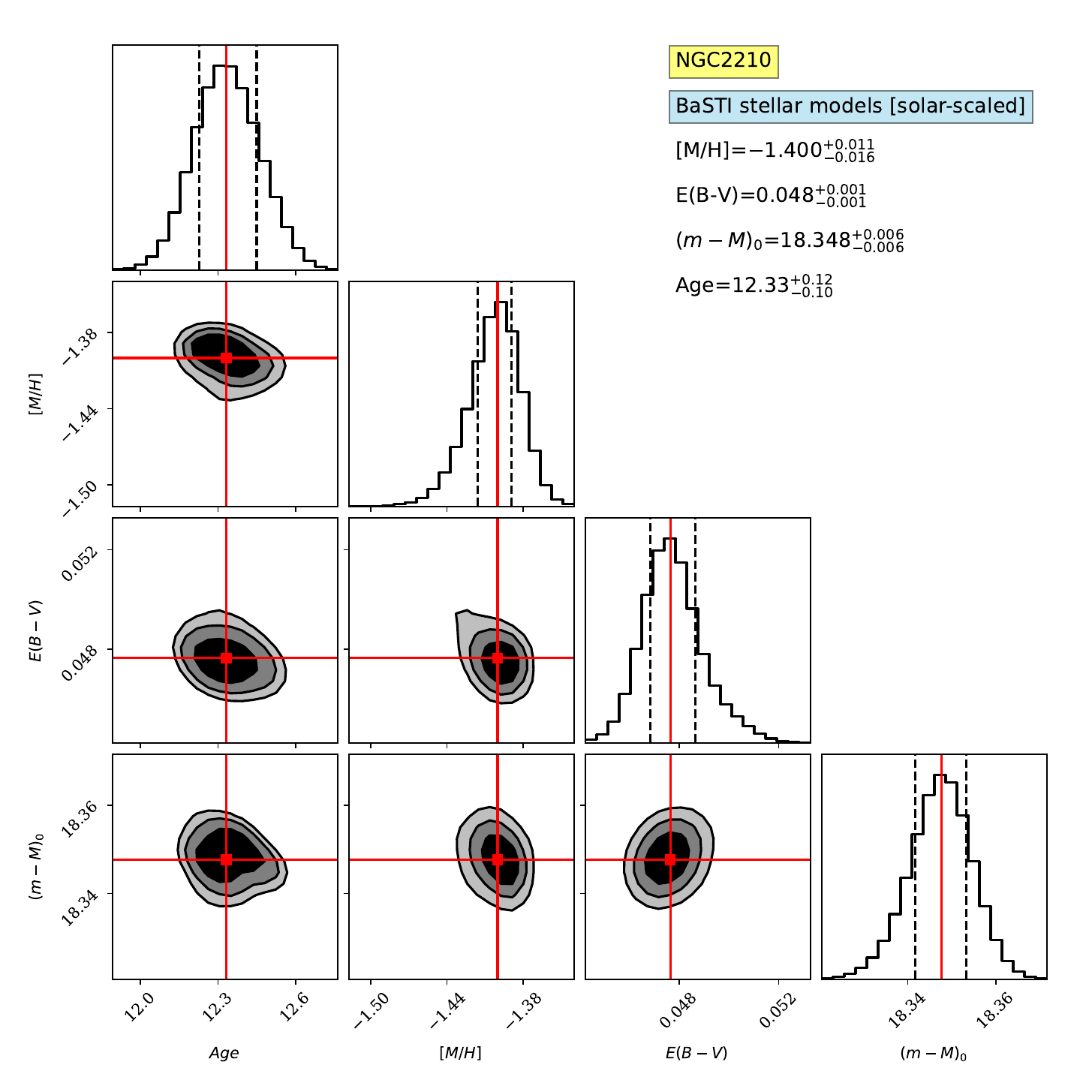}
         \caption{}
     \end{subfigure}
     \hfill
     \begin{subfigure}[b]{\columnwidth}
         \centering
         \includegraphics[width=\textwidth]{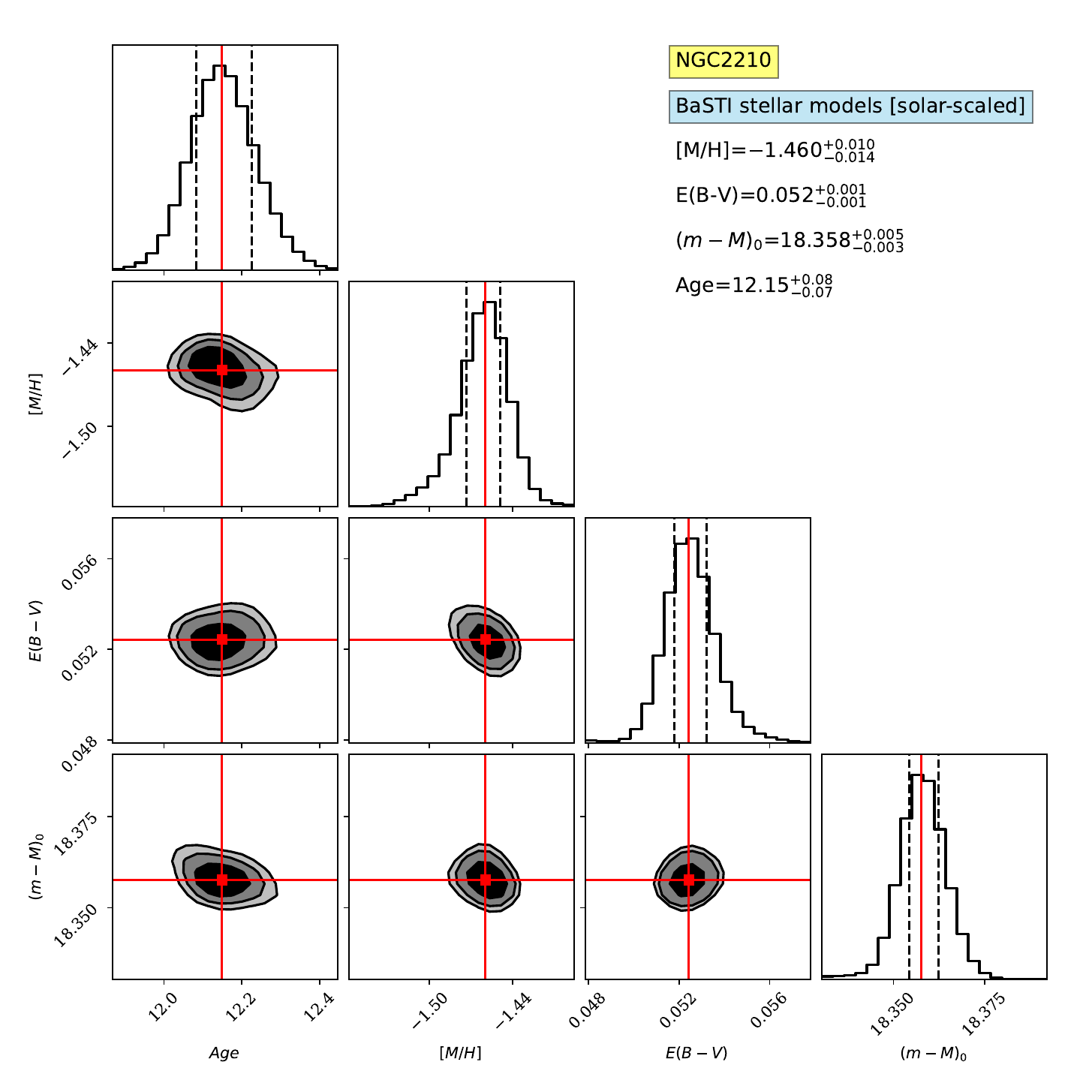}
         \caption{}
     \end{subfigure}
        \caption{Same as Fig.~\ref{fig:hodge11_isofit}, but now for NGC~2210.
        }
        \label{fig:ngc2210_isofit}
\end{figure*}

\begin{figure*}
     \centering
     \begin{subfigure}[b]{\columnwidth}
         \centering
         \includegraphics[width=\textwidth]{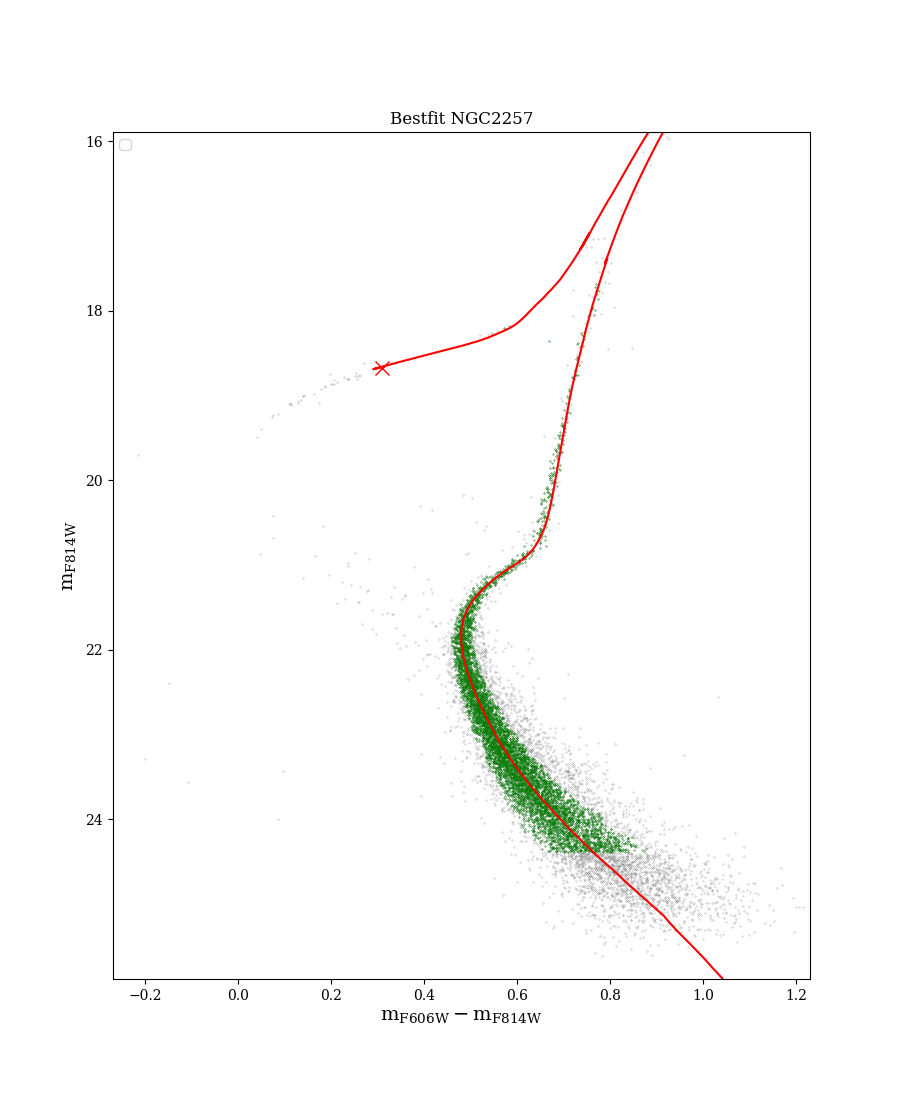}
         \caption{}
     \end{subfigure}
     \hfill
     \begin{subfigure}[b]{\columnwidth}
         \centering
         \includegraphics[width=\textwidth]{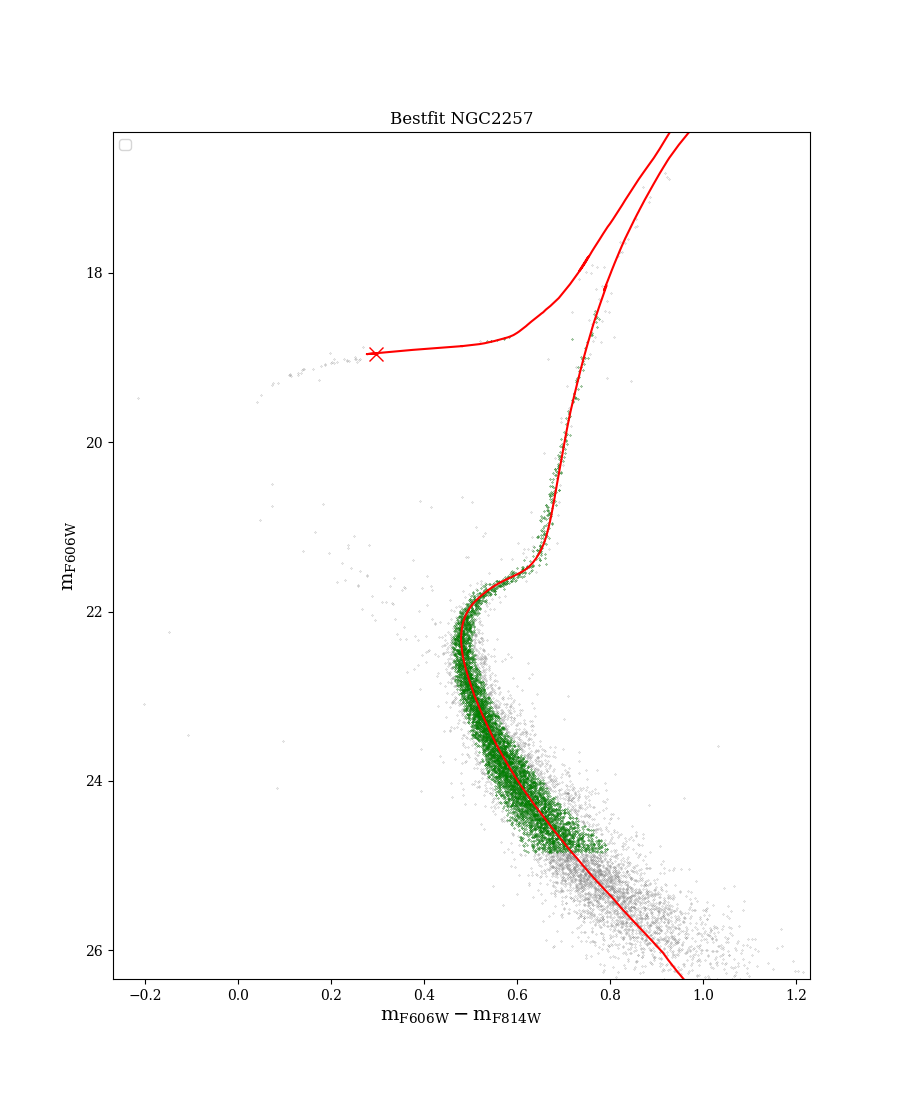}
         \caption{ }
     \end{subfigure}

     \begin{subfigure}[b]{\columnwidth}
         \centering
         \includegraphics[width=\textwidth]{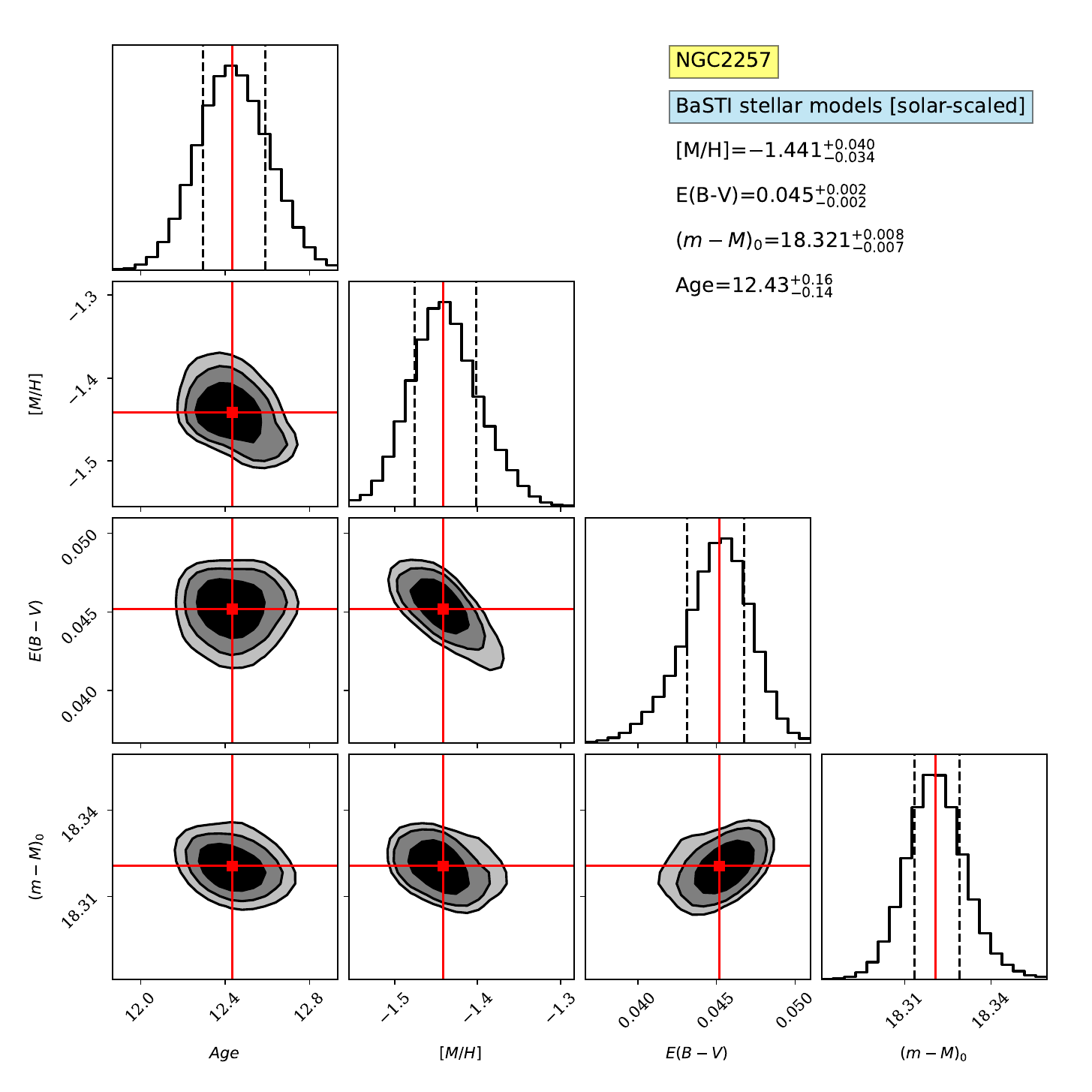}
         \caption{}
     \end{subfigure}
     \hfill
     \begin{subfigure}[b]{\columnwidth}
         \centering
         \includegraphics[width=\textwidth]{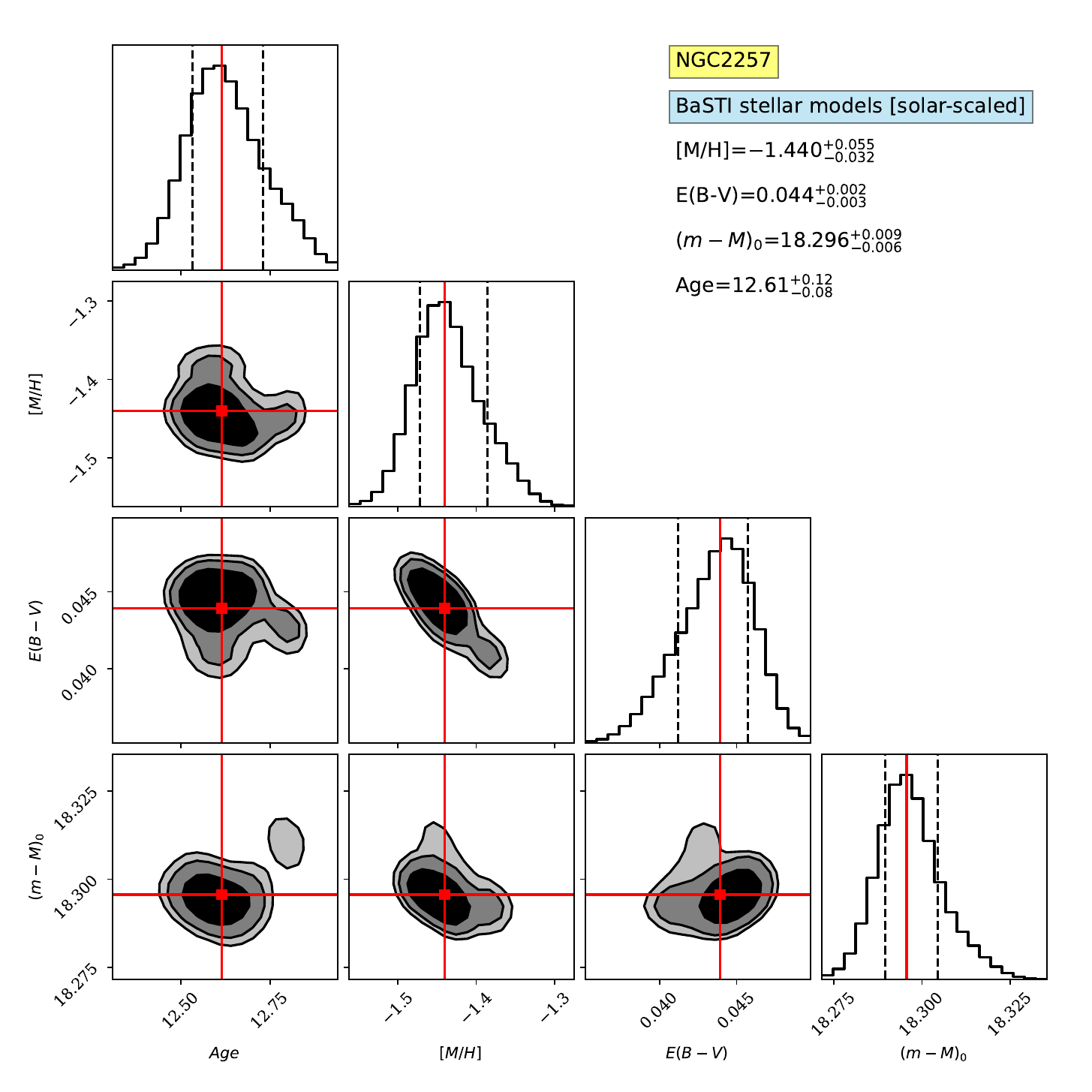}
         \caption{}
     \end{subfigure}
        \caption{Same as Fig.~\ref{fig:hodge11_isofit}, but now for NGC~2257.
        }
        \label{fig:ngc2257_isofit}
\end{figure*}

\begin{figure*}
     \centering
     \begin{subfigure}[b]{\columnwidth}
         \centering
         \includegraphics[width=\textwidth]{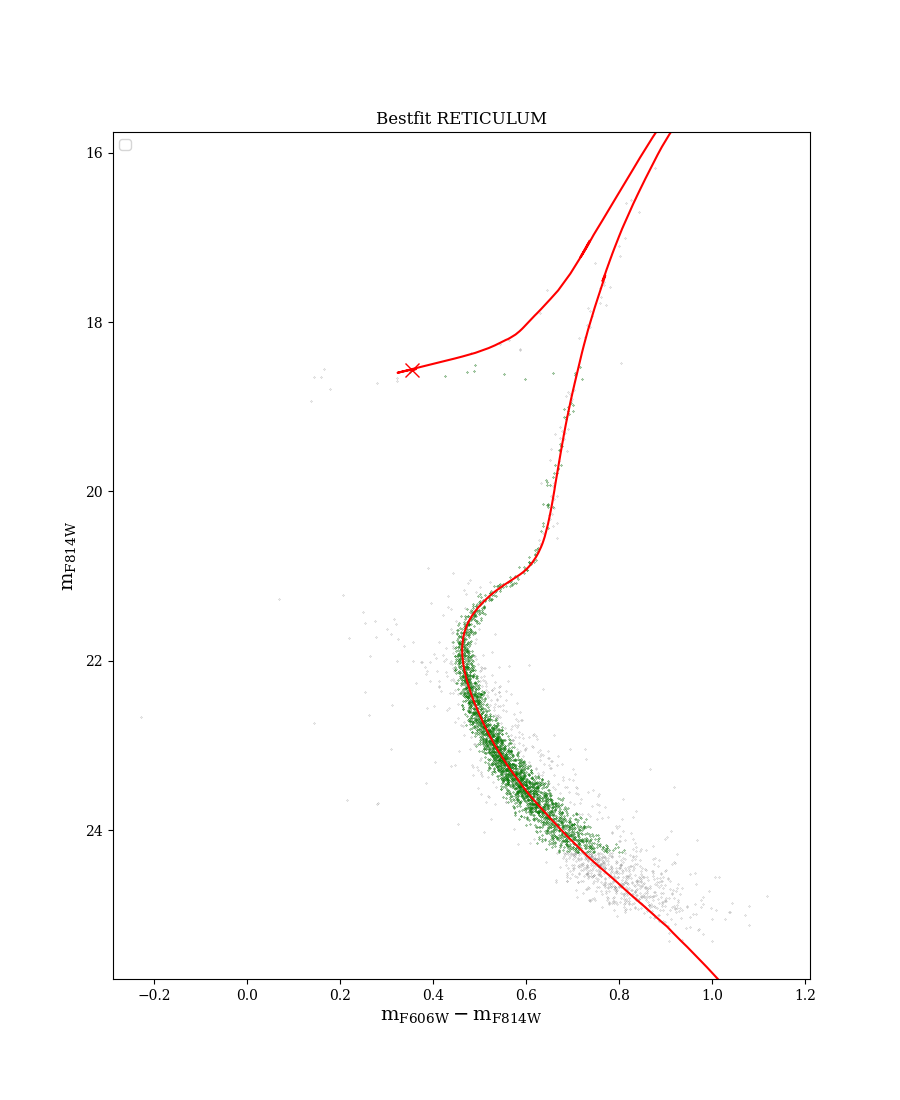}
         \caption{}
     \end{subfigure}
     \hfill
     \begin{subfigure}[b]{\columnwidth}
         \centering
         \includegraphics[width=\textwidth]{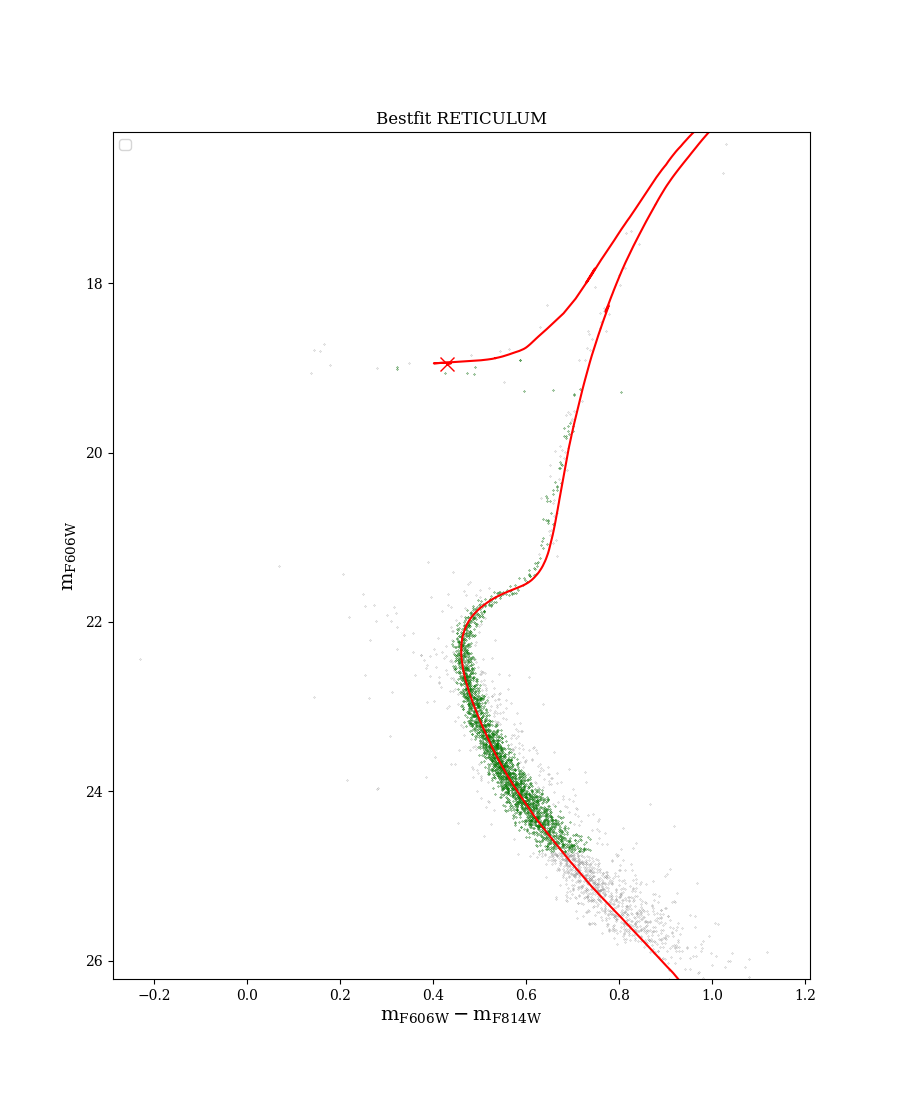}
         \caption{ }
     \end{subfigure}

     \begin{subfigure}[b]{\columnwidth}
         \centering
         \includegraphics[width=\textwidth]{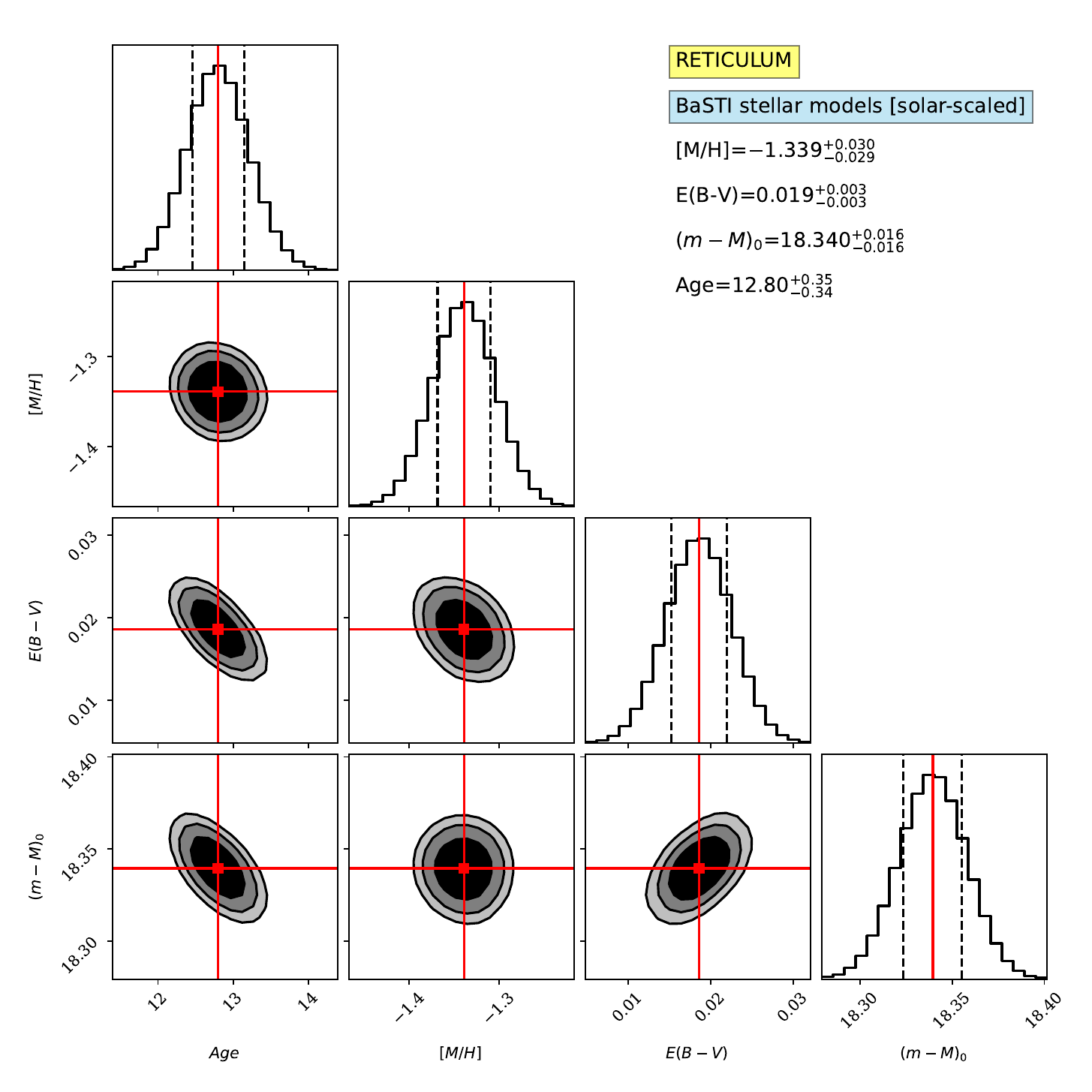}
         \caption{}
     \end{subfigure}
     \hfill
     \begin{subfigure}[b]{\columnwidth}
         \centering
         \includegraphics[width=\textwidth]{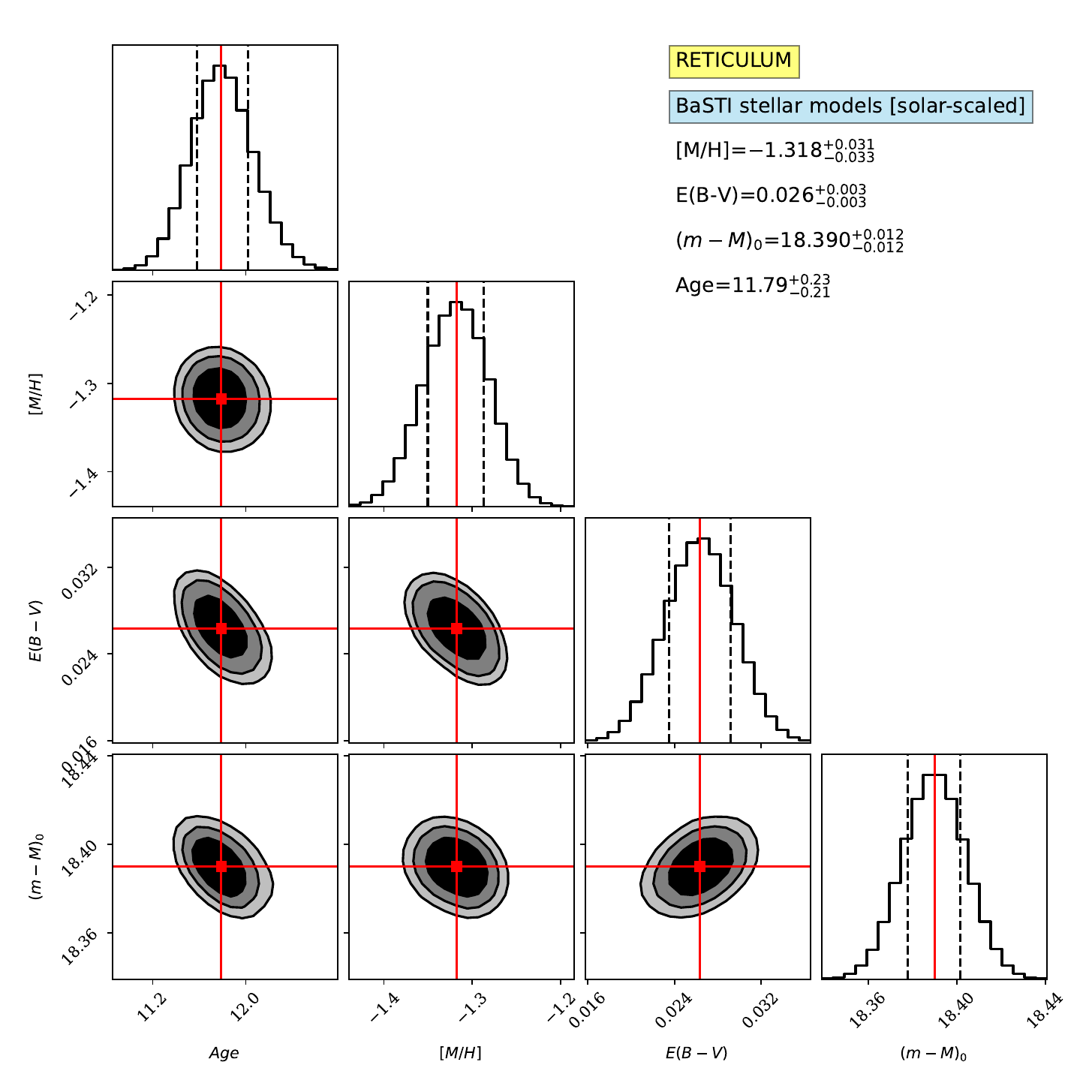}
         \caption{}
     \end{subfigure}
        \caption{Same as Fig.~\ref{fig:hodge11_isofit}, but now for Reticulum.
        }
        \label{fig:reticulum_isofit}
\end{figure*}

Figs.~\ref{fig:hodge11_isofit}--\ref{fig:reticulum_isofit} illustrate the results of the isochrone fitting routine for the seven old LMC clusters analysed in this study. For each cluster, the upper two panels show their CMDs, resulting from the two different combinations of the available filters. Overplotted as a red solid line is the best-fitting isochrone model for that respective colour-magnitude space. Stars actually used for the fit are highlighted in green. The lower two panels show corner plots of the posterior distribution of the model parameters and their pairwise correlations.

\section{Isochrone comparisons}\label{sec:isoc_comparsion}

\begin{figure*}
\begin{tabular}{ccc}
\includegraphics[width=0.32\textwidth]{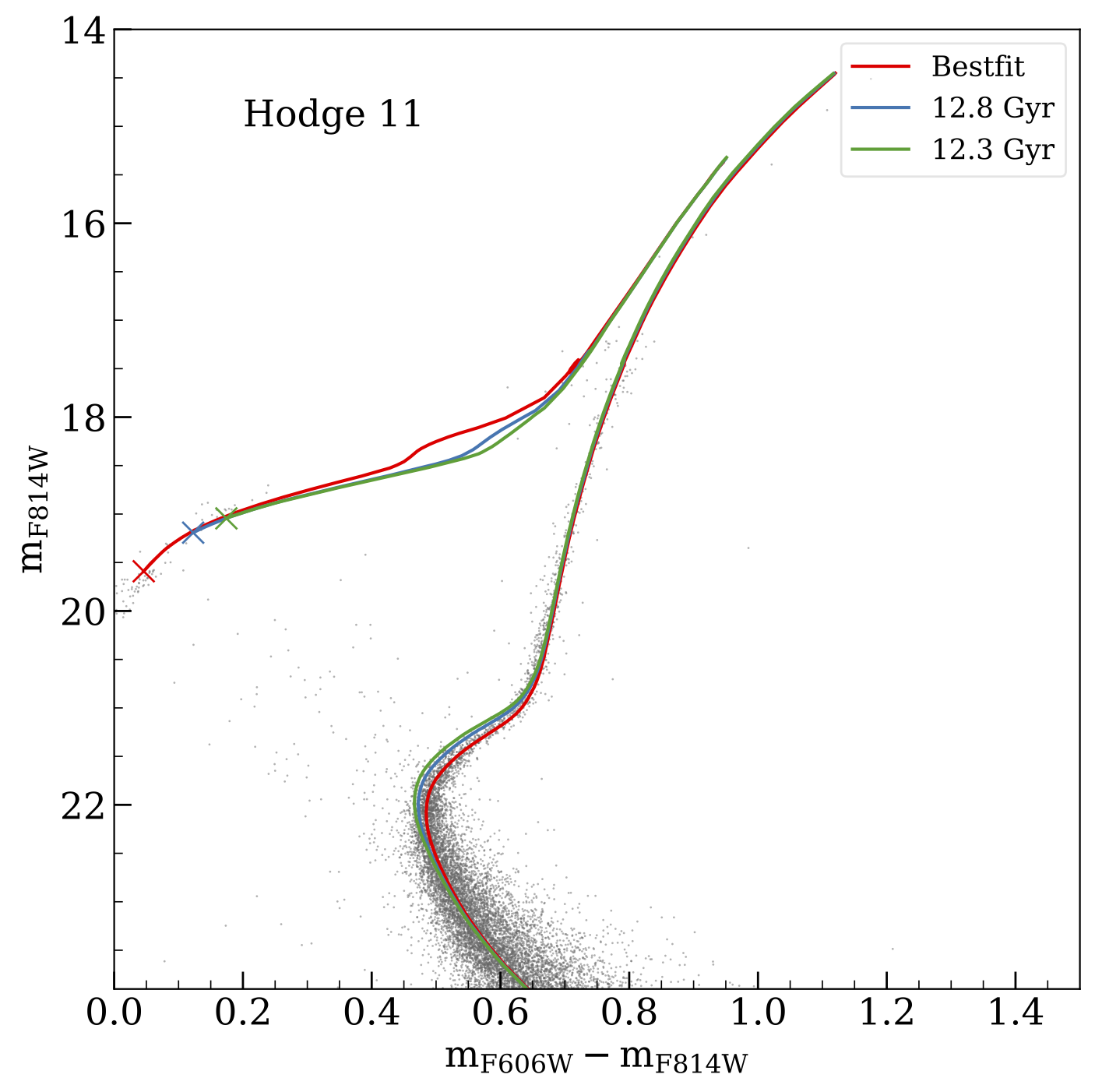} &
\includegraphics[width=0.32\textwidth]{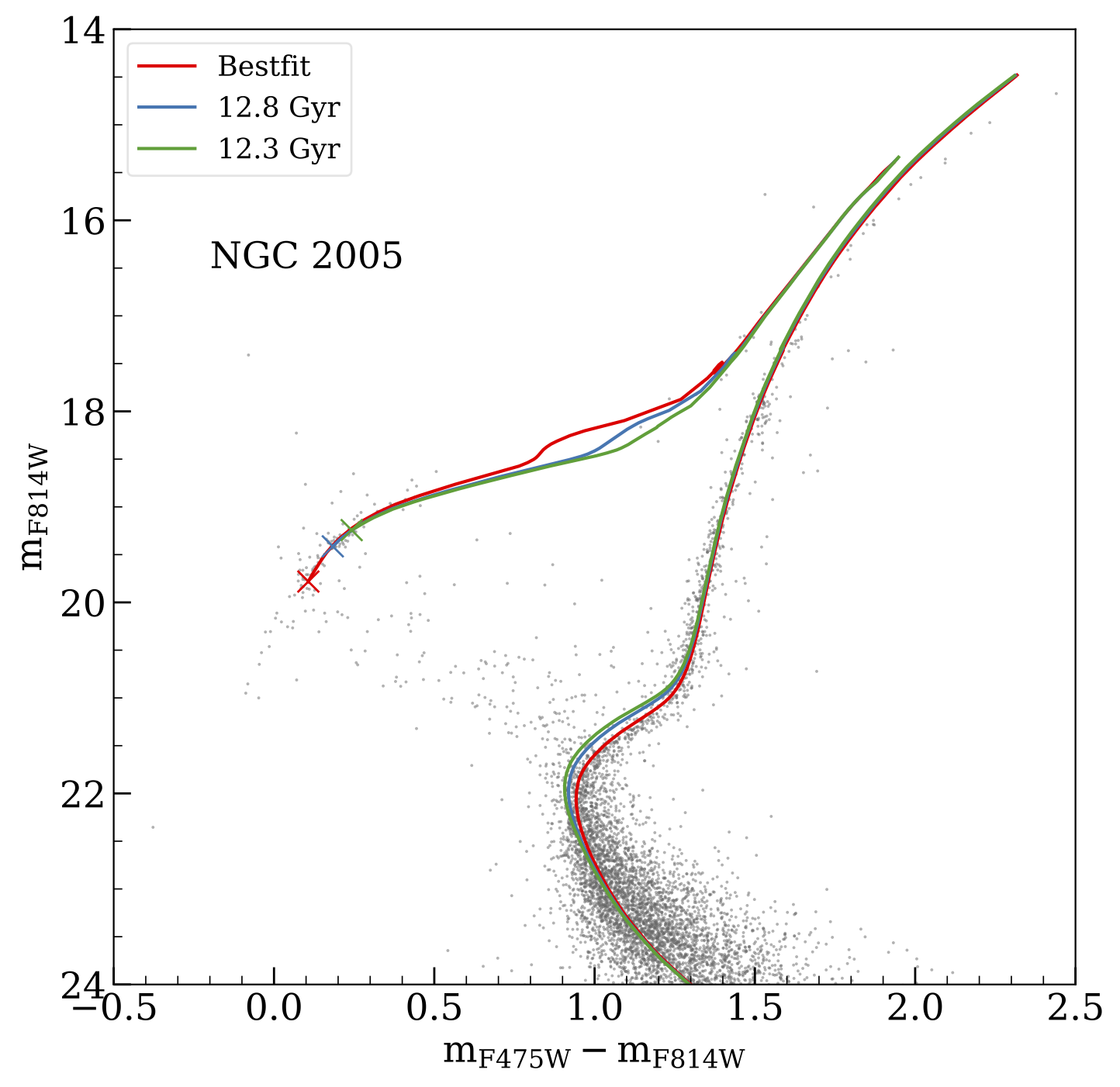} &
\includegraphics[width=0.32\textwidth]{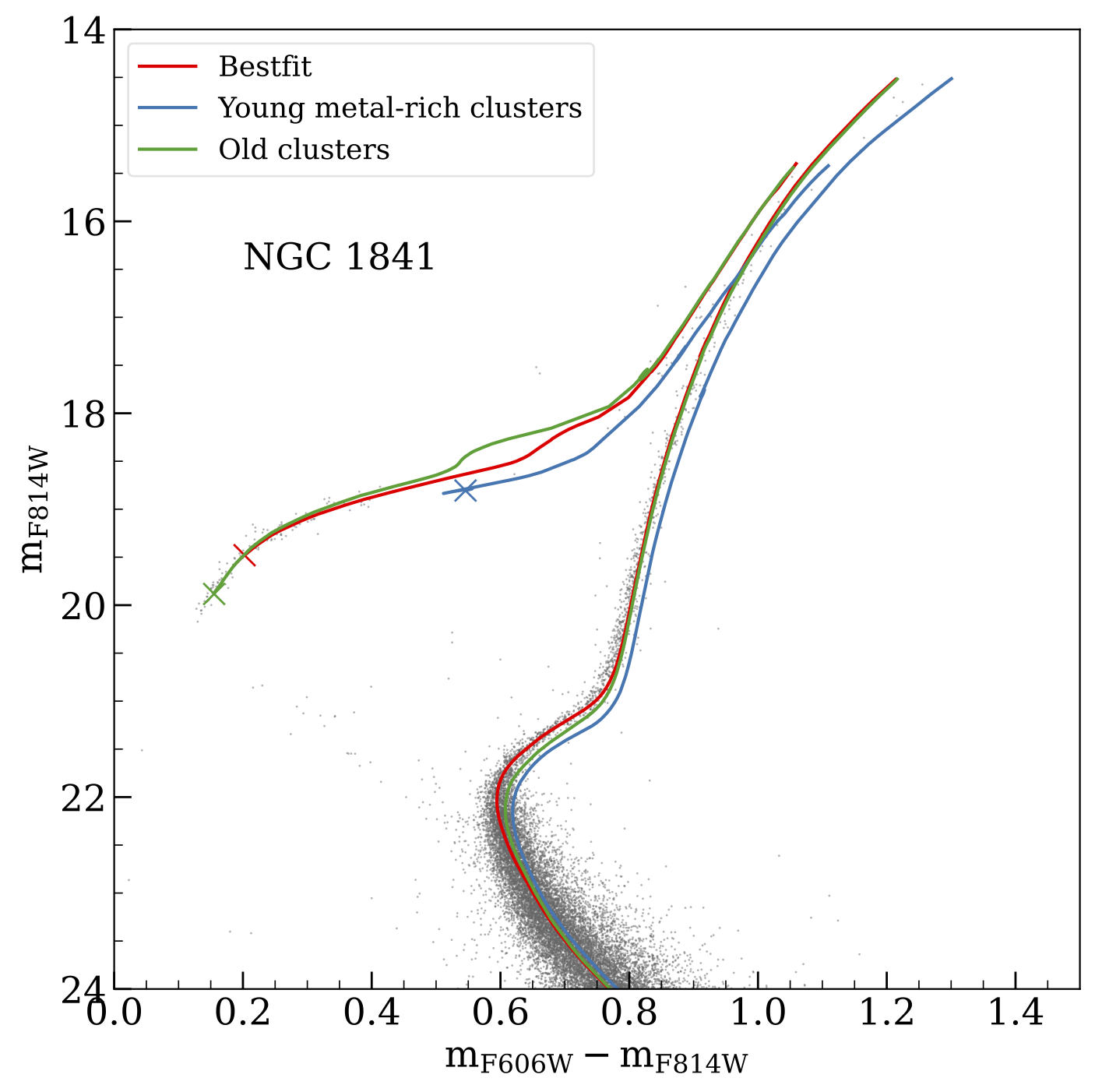} \\
\end{tabular}
\caption{Observed CMDs of Hodge~11 (left panel), NGC~2005 (middle panel) and NGC~1841 (right panel), together with isochrones of different ages and metallicities (see text).}
\label{fig:cluster_isoc_compar}

\end{figure*}

In Fig.~\ref{fig:cluster_isoc_compar} we illustrate the observed CMDs of Hodge~11, NGC~2005 and NGC~1841 together with isochrones of different ages and metallicities. In the left and middle panels, we compare the CMDs of Hodge~11 and NGC~2005, respectively, with isochrones of three different ages: one with the best-fit parameters (red curves) and two that have the same parameters as the best-fit one but are younger, 12.3~Gyr (green curve) and 12.8~Gyr (blue curve). For both clusters, the younger isochrones are able to fit the red-giant branch as well as the horizontal branch, but miss the main-sequence turn-off, suggesting that Hodge~11 and NGC 2005 are indeed older.  
In the right panel, we compare for NGC~1841 the best-fitting isochrone (red curve) with an isochrone that resembles the old-metal-poor clusters, i.e. same metallicity as NGC~1841 but with an age similar to Hodge~11 (green line) and with an isochrone that resembles the younger metal-rich clusters in the disc, i.e. same age as NGC~1841 but higher metallicity (blue line). The old, metal-poor isochrone fits the red-giant branch and the horizontal branch of the cluster but misses the turn-off, suggesting a younger age. The metal-rich isochrone misses the turn-off as well as the slope of the red-giant branch, indicating that NGC~1841 is more metal-poor.

\end{appendix}

\end{document}